\documentclass[11pt,a4paper]{article}
\pdfoutput=1

\usepackage{jheppub}

\usepackage{amssymb}
\usepackage{amsfonts}
\usepackage{graphicx}
\usepackage{epstopdf}
\usepackage{dcolumn}
\usepackage{amsmath}
\usepackage{latexsym,bm}
\usepackage{amsthm}
\usepackage{slashed}
\usepackage{float}
\usepackage{color}
\usepackage{url}
\usepackage{longtable}
\usepackage{subfig}
\usepackage{tikz}
\usepackage[all,cmtip]{xy}

\usepackage{multirow}
\usepackage{longtable}

\usepackage[titletoc]{appendix}
\makeatletter
\newtheorem*{rep@theorem}{\rep@title}
\newcommand{\newreptheorem}[2]{%
\newenvironment{rep#1}[1]{%
 \def\rep@title{#2 \ref{##1}}%
 \begin{rep@theorem}}%
 {\end{rep@theorem}}}
\makeatother

\newreptheorem{lemma}{Lemma}

\newreptheorem{conj}{Conjecture}

\theoremstyle{definition}

\newcommand \xoverline[2][0.75]{
    \sbox{\myboxA}{$\m@th#2$}
    \setbox\myboxB\null
    \ht\myboxB=\ht\myboxA
    \dp\myboxB=\dp\myboxA
    \wd\myboxB=#1\wd\myboxA
    \sbox\myboxB{$\m@th\overline{\copy\myboxB}$}
    \setlength\mylenA{\the\wd\myboxA}
    \addtolength\mylenA{-\the\wd\myboxB}
    \ifdim\wd\myboxB<\wd\myboxA
       \rlap{\hskip 0.5\mylenA\usebox\myboxB}{\usebox\myboxA}%
    \else
        \hskip -0.5\mylenA\rlap{\usebox\myboxA}{\hskip 0.5\mylenA\usebox\myboxB}%
    \fi}
\makeatother

\newcommand{\ba}{\begin{aligned}}
\newcommand{\ea}{\end{aligned}}
\def\be{\begin{equation}}
\def\ee{\end{equation}}
\def\bsp{\begin{split}}
\def\esp{\end{split}}
\def\bea{\begin{eqnarray}}
\def\eea{\end{eqnarray}}

\def\mc{\mathcal}
\def\ptl{\partial}
\def\mb{\mathbb}

\def \bp{\begin{pmatrix}}
\def\ep{\end{pmatrix}}

\def\R{\mathbb{R}}
\def\N{\mathcal{N}}
\def\P{\mathbb{P}}
\def\C{\mathbb{C}}
\def\Z{\mathbb{Z}}

\def\O{\mathcal{O}}

\def\im{\mathrm{im}}

\def\br{\breve}
\newcommand{\vol}{\mathrm{vol}}

\usepackage{tikz}
\usepackage{diagbox}
\usetikzlibrary{positioning}
\usetikzlibrary{calc}
\usetikzlibrary{decorations.pathreplacing,calligraphy}

\usepackage{xstring}
\usetikzlibrary{decorations.pathmorphing} 
\usetikzlibrary{decorations.markings} 
\usetikzlibrary{snakes}
\usetikzlibrary{arrows} 
\usetikzlibrary{shapes} 
\usetikzlibrary{matrix} 
\usetikzlibrary{positioning} 
\usepackage[english]{babel} 
\usepackage[autostyle]{csquotes}

\usepackage{tikz}
\usetikzlibrary{arrows}
\usetikzlibrary{arrows.meta}
\usetikzlibrary{shapes.geometric,calc,arrows, positioning,shapes.misc,decorations.markings}
\tikzset{
  big arrow/.style={
    decoration={markings,mark=at position 1 with {\arrow[scale=2,#1]{>}}},
    postaction={decorate},
    shorten >=0.4pt},
  big arrow/.default=black}
  
\pgfdeclarelayer{edgelayer}
\pgfdeclarelayer{nodelayer}
\pgfsetlayers{edgelayer,nodelayer,main} 
\tikzstyle{none}=[inner sep=0pt] 

\tikzstyle{NodeCross}=[draw, shape=circle, cross out, inner sep=0pt, minimum size=6pt,line width=0.25mm]
\tikzstyle{Circle}=[draw, shape=circle, black,  fill=black, inner sep=0pt, minimum size=6pt]
\tikzstyle{Star}=[draw, shape=star, fill=black, star points=8, inner sep=0pt, minimum size=8pt]

\tikzstyle{DashedLine}=[-, densely dashed, line width=0.25mm]
\tikzstyle{DottedLine}=[-, dotted, line width=0.25mm]
\tikzstyle{ThickLine}=[-, line width=0.25mm]
\tikzstyle{ArrowLineRight}=[-, -{Stealth[scale=1.75]}, line width=0.1mm, scale=5]
\tikzstyle{RedLine}=[-, draw={rgb,255: red,191; green,0; blue,0}, fill=none, line width=0.25mm]
\tikzstyle{DottedRed}=[-, dotted, draw={rgb,255: red,191; green,0; blue,0}, fill=none, line width=0.25mm]
\tikzstyle{DashedLineThin}=[-, densely dashed, line width=0.125mm, fill=none, draw=black]
\tikzstyle{ArrowLineRed}=[-, -{Stealth[scale=1.75]}, draw={rgb,255: red,191; green,0; blue,0}, line width=0.1mm, scale=5]
\tikzstyle{brane}=[draw]
\tikzset{D7/.style={circle, draw=black, inner sep=0pt, fill=white, minimum size=3mm}}
\tikzset{hasse/.style={circle, fill,inner sep=2pt}}
\tikzset{flavor/.style={regular polygon,fill=white,regular polygon sides=4,inner sep=2.5pt, draw}}
\tikzset{gauge/.style={circle, draw,inner sep=2.5pt}}
\tikzset{gaugeb/.style={circle, draw,fill=black,inner sep=2.5pt}}
\tikzset{gauger/.style={circle, draw,fill=cyan,inner sep=2.5pt}}
\tikzset{gaugeg/.style={circle, draw,fill=red,inner sep=2.5pt}}
\tikzset{SUd/.style={circle, draw=black, inner sep=0pt, fill=yellow, minimum size=2mm}}
\tikzset{bd/.style={circle, draw=black, inner sep=0pt, fill=black, minimum size=2mm}}
\tikzset{wd/.style={circle, draw=black, inner sep=0pt, fill=white, minimum size=2mm}}
\tikzset{Dynkin/.style={circle, draw=black, inner sep=0pt, fill=white, minimum size=2mm}}
\tikzstyle{ligne}=[draw, thick] 
\tikzset{doublearrow/.style={ draw=black!75, color=black!75, thick, double distance=3pt, }} 

\newcommand\restr[2]{{
  \left.\kern-\nulldelimiterspace 
  #1 
  \littletaller 
  \right|_{#2} 
  }}

\newcommand{\littletaller}{\mathchoice{\vphantom{\big|}}{}{}{}}

\usepackage{mathrsfs}

\usetikzlibrary{positioning}

\usepackage{tikz-cd}

\usepackage{tikz}
\usepackage{tikz-3dplot}

\def\CC{\mathcal{C}}

\usepackage{enumitem} 

\usepackage{physics}

\usepackage{chemfig}
\usepackage{import}

\makeatletter
\newcommand\xleftrightarrow[2][]{%
  \ext@arrow 9999{\longleftrightarrowfill@}{#1}{#2}}
\newcommand\longleftrightarrowfill@{%
  \arrowfill@\leftarrow\relbar\rightarrow}
\makeatother

\usepackage{booktabs}

\usepackage{braket}

\graphicspath{ {figs/} }

\title{Confinement of 3d $\mathcal{N}=2$ Gauge Theories from M-theory on CY4}

\preprint{\today \hspace*{0.1in} }

\author[\,\clubsuit]{Marwan Najjar}
\emailAdd{marwan.najjar@pku.edu.cn}
\author[\spadesuit,\clubsuit]{Yi-Nan Wang}
\emailAdd{ynwang@pku.edu.cn}

\affiliation[\clubsuit]{Center for High Energy Physics, Peking University, \protect\\
Beijing 100871, China}

\affiliation[\spadesuit]{School of Physics, Peking University, \protect\\
Beijing 100871, China}

\abstract{In this work, we present a new geometric transition in non-compact Calabi-Yau 4-folds, specifically for the cone over the 7d Sasaki-Einstein manifold $Q^{\scriptscriptstyle(1,1,1)}/\mathbb{Z}_{N}$. We discover a new smoothing of such Calabi-Yau 4-fold singularity via a partial resolution+deformation, which can be interpreted as a confined phase for a 3d $\mathcal{N}=2$ $SU(N)$ gauge theory. The confining strings are realized as M2-branes wrapping the torsional 1-cycles in this new geometric phase. We have also computed the generalized global symmetries, including finite $(-1)$-form symmetries, and SymTFT action using the link topology and intersection numbers of the resolved Calabi-Yau 4-fold.}

\begin{document}

\maketitle

\section{Introduction and summary}

The understanding of confinement in gauge theory is one of the most fundamental questions in theoretical physics. Regarding the confinement phenomena in supersymmetric field theories, the string/M-theory framework provides an elegant way to realize the confinement/deconfinement phase transition in form of geometric transitions. A famous example is the realization of confinement in 4d $\mc{N}=1$ super Yang-Mills theory, via M-theory on spaces with $G_2$ holonomy~\cite{Acharya:2000gb,Atiyah:2001qf,Acharya:2001hq,Acharya:2004qe}. Shortly speaking, M-theory on $\mathbb{S}^{3}\times\mb{R}^4/\Gamma_{ADE}$ ($\Gamma_{ADE}$ is a finite subgroup of $SU(2)$) gives rise to the $G=A,D,E$ gauge theory phase, and after a geometric transition, M-theory on $\mb{R}^4\times \mathbb{S}^{3}/\Gamma_{ADE}$ leads to a confined, non-gauge theory phase, where the confining strings in 4d spacetime come from M2-branes wrapping the torsional 1-cycle in $\mathbb{S}^{3}/\Gamma_{ADE}$. More recently, the philosophy has been applied to 5d $\N=1$ gauge theory, see \cite{Acharya:2024bnt}. Similar strategies have been applied in the context of AdS/CFT constructions, as demonstrated, for example, in \cite{Oh:2001bf,Dasgupta:2001fg}.

In this paper, we give the first attempt to realize the confinement/deconfinement phase transition in 3d $\mc{N}=2$ gauge theories that are constructed from M-theory on local Calabi-Yau fourfolds. Such models were initially studied in \cite{Diaconescu:1998ua,Gukov:1999ya,Intriligator:2012ue}, and the dictionary between the CY4 geometry and field theory was made more precise in \cite{Najjar:2023hee}\footnote{For 3d $\mc{N}=2$ rank-0 theories one can refer to \cite{Sangiovanni:2024nfz}.}.

We start with a canonical fourfold singularity $X_4$, whose crepant resolution $\widetilde{X}_4$ has a $\mb{P}^1$ ruling structure, which would lead to a 3d $\mc{N}=2$ $SU(N)$ gauge theory description~\cite{Diaconescu:1998ua,Intriligator:2012ue,Najjar:2023hee}. Now we propose a new desingularization of $X_4$, denoted as $\overline{X}_4$, that is called ``DR-phase'', because it consists of a partial resolution followed by a partial deformation. On $\overline{X}_4$ there exists non-supersymmetric compact 3-cycles and 5-cycles instead of the compact 6-cycles as in the resolved geometry $\widetilde{X}_4$. As a consequence, M-theory on $\overline{X}_4$ has no gauge theory description, and we conjecture it to be in a confined phase of the 3d $\mc{N}=2$ $SU(N)$ gauge theory.

As the concrete example, we consider $X_4$ to be the cone over the Sasaki-Einstein manifold $Q^{\scriptscriptstyle (1,1,1)}$ defined in \cite{DAuria:1983sda}. $X_4$ has a toric description as well as the non-complete-intersection description~\cite{Oh:1998qi,caibar1999minimal}
\be
\ba
&z_1 z_2-z_3 z_4=0\ ,\ z_5 z_6-z_7 z_8=0\ ,\ z_1 z_7-z_3 z_5=0\cr
&z_4 z_6-z_2 z_8=0\ ,\ z_1 z_4-z_5 z_8=0\ ,\ z_1 z_6-z_3 z_8=0\cr
&z_2 z_3-z_6 z_7=0\ ,\ z_2 z_5-z_4 z_7=0\ ,\ z_1 z_2-z_5 z_6=0\,.
\ea
\ee
Apart from the crepant resolution $\widetilde{X}_4$ with a compact 4-cycle, we discovered the new DR phase desingularization $\overline{X}_4$ defined as
\be
\ba
        z_{1}z_{2}-z_{3}z_{4}&=\epsilon\neq 0
        \cr
        z_5 z_6-z_7 z_8&=-\epsilon\cr
        z_{1}\mu_1+z_8\mu_2&=0\cr
        z_{5}\mu_1+z_4\mu_2&=0\cr
        z_{3}\mu_1+z_6\mu_2&=0\cr
        z_{7}\mu_1+z_2\mu_2&=0
        \,,
\ea
\ee
where $[\mu_1:\mu_2]$ are the projective coordinates of an exceptional $\mb{P}^1$. The exceptional locus of $\overline{X}_4$ is proven to be $\mathbb{S}^{2}\times \mathbb{S}^{3}$.

Then we studied the cone $\mc{C}(Q^{\scriptscriptstyle (1,1,1)}/\mb{Z}_N)$ over the free quotient space $Q^{\scriptscriptstyle (1,1,1)}/\mb{Z}_N$, where the $\mb{Z}_N$ action acts on $z_i$ as ($\lambda\equiv\exp(2\pi i/N)$)
\begin{equation}
    \begin{split}               &(z_{1},z_{2},z_{3},z_{4},z_{5},z_{6},z_{7},z_{8}) 
        \\
        &\qquad \qquad \qquad \qquad \qquad \sim (\lambda \,z_{1},\lambda^{-1} \,z_{2},\lambda \,z_{3},\lambda^{-1} \,z_{4},\,\lambda^{-1} z_{5}, \lambda\,z_{6},\lambda^{-1}\,z_{7}, \lambda\,z_{8})\,.
    \end{split}
\end{equation}
The resolution phase of $\mc{C}(Q^{\scriptscriptstyle (1,1,1)}/\mb{Z}_N)$ gives a 3d $\mc{N}=2$ $SU(N)$ gauge theory. As before we define a the DR-phase smooth geometry $\overline{\mc{C}(Q^{\scriptscriptstyle (1,1,1)}/\mb{Z}_N)}$, which has the exceptional locus $\mathbb{S}^{2}\times \mathbb{S}^{3}/\mb{Z}_N$, where $\mathbb{S}^{3}/\mb{Z}_N$ is the lens space. M-theory on $\overline{\mc{C}(Q^{\scriptscriptstyle (1,1,1)}/\mb{Z}_N)}$ gives rise to a non-gauge theory description that is interpreted as the confined phase of the $SU(N)$ gauge theory.

We study in detail the 3d physical ingredients of such confined phase, including objects from M2, M5-branes wrapping various torsional and free cycles, and the $\mb{Z}_N$ 0-form/2-form gauge fields coming from the reduction of M-theory $C_3$ gauge field on torsional cocycles in $\mathbb{S}^{2}\times \mathbb{S}^{3}/\mb{Z}_N$~\cite{Berasaluce-Gonzalez:2012abm}. In particular, the confining strings are realized as M2-brane wrapped over the $\mb{Z}_N$ torsional 1-cycles on the lens space part $\mathbb{S}^{3}/\mb{Z}_N$, in analog to the cases of $G_2$ spaces~\cite{Acharya:2001hq}.

Following \cite{vanBeest:2022fss}, we have also computed the (invertible) generalized global symmetries of the 3d $\mc{N}=2$ theory from the link topology of $Q^{\scriptscriptstyle (1,1,1)}/\mb{Z}_N$, as well as the 4d SymTFT action from the intersection numbers of the resolved geometry. Following \cite{Garcia-Valdecasas:2023mis,Najjar:2024vmm}, we defined the topological generator of such generalized global symmetries using the Page charge of flux branes.

The torsional gauge fields from $C_3$ over compact torsional cycles as well as the remnant of the 4d SymTFT action will potentially give rise to the gapped TQFT description of the confined phase of the 3d $\mc{N}=2$ theory engineered from M-theory on the DR-phase $\overline{\mc{C}(Q^{\scriptscriptstyle (1,1,1)}/\mb{Z}_N)}$. Nonetheless, the fully detailed 3d TQFT action is subject to future work.

The structure of this paper is as follows:
\begin{itemize}
    \item In Section \ref{sec:CY4-Q111-quotient-physics}, we review the CY4 cone over the 7-dimensional Sasaki-Einstein manifold $Q^{\scriptscriptstyle (1,1,1)}$, which serves as the link space. We perform a specific $\mathbb{Z}_{N}$ quotient on $Q^{\scriptscriptstyle (1,1,1)}$ such that the resulting cone space, $\mathcal{C}(Q^{\scriptscriptstyle (1,1,1)}/\mathbb{Z}_{N})$, retains its Calabi-Yau structure. Additionally, we discuss the physics associated with $\CC(Q^{\scriptscriptstyle (1,1,1)}/\mathbb{Z}_{N})$ in the context of M-theory and its dual description in terms of $(p,q,r)$ 4-branes.
    \item In Section \ref{sec:symtft}, we construct the 4d bulk SymTFT, which encapsulates the possible $p$-form symmetries of our 3d $\N=2$ theory, as well as the associated 't Hooft anomalies. We provide specific examples of the SymTFT to illustrate its structure and properties. Furthermore, we discuss the branes that generate $p$-dimensional charged defects and the corresponding symmetry topological operators.
    \item In Section \ref{sec:geometrical-transition}, we discuss the crepant resolution and the new DR-phase of the cone spaces $\CC(Q^{\scriptscriptstyle (1,1,1)})$ and $\CC(Q^{\scriptscriptstyle(1,1,1)}/\mathbb{Z}_{N})$. Additionally, we explore the physics associated with the new DR-phase, including its deep IR topological description.
\end{itemize}
In addition, the paper includes two appendices. In Appendix \ref{app:arguments-for-the-DR-phase}, we present physical arguments that provide evidence for the existence of the new geometric transition. In Appendix \ref{app:interlacing-geometry}, we describe the geometry of $\mathcal{C}(Q^{\scriptscriptstyle(1,1,1)}/\mathbb{Z}_{N})$ as an interlacing structure of two orthogonal $\mathcal{C}(T^{\scriptscriptstyle(1,1)}/\mathbb{Z}_{N})$ spaces.

\section{The CY4 cone over \texorpdfstring{$Q^{\scriptscriptstyle(1,1,1)}/\Z_{N}$}{Q(1,1,1)/ZN} in M-theory}\label{sec:CY4-Q111-quotient-physics}

In this section, we explore the geometric engineering of the CY4 cone over $Q^{\scriptscriptstyle(1,1,1)}/\mathbb{Z}_{N}$ in M-theory, alongside its dual $(p,q,r)$ 4-brane description.    

\subsection{The cone over \ \texorpdfstring{$Q^{\scriptscriptstyle(1,1,1)}$}{Q(1,1,1)}}

We review the Sasaki-Einstein (SE) 7-manifold $Q^{\scriptscriptstyle(1,1,1)}$, the associated CY4 cone, and their toric description.  

\paragraph{The SE 7-manifold $Q^{\scriptscriptstyle(1,1,1)}$ and the CY4 cone.} In the following, we consider the CY4 cone over the Sasaki-Einstein 7-manifold (SE7), which is constructed as a  $U(1)$ bundle over $\C\P^{1}\times \C\P^{1}\times \C\P^{1}$.  For further discussions on SE7-manifolds, the reader may consult \cite{DAuria:1983sda,Nilsson:1984bj, Sorokin:1984ca, Sorokin:1985ap,DUFF19861,Gauntlett:2004hh,Franco:2009sp}. 

Since a $U(1)$ bundle over a $\C\P^{1}$ is classified by $\pi_{1}(U(1)) = \Z$, then the $U(1)$ bundle of the fore-mentioned 7-dimensional space corresponds to a point $(p,q,r)$ in the three-dimensional $\Z^{3}$ lattice. Without loss of generality, we focus on the positive points in $\Z^{3}$. Furthermore, to ensure non-trivial topology, we exclude points lying on the axes of the $\Z^{3}$ lattice. 

The first non-trivial space that meet these requirements is located at the $(p,q,r)=(1,1,1)$ point, which admits two parallel spinors, i.e. $SU(3)$ holonomy \cite{DUFF19861}. We emphasize that the non-trivial topology of this space is closely tied to the existence of an $SU(3)$ structure on the 7-manifold \cite{DUFF19861}.

The isometry group for the above SE7 manifold can be read as $SU(2)\times SU(2)\times SU(2)\times U(1)$. In turn, the SE7 can be best captured by the following coset manifold \cite{DAuria:1983sda,FRIEDRICH1997259,Acharya:1998db,Gauntlett:2004hh,Franco:2009sp}
\begin{equation}\label{defQ111space}
  Q^{\scriptscriptstyle(1,1,1)} \ = \ \frac{\left(SU(2)\times SU(2)\times SU(2)\right)\times U(1)}{\left(U(1)\times U(1)\right)\times U(1)}\,.
\end{equation}
The point $(p,q,r)\in\Z^{3}$ now corresponds to the diagonal elements in the maximal torus of $SU(2)\times SU(2)\times SU(2)$ and the $U(1)\times U(1)$ quotient are taken to be orthogonal to that direction.
  
In the following, we will focus exclusively on the $Q^{\scriptscriptstyle(1,1,1)}$ space, as it is the only member of the aforementioned family that admits two parallel spinors\footnote{The space $Q^{\scriptscriptstyle (2,2,2)}$ admits two parallel spinors; however, it is given as $Q^{\scriptscriptstyle(2,2,2)}=Q^{\scriptscriptstyle(1,1,1)}/\Z_{2}$ \cite{Franco:2009sp}.} \cite{DUFF19861,Franco:2009sp}. Its homology groups are given as follows:
\begin{equation}\label{homologyQ111}
    H_{\bullet}(Q^{\scriptscriptstyle(1,1,1)},\Z) = \{\Z,0,\Z^{2},\mb{Z}_2,0,\Z^{2},0,\Z\}.
\end{equation}
Here, the 2-cycles are two copies of $\C\P^{1}$'s and the five-cycles are two copies of $U(1)\hookrightarrow T^{\scriptscriptstyle(1,1)}\rightarrow \C\P^{1}\times\C\P^{1}$ \cite{Fabbri:1999hw}. In addition, the cohomology groups are given as \cite{Fabbri:1999hw} 
\begin{equation}
    H^{\bullet}(Q^{\scriptscriptstyle(1,1,1)},\Z) = \{\Z,0,\Z\cdot\omega_{1}\oplus\Z\cdot\omega_{2},0,\Z_{2}\cdot(\omega_{1}\omega_{2}+\omega_{2}\omega_{3}+\omega_{1}\omega_{3}),\Z\cdot\alpha\oplus\Z\cdot\beta,0,\Z\}.
\end{equation}
With $\omega_{i}$ are the generators of the second cohomology group of the $\C\P^{1}$’s and $\pi_{\ast}\alpha = \omega_{1}\omega_{2}-\omega_{1}\omega_{3}, \,\pi_{\ast}\beta = \omega_{1}\omega_{2}-\omega_{2}\omega_{3}$.

The metric on the cone over $Q^{\scriptscriptstyle(1,1,1)}$, which we denote as $\CC(Q^{\scriptscriptstyle(1,1,1)})$, can be written with the help of 3 copies of the left $SU(2)$ invariants 1-forms. In general, each copy of the $SU(2)$ left-invariant 1-forms are denoted by $\{\sigma_{i}\}$ with $i=1,2,3$ and satisfying \cite{EGUCHI197982,Eguchi:1980jx}
\begin{equation}
    d\sigma_{i} = \frac{1}{2}\epsilon_{ijk} \ \sigma_{j}\wedge\sigma_{k}.
\end{equation}
In terms of polar coordinates, $0\leq\theta\leq\pi,\, 0\leq\phi\leq  2\pi$, and $0\leq\psi\leq 4\pi$, the 1-forms can be given as the following 
\begin{equation}\label{eq:def-left-invariant-1-form}
    \begin{split}
        \sigma_{1} &= -\sin\psi\, d\theta + \cos\psi\sin\theta\,d\phi,
        \\
        \sigma_{2} &= \cos\psi\, d\theta + \sin\psi\sin\theta\,d\phi,
        \\
        \sigma_{3} &= d\psi + \cos\theta\,d\phi. 
    \end{split}
\end{equation}
Let us denote the three copies of the 1-forms by $\sigma$, $\Sigma$, and $\widetilde{\Sigma}$ with polar angles $(\theta_{\scriptscriptstyle I},\phi_{\scriptscriptstyle I},\psi_{\scriptscriptstyle I})$ with $I=1,2,3$ that runes over the forementioned 1-forms. The action of the $U(1)\times U(1)$ quotient in (\ref{defQ111space}) on the 3 copies of the above 1-forms is imposing the identification: $\psi_{1}\sim\psi_{2}\sim\psi_{3}$, which we denote simply by $\psi$.

Hence, the metric on the space (\ref{defQ111space}) can be written as \cite{Franco:2009sp}
\begin{equation}\label{eq:metric-over-Q111}
\begin{split}
        ds^{2}(Q^{\scriptscriptstyle(1,1,1)}) &= \frac{1}{8}\,(\sigma_{1}^{2} + \sigma_{2}^{2}  + \Sigma_{1}^{2} + \Sigma_{2}^{2}  + \widetilde{\Sigma}_{1}^{2} + \widetilde{\Sigma}_{2}^{2} )
        \\
        & \qquad + \frac{1}{16}\, (\sigma_{3}+\Sigma_{3}+\widetilde{\Sigma}_{3})^{2}.
        \\
         &=\frac{1}{8}\,\sum_{I=1}^{3} (d\theta_{I}^{2} + \sin^{2}\theta_{I}d\phi_{I}^{2})  + \frac{1}{16}\, (d\psi+\sum_{I=1}^{3}\cos\theta_{I}d\phi_{I})^{2}.
\end{split}
\end{equation}
The topology of the space $Q^{\scriptscriptstyle(1,1,1)}$ is embedded within the structure of this metric, reflecting its underlying geometric features. Furthermore, the above metric reflects the isometry group of (\ref{defQ111space}).

\paragraph{Describing the cone $\mathcal{C}(Q^{\scriptscriptstyle(1,1,1)})$.} Consider $A_{i},B_{i},C_{i}$ with $i=1,2$ as doublets of the three $SU(2)$ factors that appears in the isometry group of the link $Q^{\scriptscriptstyle(1,1,1)}$. To describe the cone over $Q^{\scriptscriptstyle(1,1,1)}$, $\mathcal{C}(Q^{\scriptscriptstyle(1,1,1)})$, we shall define gauge invariant combinations in $\C^{8}$ as the following \cite{Oh:1998qi,Fabbri:1999hw,DallAgata:1999ivu,Herzog:2000rz}
\begin{equation}\label{theC8coordinates}
 \begin{split}
&z_{1} = A_{1}B_{2}C_{1}, \quad z_{2} = A_{2}B_{1}C_{2}, \quad z_{3} = A_{2}B_{2}C_{1}, \quad z_{4} = A_{1}B_{1}C_{2},
        \\
&z_{5} = A_{1}B_{1}C_{1}, \quad z_{6} = A_{2}B_{2}C_{2}, \quad z_{7} = A_{2}B_{1}C_{1}, \quad z_{8} = A_{1}B_{2}C_{2}.
 \end{split}
\end{equation}
In particular, the above relations describe the embedding of $\mathcal{C}(Q^{\scriptscriptstyle(1,1,1)})$ in the $\C^{8}$ space. To describe the CY4 cone $\mathcal{C}(Q^{\scriptscriptstyle(1,1,1)})$, we demand the following constraints \cite{Oh:1998qi,caibar1999minimal}
\be
\ba
\label{eq:Q111}
&z_1 z_2-z_3 z_4=0\ ,\ z_5 z_6-z_7 z_8=0\ ,\ z_1 z_7-z_3 z_5=0\cr
&z_4 z_6-z_2 z_8=0\ ,\ z_1 z_4-z_5 z_8=0\ ,\ z_1 z_6-z_3 z_8=0\cr
&z_2 z_3-z_6 z_7=0\ ,\ z_2 z_5-z_4 z_7=0\ ,\ z_1 z_2-z_5 z_6=0\,.
\ea
\ee
We observe that, each of the first two equations can be used to describe an independent CY3 conifold, which is the cone over $T^{\scriptscriptstyle(1,1)}$ \cite{Candelas:1989js}. Whereas, the rest of the equations can be regarded as constraints to obtain a CY 4-fold.

\paragraph{Resolution.} Here, we briefly discuss the resolution of the cone space $\mathcal{C}(Q^{\scriptscriptstyle(1,1,1)})$. Further details on the resolution process and its implications are deferred to Section \ref{sec:geometrical-transition}.

\begin{figure}[H]
\centering{
\includegraphics[scale=0.6]{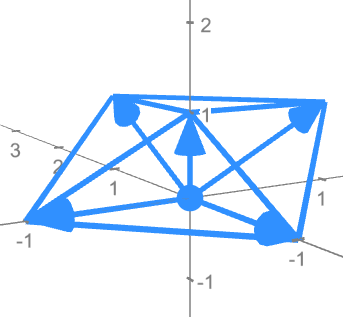}
}
\caption{The toric digram of $\mathcal{C}(Q^{\scriptscriptstyle(1,1,1)})$. Remark, there are 3 different flops upon resolving the above geometry. The toric 3d diagram can be found \href{https://www.math3d.org/nxroeIYdC}{here}.}
\label{Fig:toricQ111}
\end{figure}

The resolution of toric diagrams is generally achieved via the triangulation of the toric fan, as discussed in, e.g., \cite{Closset:2009sv}. Figure \ref{Fig:toricQ111} illustrates the resolved cone $\mathcal{C}(Q^{\scriptscriptstyle(1,1,1)})$, where the exceptional locus is identified as $\mathbb{P}^{1} \times \mathbb{P}^{1}$. Notably, there are three distinct ways to perform the triangulation, corresponding to three equivalent geometries. The operation of transitioning between these geometries is known as a \textit{flop}. 

This geometry does not admit compact divisors, i.e., compact six-cycles. Instead, it contains the following non-compact divisors:
\begin{equation}
\begin{split}
    &S_1:\ (0,0,0)\ ,\ S_2:\ (-1,0,0)\ ,\ S_3:\ (0,-1,0)\ ,
    \\
    &S_4:\ (0,0,1)\ ,\ S_5:\ (0,1,1)\ ,\ S_6:\ (1,0,1)\,. 
\end{split}  
\end{equation}
These divisors satisfy the following relations:
\begin{equation}
    \begin{split}
        -S_{2}+S_{6}&=0\,,
        \\
        -S_{3}+S_{5}&=0\,,
        \\
        S_{4}+S_{5}+S_{6}&=0\,,
        \\
        S_{1}+S_{2}+S_{3}+S_{4}+S_{5}+S_{6}&=0\,.
    \end{split}
\end{equation}
The compact four-cycle $\P^{1} \times \P^{1}$ can be identified with $S_1\cdot S_4$, and the two $\P^1$s are identified with the following curves:
\begin{equation}
   C_{1}\,:= \, S_{1}\cdot S_{2}\cdot S_{4}\,,\qquad C_{2} \,:=\, S_{1}\cdot S_{3}\cdot S_{4}\,. 
\end{equation}

\subsection{A \texorpdfstring{$\Z_{N}$}{ZN} quotient of \texorpdfstring{$\mathcal{C}(Q^{\scriptscriptstyle(1,1,1)})$}{C(Q(1,1,1))} and its toric description}

\paragraph{A $\Z_{N}$ quotient of the cone $\mathcal{C}(Q^{\scriptscriptstyle(1,1,1)})$.} 

Consider a $\Z_{N}$ discrete group quotient of the cone space $\mathcal{C}(Q^{\scriptscriptstyle(1,1,1)})$, generated by $\lambda = e^{2\pi\,i/N}$, with action on the $z_{i}\in\C^{8}$ coordinates is specified as follows, 
\begin{equation}\label{quotientCQZNonzi}
    \begin{split}               &(z_{1},z_{2},z_{3},z_{4},z_{5},z_{6},z_{7},z_{8}) 
        \\
        &\qquad \qquad \qquad \qquad \qquad \sim (\lambda \,z_{1},\lambda^{-1} \,z_{2},\lambda \,z_{3},\lambda^{-1} \,z_{4},\,\lambda^{-1} z_{5}, \lambda\,z_{6},\lambda^{-1}\,z_{7}, \lambda\,z_{8}).
    \end{split}
\end{equation}

To determine the toric diagram that describe the quotient space above, we proceed with the following algorithm. For each of these doublets $A_{i},B_{i},C_{i}$ with $i=1,2$, we associate a pair of local toric coordinates $t_{a}\in\C^{*}$, with $a = 1,\cdots,6$. The quotient above can be extended to the local toric coordinates as,
\begin{equation}\label{eq:ZN-action-on-ti}
    (t_{1},t_{2},t_{3},t_{4},t_{5},t_{6}) \  \rightarrow \  (t_{1},t_{2},\lambda^{-1}\,t_{3},\lambda\, t_{4}, t_{5},t_{6}) \,.
\end{equation}

The relation between the toric coordinates and the $\C^{8}$ coordinates are given by,
\begin{equation}
 \begin{split}
&z_{1} = \frac{t_{4}t_{5}}{t_{1}}, \quad z_{2} = \frac{t_{3}t_{6}}{t_{2}},\quad z_{3} = \frac{t_{4}t_{5}}{t_{2}}, \quad z_{4} = \frac{t_{3}t_{6}}{t_{1}},
        \\
&z_{5} = \frac{t_{3}t_{5}}{t_{1}}, \quad z_{6} = \frac{t_{4}t_{6}}{t_{2}}, \quad z_{7} =\frac{t_{3}t_{5}}{t_{2}}, \quad z_{8} =\frac{t_{4}t_{6}}{t_{1}}\, .
 \end{split}
\end{equation}
One can verify that this identification reproduces the quotient in (\ref{quotientCQZNonzi}) and satisfies the defining equation in (\ref{eq:Q111}). Therefore, these coordinates describe the $\Z_{N}$ quotient geometry.

Following \cite[Lemma 3.3]{Davies:2013pna} and the general results of \cite{cox2011toric}, the quotient of the $\Z_{N}$ subgroup on the local toric coordinates $t_{a}$ retain the same fan with a subdivided lattice. The original lattice, denoted $\mc{N}$, is given by 
\begin{equation}\label{eq:lattice-N}
    \begin{split}
       \mc{N}\,=\, \{&(1,0,0,0,0,0), (0,1,0,0,0,0), (0,0,1,0,0,0),
        \\
        &(0,0,0,1,0,0),(0,0,0,0,1,0),(0,0,0,0,0,1) \}\,.
\end{split}
\end{equation}
After applying the $\Z_{N}$ quotient, the lattice $\mc{N}$ is subdivided into a new lattice $\mc{N}'$, given by
\begin{equation}
\begin{split}
    \mc{N}'\,=\,  \{&(1,0,0,0,0,0), (0,1,0,0,0,0), (0,-\frac{1}{N},\frac{1}{N},0,0,0),
        \\
        &(0,0,0,1,0,0),(0,0,0,0,1,0),(0,0,0,0,0,1) \}\,.
\end{split}
\end{equation}

The transformation matrix relating $\mc{N}$ to $\mc{N}'$ is:
\begin{equation}
 \begin{pmatrix}
1 & 0 & 0 &0 &0 &0\\
0 & 1 & 0 &0 &0 &0\\
0 & 1 & N& 0 &0 &0\\
0 & 0 & 0& 1 &0 &0\\
0 & 0 & 0& 0 &1 &0\\
0 & 0 & 0& 0 &0 &1\\
\end{pmatrix}\,.
\end{equation}
This matrix acts effectively on the upper-left $3\times 3$ sub-matrix, which is indeed the one acting on the 3d toric diagram.

The vertices of the cone $\mathcal{C}(Q^{\scriptscriptstyle (1,1,1)})$ are given by
\begin{equation}\label{eq:vertices-C-Q111}
\begin{split}
& v_{1} = (0,0,0,1;1,1),\quad v_{2} = (-1,0,0,1;1,1), \quad v_{3} = (0,-1,0,1;1,1),
        \\
&  v_{4} = (0,0,1,1;1,1),\quad  v_{5} = (0,1,1,1;1,1),\quad   v_{6} = (1,0,1,1;1,1)\,.
 \end{split}
\end{equation}
Here, the first four entries are taken from Figure \ref{Fig:toricQ111}, while the last two represent their embedding in the lattice $\mc{N}$ defined in (\ref{eq:lattice-N}).

Upon applying the transformation matrix and a shift by $v_{0} = (0,-1,0,0)$, the toric vertices of the quotient space $\mathcal{C}(Q^{\scriptscriptstyle (1,1,1)}/\Z_{N})$ are given by  
\begin{equation}\label{eq:vertices-C-Q111-ZN}
 \begin{split}
& v_{1} = (0,0,0,1),\quad v_{2} = (-1,0,0,1), \quad v_{3} = (0,-1,0,1),
        \\
&  v_{4} = (0,0,N,1),\quad  v_{5} = (0,1,N,1),\quad   v_{6} = (1,0,N,1)\,.
 \end{split}
\end{equation}
Here, we have dropped the fifth and the sixth entries which remain equivalent to that in (\ref{eq:vertices-C-Q111}).

\begin{figure}[H]
\centering{
\includegraphics[scale=0.19]{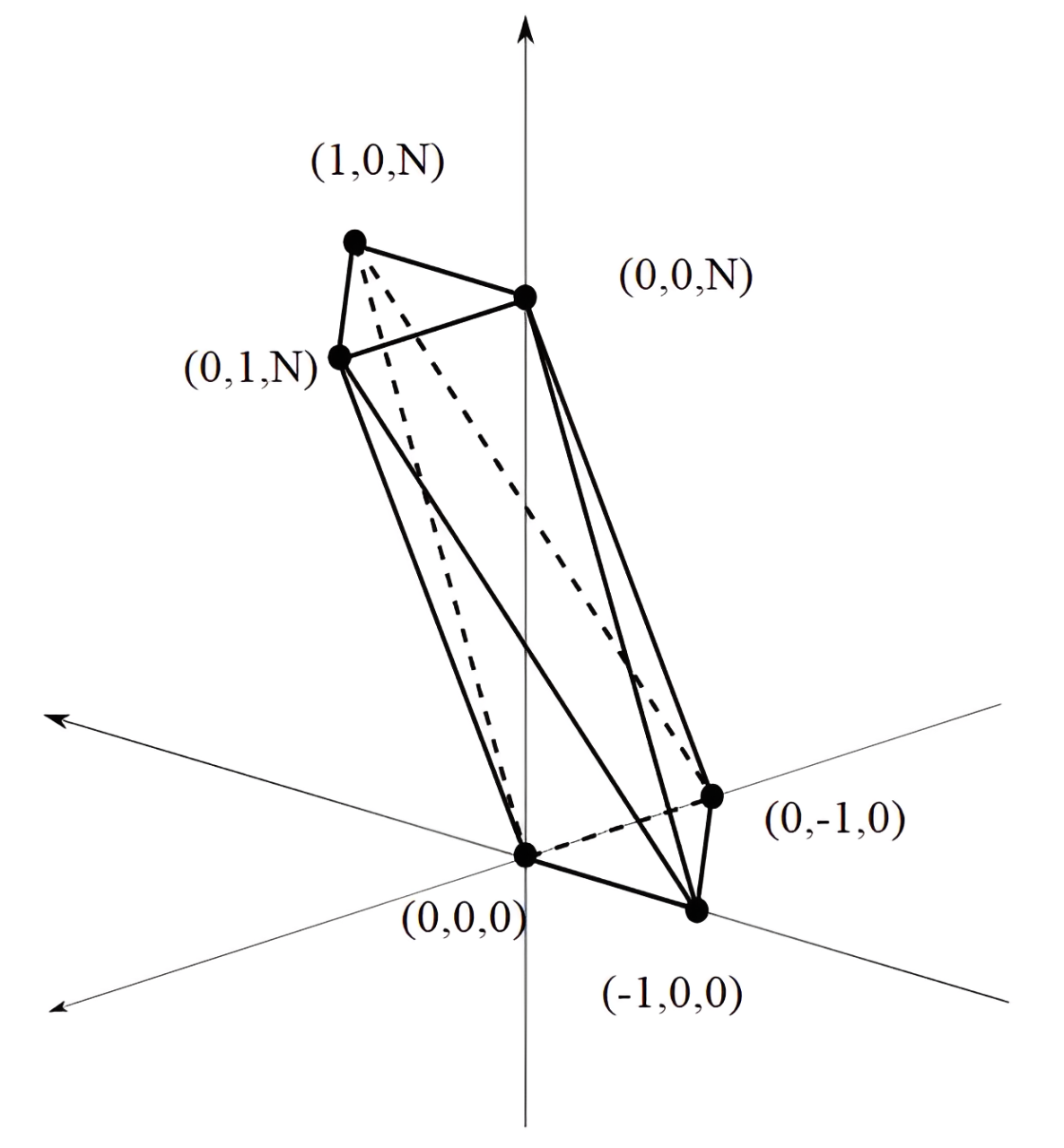}
}
\caption{The toric diagram for the $\Z_{N}$ quotient of $\mathcal{C}(Q^{\scriptscriptstyle(1,1,1)})$. Note that the above geometry is known in the literature as the $Y^{N,N}(\C\P^{1}\times\C\P^{1})$ space, which is part of a larger $Y^{N,P}(\C\P^{1}\times\C\P^{1})$ family as described, for instance, in \cite{Martelli:2008rt}. }
\label{Fig:toricQ111ZN}
\end{figure}

The toric diagram for $\mathcal{C}(Q^{\scriptscriptstyle (1,1,1)}/\Z_{N})$ is shown in Figure \ref{Fig:toricQ111ZN}.

Aside note, one could consider a more general class of quotients, as outlined in \cite[Theorem 3.1]{Davies:2013pna}. However, a detailed exploration of this generalization lies beyond the scope of the present paper. 

\paragraph{The $\Z_{N}$ quotient, supersymmetry, and isometry.}

The $\Z_{N}$ quotient, described in (\ref{quotientCQZNonzi}) and (\ref{eq:ZN-action-on-ti}), acts on the doublets $A_{i},B_{i}$, and $C_{i}$ as follows: $A_{i}$ and $C_{i}$ remain invariant, while $B_{i}$ transform non-trivially according to
\begin{equation}\label{eq:ZN-action-Bi}
    (B_{1},B_{2})\,\sim\,(\lambda^{-1}\,B_{1},\lambda\,B_{2})\,.
\end{equation}
For general $N$, this non-trivial transformation may reduce the isometry group of $Q^{\scriptscriptstyle(1,1,1)}$ to
\begin{equation}
    SU(2)\times SU(2)\times U(1)\times U(1)\,,
\end{equation}
see, e.g., \cite{Martelli:2008rt}. Similar reductions for other quotients are explored in \cite{Franco:2009sp}. However, we will shortly argue that the cone metric retains its structure even after taking the quotient, and the isometry group is at least $\left(SU(2)\right)^{3}$.

Interestingly, for $N=2$, the isometry group remains unchanged from that in (\ref{defQ111space}), as the $\Z_{2}$ group coincides with the centre of $SU(2)$, under which the $B_{i}$ doublets transform.

The SE 7-manifold $Q^{\scriptscriptstyle(1,1,1)}$ can be described using the doublets $A_{i},B_{i}$, and $C_{i}$, constrained by their norm conditions \cite{Oh:1998qi,Fabbri:1999hw,Herzog:2000rz},
\begin{equation}
        |A_{1}|^{2} + |A_{2}|^{2}\, =\, 1\,, \quad |B_{1}|^{2} + |B_{2}|^{2}\, =\, 1\,,\quad |C_{1}|^{2} + |C_{2}|^{2}\, =\, 1\,.
\end{equation}
Each doublet defines a three-sphere $\mathbb{S}^{3}\cong SU(2)$. 

The $\Z_{N}$ action in (\ref{eq:ZN-action-Bi}) can then be understood as an action on the associated $SU(2)$, expressed as:
\begin{equation}
     SU(2) \,\ni\, \begin{pmatrix}
        B_{1}&-\bar{B}_{2}\\
        B_{2}&\bar{B}_{1}
    \end{pmatrix} \quad \to \quad 
\begin{pmatrix}
        \lambda^{-1} B_{1}&-\lambda^{-1} \bar{B}_{2}\\
        \lambda B_{2}&\lambda\bar{B}_{1}
    \end{pmatrix}\,.
\end{equation}
This transformation defines a lens space $L(N,1)=\mathbb{S}^{3}/\Z_{N}$, where the $\Z_{N}$ acts on the Hopf fiber. Using the parametrization, 
\begin{equation}
    B_{1} \,=\, \cos(\frac{\theta}{2})\,e^{i(\psi+\phi)/2}\,,\qquad  B_{2} \,=\, \sin(\frac{\theta}{2})\,e^{i(\psi-\phi)/2}\,.
\end{equation}
the quotient acts on the Hopf fiber angle $\psi$ as,
\begin{equation}
    \psi\,\sim\,\psi + 4\pi/N\,.
\end{equation}

Under the $U(1)\times U(1)$ action in (\ref{defQ111space}), the Euler angles satisfy the identification $\psi_{1}\sim\psi_{2}\sim\psi_{3}\equiv \psi$. Hence, the quotient geometry is described as:
\begin{equation}\label{eq:Q111-ZN-bundle}
  \left(U(1)/\Z_{N}\right)\,\,\hookrightarrow\,\,   Q^{\scriptscriptstyle(1,1,1)}/\Z_{N} \,\,\to\,\,\C\P^{1}\times\C\P^{1}\times\C\P^{1}\,.
\end{equation}

The supersymmetric nature of the quotient can be verified through the holomorphic top form $\Omega^{\scriptscriptstyle(4,0)}$ on the CY4 cone over $Q^{\scriptscriptstyle(1,1,1)}$, as given in \cite[(2.3)]{Franco:2009sp}:
\begin{equation}
   \begin{split}
\Omega^{\scriptscriptstyle(4,0)}\,\,\sim\,\,&r^{4}e^{i\psi}\left(\frac{dr}{r}+\frac{i}{4}(d\psi+\Sigma_{j=1}^{3}\cos(\theta_{j})d\phi_{j} )\right) \wedge (d\theta_{1}+i\sin(\theta_{1})d\phi_{1})
\\
&\qquad \wedge(d\theta_{2}+i\sin(\theta_{2})d\phi_{2})\wedge (d\theta_{3}+i\sin(\theta_{3})d\phi_{3})\,.
   \end{split}
\end{equation}
The $\Z_{N}$ quotient does not affect  $\Omega^{\scriptscriptstyle(4,0)}$, confirming that the singular quotient geometry $\CC(Q^{\scriptscriptstyle(1,1,1)}/\Z_{N})$ is indeed a CY4 cone.

Furthermore, the local structure of the cone metric over $Q^{\scriptscriptstyle(1,1,1)}$ remains unchanged by the $\Z_{N}$ quotient, which only modifies the periods of the $U(1)$ bundle in (\ref{eq:Q111-ZN-bundle}). Consequently, the isometry group of the singular cone metric $\CC(Q^{\scriptscriptstyle(1,1,1)}/\Z_{N})$ is preserved as
\begin{equation}\label{eq:isometry-Q111-ZN-from-metric}
    SU(2)\times SU(2)\times SU(2)\times U(1)\,.
\end{equation}

\paragraph{Crepant resolutions and intersection numbers.}

We consider the singular cone space $\mc{C}(Q^{\scriptscriptstyle(1,1,1)}/\mb{Z}_N)$, whose toric diagram is plotted in Figure~\ref{Fig:toricQ111ZN}. After the crepant resolution (maximal triangulation of the polytope), we add the lattice points $(0,0,1)$, $\dots$, $(0,0,N-1)$ into the toric diagram corresponding to the compact exceptional divisors $D_1$, $\dots$, $D_{N-1}$. We also label the non-compact divisors as
\begin{equation}\label{eq:non-compact-divisors}
    \begin{split}
       &S_1:\ (0,0,0)\ ,\ S_2:\ (-1,0,0)\ ,\ S_3:\ (0,-1,0)\ ,
       \\
       &S_4:\ (0,0,N)\ ,\ S_5:\ (1,0,N)\ ,\ S_6:\ (0,1,N)\,. 
    \end{split}
\end{equation}
The 4D cones are
\be\label{eq:4d-cone}
\ba
&\{D_1 S_1 S_2 S_3\ ,\ D_1 S_1 S_3 S_5\ ,\ D_1 S_1 S_5 S_6\ ,\ D_1 S_1 S_6 S_2,\cr
&\,\,D_{N-1} S_4 S_2 S_3\ ,\ D_{N-1} S_4 S_3 S_5\ ,\ D_{N-1} S_4 S_5 S_6\ ,\ D_{N-1} S_4 S_6 S_2,\cr
&\,\,D_i D_{i+1} S_2 S_3\ ,\ D_i D_{i+1} S_3 S_5\ ,\ D_i D_{i+1} S_5 S_6\ ,\ D_i D_{i+1} S_6 S_2\ (i=1,\dots,{N-2})\}\,.
\ea
\ee

The linear equivalence relations read
\be\label{eq:equivalence-relations}
\ba
&S_2=S_5\ ,\ S_3=S_6\,,\cr
&\sum_{i=1}^{N-1}D_i+S_1+S_2+S_3+S_4+S_5+S_6=0\,,\cr
&\sum_{i=1}^{N-1}iD_i+NS_4+NS_5+NS_6=0\,.
\ea
\ee

Hence the non-vanishing intersection numbers are
\be\label{eq:intersection-numbers}
\ba
&D_1 S_1 S_2 S_3=1\ ,\ D_{N-1} S_4 S_2 S_3=1\ ,\ D_i D_{i+1} S_2 S_3=1\ (i=1,\dots,(N-2))\,,\cr
& S_1^2 D_1 S_3=(N-2)\ ,\ S_1 D_1^2 S_3=-N\ ,\ S_1^2 D_1 S_2=(N-2)\ ,\ S_1 D_1^2 S_2=-N\,,\cr
&S_4^2 D_{N-1} S_3=(N-2)\ ,\ S_4 D_{N-1}^2 S_3=-N\ ,\ S_4^2 D_{N-1} S_2=(N-2)\ ,\ S_4 D_{N-1}^2 S_2=-N\,,\cr
&D_i^2 S_2 S_3=-2\ (i=1,\dots,(N-1))\,,\cr
&D_i^2 D_{i+1} S_3=D_i^2 D_{i+1} S_2=N-2i-2\ ,\ D_i D_{i+1}^2 S_3=D_i D_{i+1}^2 S_2=2i-N\ (i=1,\dots,(N-2))\,,\cr
&D_1 S_1^3=D_{N-1}S_4^3=2(N-2)^2\ ,\ D_1^2 S_1^2=D_{N-1}^2 S_4^2=-2N(N-2)\ ,\ D_1^3 S_1=D_{N-1}^3 S_4=2N^2\,,\cr
&D_i^3 D_{i+1}=2(N-2i-2)^2\ ,\ D_i^2 D_{i+1}^2=2(N-2i-2)(2i-N)\ ,\ D_i D_{i+1}^3=2(N-2i)^2\,,\cr
&S_2 D_i^3=S_3 D_i^3=8\ (i=1,\dots,(N-1))\ ,\ D_i^4=-48-4(N-2i)^2\,.
\ea
\ee
Note that we have used $S_5=S_2$ and $S_6=S_3$ to omit the terms with $S_5$ and $S_6$.

\subsection{The physics of M-theory on \texorpdfstring{$\CC(Q^{\scriptscriptstyle(1,1,1)}/\Z_{N})$}{C(Q(1,1,1))/ZN}}\label{sec:physics-Q111-ZN}

We discuss the physics of the 3d field theory from M-theory on $\CC(Q^{\scriptscriptstyle(1,1,1)}/\Z_{N})$, applying the general prescription in \cite{Najjar:2023hee}. 

At the singular limit where all cycles are shrunk to zero volume, we expect to have a 3d $\mc{N}=2$ SCFT due to the lack of scale parameters.

In the fully resolved phase, one can see that $\CC(Q^{\scriptscriptstyle(1,1,1)}/\Z_{N})$ is a stack of $\mb{P}^1$ fibered over the base surface $\mb{P}^1\times\mb{P}^1$, where the $\mb{P}^1$ fibers are $C_i=D_i\cdot S_2\cdot S_3$, $(i=1,\dots,N-1)$. Hence, M-theory on the resolved should describe the ``Coulomb branch'' of a 3d $\mc{N}=2$ $\mathfrak{su}(N)$ gauge theory. When the volumes of the $\mb{P}^1$ fibers $C_i$ are shrunk to zero, we have a non-abelian $\mathfrak{su}(N)$ gauge theory description.

The Cartan generators (photons) of $\mathfrak{su}(N)$ arises from the expansion of 
\be
C_3=\sum_{i=1}^{N-1}A_i\wedge\omega^{(1,1)}_i\,,
\ee
where $\omega^{(1,1)}_i$ is the Poincar\'{e} dual $(1,1)$-form of the compact divisor $D_i$.

The gauge W-bosons come from M2-brane wrapping the $\mb{P}^1$ fiber $C_i=D_i\cdot S_2\cdot S_3$, $(i=1,\dots,N-1)$, which has normal bundle $N_{C_i|X_4}=\mc{O}(0)\oplus\mc{O}(0)\oplus\mc{O}(-2)$. From the intersection numbers,
\begin{equation}
    D_{i}^{2}S_{2}S_{3} = - 2 \ (i=1,\cdots, (N-1))\ , \quad D_{j}D_{j+1}S_{2}S_{3} = 1 \ (j=1,\cdots, (N-2))\,,
\end{equation}
we indeed verify that the charge of the M2-brane wrapping $C_i$ under the $j$-th $U(1)$ is equal to the Cartan matrix element $\mc{C}_{ij}=-C_i\cdot D_j$ for the $\mathfrak{su}(N)$ Lie algebra.

The non-abelian gauge coupling is determined by the compact  base surface $\P^{1}\times\P^{1}\cong\mathbb{S}^{2}\times \mathbb{S}^{2}$. In particular, it takes the form 
\begin{equation}
    \frac{1}{g^{2}_{\mathrm{YM}}}\,\sim\, \mathrm{Vol}(\mathbb{S}^{2}\times \mathbb{S}^{2})\,.
\end{equation}
For further discussion and a detailed treatment of the gauge coupling in this context, we refer the reader to Section 2 of~\cite{Najjar:2023hee}.

\paragraph{Flavour symmetry.} For the flavour symmetry, we expect the flavour rank to be $f=2$, and the flavour symmetry generators for $\mathfrak{u}(1)^{\oplus 2}$ can be taken as the non-compact divisors $F_1=S_1-S_2$ and $F_2=S_2-S_3$. The flavour background gauge fields for $\mathfrak{u}(1)^{\oplus 2}$ are from the expansion
\be
C_3=\sum_{\alpha=1}^{2}B_\alpha\wedge\omega^{(1,1),F}_\alpha\,,
\ee
where $\omega^{(1,1),F}_\alpha$ is Poincar\'{e} dual to $F_\alpha$ $(\alpha=1,2)$.

Besides the gauge W-bosons, there are also disorder operators coming from M2-branes wrapping the curves along the base directions, such as $D_i\cdot D_{i+1}\cdot (aS_2+bS_3)$. The physical meanings and roles of such operators are not completely clear, as in the cases of local $\mb{P}^1\times\mc{S}$ in \cite{Najjar:2023hee} and we will not elaborate here. 

\subsection{The dual \texorpdfstring{$(p,q,r)$}{(p,q,r)} 4-branes description}

Following the general results of \cite{Najjar:2023hee}, the geometric engineering of M-theory on the space $\mathcal{C}(Q^{\scriptscriptstyle(1,1,1)})$ admits a dual description in the framework of maximal 8d supergravity. This duality establishes a correspondence between the toric diagrams of CY4 spaces and configurations of $(p,q,r)$ 4-branes in 8d supergravity \cite{Leung:1997tw,Najjar:2023hee}.

\begin{figure}[H]
\centering{
\includegraphics[scale=0.42]{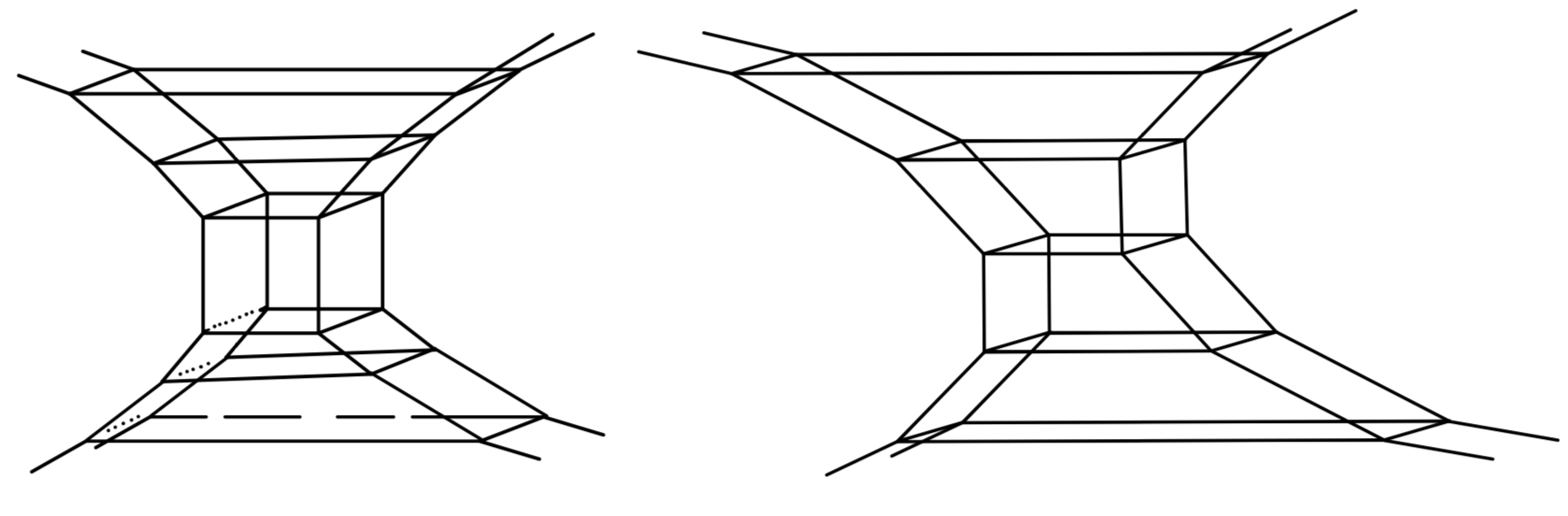}
}
\caption{The $(p,q,r)$ 4-brane description dual to the toric diagram of $\CC(Q^{\scriptscriptstyle(1,1,1)}/\Z_{N})$ illustrated at a generic point on the Colomb branch. On the left-hand side, we get $N=$ even, while on the right-hand side is for $N=$ odd. }
\label{Fig:tower-hyperconifold}
\end{figure}

The $(p,q,r)$ 4-brane configuration dual to the toric diagram in Figure \ref{Fig:toricQ111ZN} is illustrated in Figure \ref{Fig:tower-hyperconifold}. We refer to this brane configuration as the tower-hyperconifold, as its structure, both in the toric and dual brane descriptions, resembles a tower. This nomenclature extends the concept of the CY3 ladder-hyperconifold introduced in \cite{Acharya:2020vmg} to the case of CY4.

Figure \ref{Fig:tower-hyperconifold} represents a generic point in the Coulomb branch of the 3d $\mathcal{N}=2$ $\mathfrak{su}(N)$ gauge theory described earlier. The gauge fields corresponding to the maximal torus of the gauge group, i.e., the photons, are realized as strings with both ends attached to the same finite $(1,0,0)$ 4-brane. Meanwhile, the gauge fields corresponding to the non-abelian generators, i.e., the W-bosons, arise from strings stretching between distinct $(1,0,0)$ 4-branes.

Using the general procedure described in \cite{Najjar:2023hee}, one can compute the charges of these W-bosons under the $(U(1))^{N-1}$ gauge symmetry on the Coulomb branch. This analysis confirms that the charges correspond to the Cartan matrix of the $\mathfrak{su}(N)$ Lie algebra. Thus, the enhancement of the gauge symmetry from $\mathfrak{u}(1)^{\oplus(N-1)}$ to $\mathfrak{su}(N)$ is understood in terms of collapsing all the $(1,0,0)$ 4-branes into a single stack of $N$ coincident finite $(1,0,0)$ 4-branes. The infrared (IR) superconformal field theory (SCFT) description is then obtained by further shrinking these finite branes to zero size, reducing them to a single point.

Further details regarding both the abelian and non-abelian gauge theory couplings can be found in Section 5.2 of \cite{Najjar:2023hee}.

\section{SymTFT}\label{sec:symtft}

\subsection{SymTFT from the link \texorpdfstring{$Q^{\scriptscriptstyle(1,1,1)}/\Z_{N}$}{Q(1,1,1)/ZN}}

\paragraph{Topology of the link $Q^{\scriptscriptstyle(1,1,1)}/\Z_{N}$.} 

To proceed with the construction of the Symmetry Topological Field Theory (SymTFT) and analyse the brane realization of defects and symmetry operators, we first examine the topological structure of the link $Q^{\scriptscriptstyle(1,1,1)}/\Z_{N}$. This includes its homology and differential cohomology groups. According to \cite{Martelli:2008rt}, the homology cycles are given by:
\begin{equation}\label{homologyQ111ZN}
    H_{\bullet} (Q^{\scriptscriptstyle(1,1,1)}/\Z_{N}) = (\Z  , \Z_N   ,\Z^{2} , \Gamma ,  0 , \Z^2\oplus\Z_N , 0 , \Z)\,.
\end{equation}
Here, the torsional cycle $\Gamma$ is defined as
\begin{equation}
    \Gamma = \Z^{3} / \expval{\ (0,-N,-N), (N,N,0),(N,0,N)\ }.
\end{equation}
Using a Smith Normal Form calculation, we can compute
\be\label{eq:Gamma-tor-3-cycle}
\Gamma=\Z_N\oplus\Z_N\oplus\Z_{2N}\,.
\ee
It is noteworthy that the elements of $\Gamma$ are self-dual torsional 3-cycles on $Q^{\scriptscriptstyle(1,1,1)}/\Z_{N}$. Similar to (\ref{homologyQ111}), the 2-cycles are two copies of $\C\P^{1}$'s. However, the five-cycles are two copies of $\Z_{N}$ quotient of $U(1)\hookrightarrow T^{\scriptscriptstyle(1,1)}\rightarrow \C\P^{1}\times\C\P^{1}$. In particular, the quotient acts on the $U(1)$ Hopf fiber, so topologically $T^{\scriptscriptstyle(1,1)}/\Z_{N} = \mathbb{S}^{2}\times (\mathbb{S}^{3}/\Z_{N})$.

The corresponding differential cohomology, which uplifts the dual cohomology of the above homology group, is given by:
\begin{equation}\label{cohomologyQ111ZN}
  \br{H}^{\bullet}(Q^{\scriptscriptstyle(1,1,1)}/\Z_{N},\Z) =  \left\{ \br{1} \right\} , \ 0 , \ \left\{ \br{t}_2,\, \br{v}_{2}^{i} \right\} , \ \left\{0  \right\} , \  \left\{ \br{t}_{4}^{a}  \right\} ,  \ \left\{ \br{v}_{5}^{i} \right\},  \ \left\{ \br{t}_{6}^{i} \right\},  \ \left\{ \br{v}_{7} \right\},
\end{equation}
with $i=1,2$ and $a=1,2,3$. Here, $\br{t}_{4}^{a}$ are dual to the torsional 3-cycles in $\Gamma$.

\paragraph{Expansion of M-Theory field strengths.}

Following the standard procedure for constructing the SymTFT, we uplift the M-theory field strengths to differential cohomology and expand them in a basis of cohomology \cite{Apruzzi:2021nmk}. This yields:
\begin{equation}\label{expansionG4G7}
    \begin{split}
        &\br{G}_{4} = \br{F}_{4}\star \br{1} + \br{F}_{2}^{i} \star \br{v}_{2}^{i} + \br{B}_{2}\star\br{t}_{2}  + \br{B}_{0}^{a} \star \br{t}_{4}^{a},
        \\
        & \br{dG}_{7} =\br{f}_{8}\star \br{1} + \br{f}_{6}^{i} \star \br{v}_{2}^{i} + \br{\mathcal{B}}_{6}\star\br{t}_{2} + \br{\mathcal{B}}_{4}^{a} \star \br{t}_{4}^{a}   + \br{f}_{3}^{i} \star \br{v}_{5}^{i} + \br{\mathcal{B}}_{2}\star\br{t}_{6} + \br{f}_{1}\star\br{v}_{7}.
    \end{split}
\end{equation}
Along with
\begin{equation}
    \mathscr{F}(\br{f}_{p+1}^{\bullet}) = \frac{1}{2\pi}dh_{p}^{\bullet}\,,\quad c(\br{f}_{p+1}^{\bullet})\,,\quad \mathscr{F}(\br{\mathcal{B}}_{p+1}^{\bullet}) = \delta A_{p}^{\bullet}\,. 
\end{equation}

From the expansion of $\br{G}_{4}$, we learn that $\br{F}_{2}$ and $\br{F}_{2}^{i}$ are field strength of continuous $U(1)$ 2-form and 0-form symmetries \cite{Apruzzi:2021nmk}, respectively. On the other hand, $\br{B}_{2}$ and $\br{B}_{0}^{a}$ are the gauge fields of discrete 1-form and $(-1)$-form symmetries, respectively.

The interpretation of the bulk fields of $\br{G}_{7}$ would be best understood through the BF terms that we calculate shortly, at least for the discrete $p$-form symmetries. Concerning discrete symmetries, we identify $\br{A}_{1}$ as the gauge field of a $\Z_{N}$ 0-form symmetry, which is dual to the aforementioned discrete 1-form symmetry. Additionally, the fields $\br{A}_{3}^{a}$ serve as gauge fields for a 2-form symmetry, dual to the discrete $(-1)$-form symmetry. Through the analysis of topological symmetry operators, as detailed in \cite{Najjar:2024vmm}, we establish that $h_{0}$ and $h_{2}^{i}$ are associated with $(-1)$-form and $0$-form symmetries, respectively.
      
\paragraph{Mixed anomalies from M-theory.} 

To determine the potential mixed anomalies associated with the link $L_{7} \equiv Q^{\scriptscriptstyle(1,1,1)}/\Z_{N}$ in the geometrically engineered 3d $\N=2$ theory, we substitute the form of $\br{G}_{4}$ from (\ref{expansionG4G7}) into the topological Chern-Simons action in M-theory, which is uplifted to differential cohomology following \cite{Apruzzi:2021nmk,vanBeest:2022fss}:
\begin{equation}
    S_{\text{CS}}^{\text{M}} \,=\, -\frac{1}{6}\,\int_{L_{7}\,\times\, \mathcal{M}_{4}}^{\br{H}} \br{G}_{4}\star \br{G}_{4}\star \br{G}_{4}.
\end{equation}
Here, $\mathcal{M}_{4}$ are the bulk 4-dimensional manifold on which the SymTFT resides. 

By direct computation we arrive at
\begin{equation}\label{CSG4forQ111ZN}
    \begin{split}
        S_{\text{CS}}^{\text{M}}   \ \supset  \ & \sum_{a,b} \int_{L_{7}}^{\br{H}}  \br{t}_{4}^{a}\star\br{t}_{4}^{b}\star\br{1} \,\, \int_{\mathcal{M}_{4}} {B}_{0}^{a}\smile{B}_{0}^{b}\smile \frac{{F}_{4}}{2\pi}
        \\
        & + \sum_{a,i,j} \int_{L_{7}}^{\br{H}}  \br{t}_{4}^{a}\star\br{v}_{2}^{i}\star\br{v}_{2}^{j}\,\,  \int_{\mathcal{M}_{4}} {B}_{0}^{a}\smile \frac{{F}_{2}^{i}}{2\pi}\smile \frac{{F}^{j}_{2}}{2\pi}
        \\
        & + \sum_{a} \int_{L_{7}}^{\br{H}}  \br{t}_{4}^{a}\star\br{t}_{2}\star\br{t}_{2} \,\, \int_{\mathcal{M}_{4}} {B}_{0}^{a}\smile {B}_{2}\smile {B}_{2}
        \\
        & + \sum_{a} \int_{L_{7}}^{\br{H}}  \br{t}_{4}^{a}\star\br{t}_{2}\star\br{v}_{2}^{i}  \,\,\int_{\mathcal{M}_{4}}  {B}_{0}^{a}\smile {B}_{2}\smile \frac{{F}_{2}^{i}}{2\pi}\,.
    \end{split}
\end{equation}

We note that the anomaly polynomial 
\begin{equation}
\mathcal{A}[B_{0},B_{2}] \ = \ \sum_{a}\,I(\br{t}^{a}_{4},\br{t}_{2},\br{t}_{2}) \,\,   \int_{\mathcal{M}_{4}}  {B}_{0}^{a}\smile {B}_{2}\smile {B}_{2} \quad  (\text{mod \, $1$})\,,
\end{equation}
has been identified for various similar examples in \cite{vanBeest:2022fss}, where it is interpreted as the obstruction to gauging certain subgroups of the 1-form symmetry. Here, we define $I(\br{t}_{4}^{a},\br{t}_{2},\br{t}_{2}) = \int_{\scriptscriptstyle L_{7}}^{\scriptscriptstyle\br{H}}\br{t}_{4}^{a}\star \br{t}_{2}\star\br{t}_{2}$. Shortly, we will compute the precise coefficient for several values of $N$. As it will turn out, see (\ref{symtftN=2}), (\ref{symtftN=3}), and (\ref{symtftN=4}).

This anomaly can also be understood as an obstruction to gauging the $\br{B}_{0}^{a}$ fields, i.e. make them spacetime dependent. This interpretation is analogous to the anomaly theory involving the Yang-Mills $\theta$-angle and the discrete 1-form symmetry in 4D, as discussed in \cite{Cordova:2019uob}, see also \cite{Gaiotto:2014kfa,Kapustin:2014gua,Najjar:2024vmm}.

\paragraph{The BF terms.} By inserting the expansion of $\br{G}_{4}$ and $\br{dG_{7}}$ from (\ref{expansionG4G7}) into the differential cohomology refinement of the M-theory kinetic term,
\begin{equation}
   S^{\text{M}}_{\text{kin}} \,=\, \int_{L_{7}\times\, \mathcal{M}_{4}}^{\br{H}} \br{G}_{4}\star \br{dG}_{7}\,,
\end{equation}
we compute the BF terms for discrete and continuous $p$-form symmetries.

Following \cite{Najjar:2024vmm} and references therein, the BF terms are obtained as:
\begin{equation}\label{eq:BF-term}
    \begin{split}
        \int_{L_{7}\times\, \mathcal{M}_{4}}^{\br{H}} \br{G}_{4}\star d\br{G}_{7} \,\supset\,
        & \sum_{a,b} \int_{L_{7}}^{\br{H}}  \br{t}_{4}^{a}\star\br{t}_{4}^{b}  \,\, \int_{\mathcal{M}_{4}} {B}_{0}^{a}\smile \delta {A}_{3}^{b} + \int_{L_{7}}^{\br{H}}  \br{t}_{2}\star\br{t}_{6} \,\,\int_{\mathcal{M}_{4}} {B}_{2}\smile \delta{A}_{1}
        \\
         & +\sum_{i,j} \int_{L_{7}}^{\br{H}}  \br{v}_{2}^{i}\star\br{v}_{5}^{j}\,\, \int_{\mathcal{M}_{4}} \frac{{F}_{2}^{i}}{2\pi}\wedge \frac{{h}_{2}^{j}}{2\pi} +\int_{L_{7}}^{\br{H}}  \br{1}\star\br{v}_{7}\,\,  \int_{\mathcal{M}_{4}} \frac{{F}_{4}}{2\pi}\wedge \frac{{h}_{0}}{2\pi}\,.
    \end{split}
\end{equation}
For simplicity, the integrals of the Poincaré duals of free cycles over $L_{7}$ are normalized to one. Using the linking pairing of torsional cycles \cite[App.A]{Najjar:2024vmm}, we find
\begin{equation}
    \int_{L_{7}}^{\br{H}}  \br{t}_{2}\star\br{t}_{6} \,=\, -\frac{1}{N}\,,
\end{equation}
and
\begin{equation}\label{eq:t4-t4}
   \int_{L_{7}}^{\br{H}}  \br{t}_{4}^{a}\star\br{t}_{4}^{b} \,=\, \begin{cases} - \frac{1}{N}\,\delta_{ab} & \quad a=1,2\,, \\
   \\
   -\frac{1}{2N}\,\delta_{ab} & \quad a=3\,. \end{cases}
\end{equation}
Note that, the integral over the link of (\ref{CSG4forQ111ZN}) is computed through the equation above.

The resulting BF terms are:
\begin{equation}\label{eq:BF-term-result}
    \begin{split}
        S_{\text{BF}} \,\supset\,
        & -\frac{1}{N}\int_{\mathcal{M}_{4}} {B}_{0}^{1}\smile \delta {A}_{3}^{1} -\frac{1}{N}  \int_{\mathcal{M}_{4}} {B}_{0}^{2}\smile \delta {A}_{3}^{2} -\frac{1}{2N}  \int_{\mathcal{M}_{4}} {B}_{0}^{3}\smile \delta {A}_{3}^{3}
        \\
        & -\frac{1}{N} \int_{\mathcal{M}_{4}} {B}_{2}\smile \delta{A}_{1} + \int_{\mathcal{M}_{4}} \frac{{F}_{2}^{i}}{2\pi}\wedge \frac{{h}_{2}^{j}}{2\pi} + \int_{\mathcal{M}_{4}} \frac{{F}_{4}}{2\pi}\wedge \frac{{h}_{0}}{2\pi}\,.
    \end{split}
\end{equation}

\subsection{Examples}

\paragraph{Divisors, 4-cycles, and $p$-form symmetry representative.} 

We provide a representation of the $1$-form and $(-1)$-form symmetries through compact divisors and compact 4-cycles, respectively, using the equations (\ref{eq:non-compact-divisors}), (\ref{eq:equivalence-relations}), and (\ref{eq:intersection-numbers}) along with the associated discussion.

From the Smith Normal Decomposition of the intersection matrix $\{D_i\cdot C_\alpha\}$ between compact divisors and compact curves, we find that there is a $\mb{Z}_N$ 1-form symmetry represented by the linear combination of compact divisors
\be
D=\sum_{j=1}^{N-1}jD_j\,.
\ee

From the Smith Normal Decomposition of the intersection matrix $\{\mc{S}_\alpha\cdot \mc{S}_\beta\}$ between compact 4-cycles, we find that there is a $\mb{Z}_N\oplus\mb{Z}_N\oplus\mb{Z}_{2N}$ $(-1)$-form symmetry represented by certain linear combinations of 4-cycles. The precise form of these representative 4-cycles depends on the value of $N$, and will be analysed on a case-by-case basis below.

\paragraph{The computation.}

Following the computation procedures in \cite{vanBeest:2022fss,Najjar:2023hee} and the intersection numbers, we obtain the SymTFT action for different cases of $N$. Note that we always denote by $B_2$ the background gauge field for the $\mb{Z}_N$ 1-form symmetry and by ${A}_1$ the background gauge field of the dual $\mb{Z}_N$ 0-form symmetry. We denote by $B^{1}_{0}$, $B^{2}_{0}$, $B^{3}_{0}$ the background gauge field for the $\mb{Z}_N$, $\mb{Z}_N$ and $\mb{Z}_{2N}$ $(-1)$-form symmetries and by ${A}^1_3$, ${A}^{2}_{3}$, ${A}^{3}_{3}$ the background gauge fields for the dual 2-form symmetries. We also have two continuous $U(1)$ 0-form symmetries, corresponding to the non-compact divisors $\widetilde{F}_1=S_1-S_2$ and $\widetilde{F}_2=S_2-S_3$, with background field strength $F_{2}^{1}$ and $F_{2}^{2}$.

\begin{enumerate}
\item $N=2$ 

We take the compact 4-cycle representatives of the $\mb{Z}_2$, $\mb{Z}_2$ and $\mb{Z}_{4}$ $(-1)$-form symmetries to be 
\be
\mc{S}_1=D_1 S_2\ ,\ \mc{S}_2=D_1 S_1\ ,\ \mc{S}_3=D_1(-S_1-S_2+S_3)\,,
\ee
and we can compute the SymTFT action
\be\label{symtftN=2}
\ba
\frac{S_{\rm SymTFT}}{2\pi}&=\int_{\mc{M}_4}\frac{1}{2}(B_2\smile\delta A_{1}+B_{0}^{1}\delta A^{1}_{3}+B_{0}^{2}\delta {A}^{2}_{3})+\frac{1}{4}B_{0}^{3}\delta{A}^{3}_{3}\cr
&=\int_{\mc{M}_4}\frac{1}{4}(2B_{0}^{1}+2B_{0}^{2}-B_{0}^{3})B_2\smile B_2
\\
&+\int_{\mc{M}_4}\frac{1}{2}B_{0}^{1} B_2 (-F_{2}^{1}+F_{2}^{2})+\frac{1}{2}B_{0}^{2} B_2 F_{2}^{1}-\frac{1}{2}B_{0}^{3} B_2 F_{2}^{2}\cr
&+\int_{\mc{M}_4}\frac{1}{2}B_{0}^{1} F^{1}_{2} F_{2}^2+\frac{1}{2}B_{0}^{2}(F_{2}^{1} F_{2}^2+(F_{2}^2)^2)+\frac{1}{4}B_{0}^{3}(-(F_{2}^{1})^2+F_{2}^1 F_{2}^{2}+(F_{2}^2)^2)\,.
\ea
\ee

\item $N=3$

We take the compact 4-cycle representatives of the $\mb{Z}_3$, $\mb{Z}_3$ and $\mb{Z}_{6}$ $(-1)$-form symmetries to be 
\be
\ba
&\mc{S}_1=D_1 S_2-D_2 S_2\ ,\ \mc{S}_2=-D_1 S_1-D_1 S_2+D_2 S_2+D_1 S_3-D_2 S_3\,,\cr
&\mc{S}_3=-D_1 S_1+2D_1 S_2+D_2 S_2+2D_1 S_3+D_2 S_3\,,
\ea
\ee
and we can compute the SymTFT action
\be\label{symtftN=3}
\ba
\frac{S_{\rm SymTFT}}{2\pi}&=\int_{\mc{M}_4}\frac{1}{3}(B_2\smile\delta{A}_1+B_{0}^{1}\delta{A}^{1}_{3}+B_{0}^{2}\delta{A}^{2}_{3})+\frac{1}{6}B_{0}^{3}\delta{A}^{3}_{3}+\frac{1}{6}(B_{0}^{2}+2B_{0}^{3})B_2\smile B_2\cr
&+\int_{\mc{M}_4}\frac{1}{3}B_{0}^{1} B_2 (-F_{2}^{1}+2F_{2}^{2})-\frac{1}{3}B_{0}^{2} B_2 F_{2}^{2}-\frac{1}{3}B_{0}^{3} B_2 F_{2}^{1}\cr
&+\int_{\mc{M}_4}\frac{1}{6}B_{0}^{1} ((F_{2}^{1})^2-2F_{2}^{1} F_{2}^{2})+\frac{1}{6}B_{0}^{2}((F_{2}^{1})^2-4F_{2}^{1} F_{2}^{2}+2(F_{2}^{2})^{2})
\\
&+\int_{\mc{M}_4}\frac{1}{6}B_{0}^{3}(-F_{2}^{1} F_{2}^{2}+(F_{2}^{2})^{2})\,.
\ea
\ee

\item $N=4$

We take the compact 4-cycle representatives of the $\mb{Z}_4$, $\mb{Z}_4$ and $\mb{Z}_{8}$ $(-1)$-form symmetries to be 
\be
\ba
&\mc{S}_1=-D_1 S_2+2D_2 S_2+D_3 S_2+2D_1 S_3+2S_3 S_3+2D_3 S_4\,,\cr
&\mc{S}_2=D_1 S_2+2D_2 S_2-D_3 S_2+D_3 S_4\,,\cr
&\mc{S}_3=D_1 S_2+2D_2 S_2+3D_3 S_2+D_1 S_3+2D_2 S_3+3D_3 S_3-D_3 S_4\,,
\ea
\ee
and we can compute the SymTFT action
\be\label{symtftN=4}
\ba
\frac{S_{\rm SymTFT}}{2\pi}&=\int_{\mc{M}_4}\frac{1}{4}(B_2\smile\delta{A}_1+B_{0}^{1}\delta{A}^{1}_{3}+B_{0}^{2}\delta{A}^{2}_{3})+\frac{1}{8}B_{0}^{3}\delta{A}^{3}_{3}+\frac{1}{8}(6B_{0}^{2}+3B_{0}^{3})B_2\smile B_2\cr
&+\int_{\mc{M}_4}\frac{1}{8}B_{0}^{1} B_2 (-F_{2}^{1}-F_{2}^{2})+\frac{1}{8}B_{0}^{2} B_2 (2F_{2}^{1}-F_{2}^{2})-\frac{1}{8}B_{0}^{3} B_2 F_{2}^{1}\cr
&+\int_{\mc{M}_4}\frac{1}{4}B_{0}^{1} (-(F_{2}^{1})^2+F_{2}^{1} F_{2}^{2}-2(F_{2}^{2})^{2})+\frac{1}{4}B_{0}^{2}((F_{2}^{1})^2-(F_{2}^{2})^{2})
\\
&+\int_{\mc{M}_4}\frac{1}{8}B_{0}^{3}((F_{2}^1)^2-F_{2}^{1} F_{2}^{2}+(F_{2}^{2})^{2})\,.
\ea
\ee
\end{enumerate}

\subsection{Branes, charged defects, and symmetry operators}

Following the general framework outlined in \cite{Heckman:2022muc,Cvetic:2023plv,Najjar:2024vmm}, we analyse the charged defects and symmetry topological operators associated with 3d $\N=2$ theories. Specifically, our approach builds upon the discussion in \cite[Sec.2.2]{Najjar:2024vmm} and references therein.

Charged defects, in the context of both discrete and continuous symmetries, are realized through BPS M-branes wrapping torsional and free cycles, respectively, of the link $L_{7}=Q^{\scriptscriptstyle(1,1,1)}/\Z_{N}$, and extending along the radial direction of the CY4 cone. Concretely, as established in \cite{DelZotto:2015isa,Albertini:2020mdx}, these defects take the form 
\begin{equation}\label{def:defect1}
    \mathbb{D}^{m} := \bigcup_{p=2,5} \{\text{M$p$-branes on } H_{p-m}(L_{7},\Z) \times [0, \infty) \} .
\end{equation}\par

For discrete $p$-form symmetries, the associated symmetry operators originate from BPS M-branes wrapping torsional cycles \cite{Heckman:2022muc}
\begin{equation}\label{def:discrete-sym-op}
    \mathbb{U}^{m'+1}_{\text{Disc.}} := \bigcup_{p=2,5} \{\text{M$p$-branes on } \text{Tor}H_{p-m'}(L_{7},\Z) \,\text{and transverse to} \,[0,\infty)   \}.
\end{equation}
In contrast, continuous symmetry topological operators are realized through $P_{4}$ and $P_{7}$-fluxbranes, as defined in \cite[Sec.2.2]{Najjar:2024vmm}, and are expressed as
\begin{equation}\label{def:continuous-sym-op}
    \mathbb{U}^{m'}_{\text{Cont.}} := \bigcup_{p=2,5} \{\text{$P_{(p+2)}$-fluxbranes on } H^{\text{free}}_{p-m'}(L_{7},\Z) \,\text{and transverse to} \,[0,\infty)  \}.
\end{equation}

Following \cite{Najjar:2024vmm}, the $P_{(p+2)}$-fluxbranes are identified with the Page charges introduced in \cite{Page:1983mke}. The symmetry topological operator constructed via the $P_{4}$-fluxbrane is given by
\begin{equation}
    \mathcal{U}^{P_{4}\text{-flux on}\,\gamma_{k}}(\Sigma_{4-k}) = \exp{i\varphi\int_{\Sigma_{4-k}\times \gamma_{k}} \frac{G_{4}}{2\pi}}\,
\end{equation}
which aligns with the fluxbrane operator discussed in \cite{Cvetic:2023plv}.

The topological symmetry operator corresponding to the $P_{7}$-fluxbrane is given by
\begin{equation}
    \mathcal{U}^{P_{7}\text{-flux on}\,\gamma_{k}}(\Sigma_{7-k}) = \exp{i\varphi\int_{\Sigma_{7-k}\times \gamma_{k}} \frac{P_{7}}{2\pi}}\,.
\end{equation}
Here, $P_{7}$ is defined through the Hopf-Wess-Zumino (HWZ) action as derived in \cite{Bandos:1997ui,Intriligator:2000eq},
\begin{equation}\label{eq:HWZ-action}
 S_{\text{HWZ}}\,=\, \frac{1}{2\pi} \int_{\Sigma_{7}}  \phi^{\ast}G_{7} +\frac{1}{4\pi} \iota_{7,\ast} H_{3}\wedge\phi^{\ast}G_{4}\,.  
\end{equation}
Particularly, the integrand is identified with the $P_{7}$ Page charge associated with the M2-brane \cite{Page:1983mke} and is expressed as, 
\begin{equation}\label{Page-P7-charge}
    P_{7}\, := \phi^{\ast}G_{7} +\frac{1}{4\pi} \iota_{7,\ast} H_{3}\wedge\phi^{\ast}G_{4}\,.
\end{equation}
Here, $\iota_7 : \Sigma_6^{\text{M5}} \hookrightarrow \Sigma_7$ denotes the embedding of the M5-brane worldvolume into $\Sigma_7$. Using this, the pullback map is defined as
\begin{equation}
    \phi^{\ast} := \iota_{7,\ast} \circ \iota^{\ast}_{\text{M5}} \ : \ \Omega^{\bullet} (M_d \times X_{11-d}) \longrightarrow \Omega^{\bullet} (\Sigma_7)\,.
\end{equation}

\subsubsection*{Discrete symmetries from Mp-branes}

In this section, we focus on the symmetry topological operators that generate the discrete global $p$-form symmetries in our model. The brane configurations corresponding to defects and symmetry operators are summarized in Table \ref{tab:M2M5-Q111-ZN}. Generally, these defects exhibit non-trivial linking pairings with the symmetry operators under which they are charged.

\begin{table}[ht]
\centering
\begin{tabular}{|l|c|c|c|c|}
\hline 
& M2 & & M5 & \\
\hline
Tor$H_{1}(L_{7},\Z)\times [0,\infty)$ \hspace{2pt} & Wilson line & $\diamondsuit$ &  &   \\
Tor$H_{3}(L_{7},\Z)\times [0,\infty)$ \hspace{2pt} & --- & $\clubsuit$  & Domain wall & \begin{color}{blue}$\triangle$\end{color} \\
Tor$H_{5}(L_{7},\Z)\times [0,\infty)$ \hspace{2pt} & --- &  & Local operator & \begin{color}{red}$\heartsuit$\end{color} \\
\hline
Tor$H_{1}(L_{7},\Z)$ \hspace{4pt} & 0-form sym. generator & \begin{color}{red}$\heartsuit$\end{color}& &  \\
Tor$H_{3}(L_{7},\Z)$ \hspace{4pt} & 2-form sym. generator & \begin{color}{blue}$\triangle$\end{color} & $(-1)$-form sym. generator & $\clubsuit$ \\
Tor$H_{5}(L_{7},\Z)$ \hspace{4pt} & --- &  & 1-form sym. generator & $\diamondsuit$  \\
\hline
\end{tabular}
\caption{Branes wrapping torsional cycles in $L_{7} = Q^{\scriptscriptstyle(1,1,1)}/\Z_{N}$ give rise to finite symmetries. We mark with equal symbol the charged defect and the corresponding symmetry generators.}
\label{tab:M2M5-Q111-ZN}
\end{table}

\paragraph{Discrete 0/1-form symmetries.}

In 3-dimensional spacetime, the electric 1-form symmetry is dual to the magnetic 0-form symmetry. The topological operator for the electric $\Z_{N}^{[1]}$ 1-form symmetry arises from M5-branes wrapping torsional 5-cycles. The precise form of the symmetry operator is derived from the differential cohomology uplift of the M5-brane’s topological action and is given as  
\begin{equation}
\begin{aligned}
    \exp \left\{  2 \pi i S_{\text{WZ}}^{\text{M5 on }\mathsf{PD}(t_2) } (\Sigma_1) \right\} &= \exp \left\{ 2 \pi i \int^{\br{H}}_{L_7 \times \Sigma_1} \br{t}_2 \star \br{\dd G}_7 \right\} \\
    &= \exp \left\{ 2 \pi i \int^{\br{H}}_{L_7\times \Sigma_1 }\br{t}_2 \star \left[ \br{\mathcal{B}}_2 \star \br{t}_6 + \cdots \right] \right\}  \\
    &= \exp \left\{ -\frac{2 \pi i}{N} \int_{\Sigma_1 } A_1 \right\} .
\end{aligned}
\end{equation}
The symmetry operator takes the expected form, given by the holonomy of the discrete gauge field $A_{1}$ for the dual 0-form symmetry.

The dual $\Z_{N}^{[0]}$ 0-form symmetry is given by M2-brane wrapping torsional 1-cycles. The corresponding symmetry operator  is obtained from the differential cohomology uplift of the M2-brane’s topological action, namely 
\begin{equation}
    \begin{split}
         \exp \left\{  2 \pi i S_{\text{WZ}}^{\text{M2 on }\mathsf{PD}(t_{6}) } (\Sigma_2) \right\} &= \exp{2\pi i \int_{L_{7}\times \Sigma_{2}} \br{t}_{6} \star \br{G}_{4} } 
        \\
        &= \exp{2\pi i \int_{L_{7}\times \Sigma_{2}}\br{t}_{6}\star( \br{B}_{2} \star \br{t}_{2} + \cdots )}
        \\
        & =  \exp{-\frac{2\pi i}{N}  \, \int_{\Sigma_{2}}  {B}_{2}} 
    \end{split}
\end{equation}
Expressed as the holonomy of the discrete gauge field $B_{2}$ for the dual 1-form symmetry. 

\paragraph{An interpretation of the $\Z_{N}$ 0-form symmetry.}

So far, we have focused on the 3d $\N=2$ pure $SU(N)$ gauge theory. From a field-theoretic perspective, this theory can be derived via dimensional reduction on $S^{1}$ from the 4D $\N=1$ pure $SU(N)$ gauge theory. It is well-known that the 4d theory exhibits a discrete electric or magnetic $\Z_{N}$ 1-form symmetries dependent on the global form of the gauge group, where $\Z_{N}$ corresponds to the centre of the $SU(N)$ gauge group \cite{Aharony:2013hda,Gaiotto:2014kfa}. In our geometric engineering set-up, we can think of the $S^{1}$ circle as an auxiliary spatial direction.

When compactifying the 4d theory on $\R^{1,2}\times S^{1}$, the 1-form symmetries, along with the local degrees of freedom and other symmetries, must be considered under reduction. In 4d, both of $\Z_{N}^{\scriptscriptstyle[1,\,e]}$ and $\Z_{N}^{\scriptscriptstyle[1,\,m]}$ symmetries' charged defects are given by Wilson loops and 't Hooft loops, respectively. Upon reduction, these loops have two options: 
\begin{itemize}
    \item Being orthogonal to the $S^{1}$ direction.
    \item Wrapping the $S^{1}$ direction.
\end{itemize}
Since in 3d a discrete 1-form symmetry should be dual to a discrete 0-form symmetry, then a priori there are exactly two choices:
\begin{itemize}
    \item The electric $\Z_{N}^{\scriptscriptstyle[1,\,e]}$ 1-form symmetry survive, while the 4d magnetic $\Z_{N}^{\scriptscriptstyle[1,\,m]}$ turns into a $\Z_{N}^{\scriptscriptstyle[0,\,m]}$ 0-form symmetry.
    \item The magnetic $\Z_{N}^{\scriptscriptstyle[1,\,m]}$ 1-form symmetry survive, while the electric $\Z_{N}^{\scriptscriptstyle[1,\,e]}$ turns into a $\Z_{N}^{\scriptscriptstyle[0,\,e]}$ 0-form symmetry.
\end{itemize}
The first option above is equivalent to the fact that the Wilson loop is orthogonal to the $S^{1}$-direction, while the 't Hooft loops wrapping the compactification circle and appear as a local operator. The second option is the opposite choice. 

Looking back to Table \ref{tab:M2M5-Q111-ZN} and thinking of the $S^{1}$-direction as an auxiliary spatial direction, we observe that our model is consistent with the first choice above. In particular, the Wilson lines are originated from the electric M2-branes, while the local operators are originated from the electromagnetic dual M5-branes.

\paragraph{Discrete $(-1)$/$2$-form symmetries.}

Since the link $L_{7}$ admits torsional 3-cycles as given in (\ref{eq:Gamma-tor-3-cycle}), specifically $\Gamma = \Z_{N}\oplus\Z_{N}\oplus\Z_{2N}$, it is possible to construct three copies of discrete $(-1)$-form symmetries by wrapping M5-branes on these cycles. The corresponding topological operator is given by
\begin{equation}
\begin{aligned}
    \exp \left\{  2 \pi i S_{\text{WZ}}^{\text{M5 on }\mathsf{PD}(t_{4}^{a}) } (\Sigma_3) \right\} &= \exp \left\{ 2 \pi i \int^{\br{H}}_{L_7 \times \Sigma_3} \br{t}_{4}^{a} \star \br{\dd G}_7 \right\} \\
    &= \exp \left\{ 2 \pi i \int^{\br{H}}_{L_7\times \Sigma_3 }\br{t}_{4}^{a} \star \left[ \br{\mathcal{B}}^{b}_4 \star \br{t}^{b}_4 + \cdots \right] \right\}  \\
    &= \exp \left\{ -\frac{2 \pi i}{|\Gamma|^{a}} \int_{\Sigma_3 } A_{3}^{a} \right\}\,, \qquad a=1,2,3\,.
\end{aligned}
\end{equation}
This construction defines three copies of discrete $(-1)$-form symmetries\footnote{The existence of $(-1)$-form symmetry is fundamentally tied to the notion of decomposition \cite{Pantev:2005rh,Pantev:2005wj,Pantev:2005zs,Hellerman:2006zs,Sharpe:2022ene}, and subsequently developed in a range of works including \cite{Seiberg:2010qd,Tachikawa:2013hya,Sharpe:2014tca,Tanizaki:2019rbk,Sharpe:2022ene,Najjar:2024vmm,Najjar:2025htp}. We defer a more detailed discussion of decomposition in 3d $\mathcal{N}=2$ theories to future work.},
\begin{equation}
    \Z_{N}^{\scriptscriptstyle [-1]}\oplus \Z_{N}^{\scriptscriptstyle [-1]}\oplus \Z_{2N}^{\scriptscriptstyle [-1]}\,,
\end{equation}
each associated with a torsional 3-cycle of order $|\Gamma|^{a}$. Specifically, for $a=1$ and $a=2$, $|\Gamma|^{a} = N$, while for $a=3$, $|\Gamma|^{3} = 2N$. 

The 3-form gauge fields $A_{3}^{a}$ are identified as the background gauge fields for the dual 2-form symmetries. The symmetry operators of these dual symmetries are realized through M2-branes wrapping the torsional 3-cycle $\Gamma$. Explicitly, they are given as:
\begin{equation}
\begin{aligned}
        \exp \left\{  2 \pi i S_{\text{WZ}}^{\text{M2 on }\mathsf{PD}(t^{a}_{4}) } (\wp) \right\} &= \exp \left\{ 2 \pi i \int^{\br{H}}_{L_7 \times \{\wp\}} \br{t}_{4}^{a} \star \br{G}_{4} \right\} \\
        &= \exp \left\{ 2 \pi i \int^{\br{H}}_{L_7\times \{\wp\} }\br{t}_{4}^{a} \star \left[ \br{B}_{0}^{b} \star \br{t}_{4}^{b} + \cdots \right] \right\}  \\
        &= \exp \left\{ -\frac{2 \pi i}{|\Gamma|^{a}} \,\mathrm{ev}_{\wp}\, B^{a}_{0} \right\} .
\end{aligned}
\end{equation}
This is identified with the holonomy of the dual $(-1)$-form gauge fields, as expected.

\subsubsection*{Continuous abelian symmetries from fluxbranes}

As defined in \cite{Najjar:2024vmm} and references therein, fluxbranes can be used to construct the topological symmetry operators associated with $U(1)$ continuous $p$-form symmetries. By expanding $\br{G}_{4}$ as in (\ref{expansionG4G7}), we observe the emergence of a 2-form symmetry and two copies of 0-form symmetries generated by $P_{7}$-fluxbranes. Furthermore, we show that additional 0-form symmetries arises as the dual of that seen from the $\br{G}_{4}$ expansion.

\paragraph{Continuous 2-form symmetry.}

In general, the topological symmetry operator for the universal 2-form symmetry originates from the $P_{7}$-fluxbrane wrapping the entire link space for a given scenario. This symmetry universally appears in geometrically engineered theories in M-theory for spacetime dimensions $d\geq 3$.

In the present case, the symmetry operator is
\begin{equation}\label{gen:2-formContinuous}
\begin{aligned}
     \exp \left\{  i \varphi S^{P_7\text{-flux along }L_7} \right\} &=\exp \left\{ i \frac{\varphi}{2\pi} \int_{L_7 \times \{\wp\}} P_7 \right\} \\
    &= \exp \left\{ i\frac{\varphi}{2\pi} \int_{L_7 \times \{\wp\}} \left[ \widetilde{h}_0 \wedge \vol_{7} + \cdots \right] \right\}  \\
    &= \exp \left\{ i \frac{\varphi}{2\pi} \,\mathrm{ev}_{\wp} \,\phi^{\ast}(h_0 + g_0) \right\} .
\end{aligned}
\end{equation}

Here, $g_{0}$ is a correction term arising from the reduction of $H_3 \wedge G_4$. Notably, this reduction cannot be carried out in two separate steps, i.e., first reducing $H_{3}$ and $G_{4}$ and then taking their wedge product. Such an approach would require expanding $H_{3}$ and $G_{4}$ in bases associated with the 5-form $\vol_{5}$ dual to free 2-cycles. To address this, $g_{0}$ is formally defined as  
\begin{equation}\label{eq:def-g_0}
\begin{aligned}
     g_{0} := \frac{1}{4\pi} \int_{L_{7}} H_3 \wedge G_4\,.
\end{aligned}
\end{equation}

Given that we adopt this formal definition of $g_{0}$, the bulk twist theory calculated in (\ref{CSG4forQ111ZN}) will include terms arising from the reduction of $\br{G}_{4}\star\br{G_{4}}$ as a whole. The expected term is given by:
\begin{equation}\label{eq:BF-corrections-F4g0}
\begin{split}
      S_{\text{CS}}^{\text{M}}   \ \supset  &-\frac{1}{6}\int_{\mathcal{M}_{4}\times L_{7}}^{\br{H}}\,(3\cdot 2)\,\left[ (\br{F}_{4}\star \br{1})\star(\br{f}_{1}\star\br{\vol}_{7})\right]
        \\
        & =\,\,\int_{\mathcal{M}_{4}\times L_{7}}^{\br{H}} \,\left[ (\br{1} \star\br{\vol}_{7})\,\times\,(\br{F}_{4} \star \br{f}_{1}) \right]
        \\
         &= (-1)^{\text{dim}(L_{7})}\int_{\mathcal{M}_{4}}^{\br{H}} \,\,\br{F}_{4} \star \br{f}_{1}\,, \qquad \text{with}\ \, \mathscr{F}(\br{f}_{1}) = \frac{1}{2\pi}dg_{0}\,.
\end{split}
\end{equation}
To obtain the final minus sign, we have invoked \cite[(A.28)]{Najjar:2024vmm}. This term can be interpreted as a correction to the continuous BF action. Combining the above contribution, we arrive at:
\begin{equation}
    \int_{\mathcal{M}_{4}}\,\frac{F_{4}}{2\pi}\,\wedge\,\left(\frac{h_{0}}{2\pi}+\frac{g_{0}}{2\pi}\right)\,,
\end{equation}
which mirrors the structure of the symmetry operator in (\ref{gen:2-formContinuous}).

\paragraph{Continuous 0-form symmetry.}

Using the $P_{7}$-fluxbrane one can construct two copies of 0-form symmetries by wrapping the homologically distinct free 5-cycles of (\ref{homologyQ111ZN}). Explicitly, these operators are given as
\begin{equation}\label{Sym-op-(0)-form-U(1)-(1)}
\begin{aligned}
        \exp \left\{  i \varphi S^{P_7\text{-flux along }H_{5}} \right\} &= \exp \left\{ i \frac{\varphi}{2\pi} \int_{L_7 \times \Sigma_2} \vol_2^{\neq i} \wedge P_7 \right\} \\
        &= \exp \left\{ i \frac{\varphi}{2\pi} \int_{L_7 \times \Sigma_{2}} \vol_2^{\neq i} \wedge \left[ \sum_{j=\mathrm{f,b}} \widetilde{h}_2^{j} \wedge \vol_5^{j} + \cdots \right] \right\}  \\
        &= \exp \left\{ i \frac{\varphi }{2\pi}\int_{\Sigma_{2}} \phi^{\ast} \left( h_2^{i} + g_2^{i} \right) \right\}\,.
\end{aligned}
\end{equation}
Here, we formally define
\begin{equation}
    g_{2}^{i} := \frac{1}{4\pi}\,\int_{L_{5}^{i}}\,H_{3}\wedge G_{4}\,,
\end{equation}
due to the same issue discussed before (\ref{eq:def-g_0}). Alternatively, one could utilize the (co)homology group of $L_{5}^{i}$ to rigorously define $g_{2}^{i}$. Specifically, since each $L_{5}^{i}$ is a copy of $T^{\scriptscriptstyle (1,1)}/\Z_{N}$, with its homology group given in (\ref{homologyT11ZN}), $g_{2}^{i}$ can be expressed as:
\begin{equation}
   \begin{split}
        g_{2}^{i}\,&=\,\frac{1}{4\pi}\,\int_{L_{5}^{i}}\,\left(\,\overline{h}_{1}\wedge\vol_{2}+\overline{h}_{0}\wedge\vol_{3}\,\right)\wedge\left(\,\overline{g}_{2}\wedge\vol_{2}+\overline{g}_{1}\wedge\vol_{3}\,\right)
        \\
        & \sim \frac{1}{4\pi}\,\, \left(\,\overline{h}_{1}\wedge\overline{g}_{1} + \overline{h}_{0}\wedge\overline{g}_{2}\,\right)
   \end{split}
\end{equation}

To overcome the issue of defining $g_{0}$ and $g_{2}$ rigorously, we propose expanding the $P_{7}$-flux in a basis of differential forms from the outset. This ensures a consistent and well-defined framework for constructing the symmetry operators.

Assuming the above, and following (\ref{eq:BF-corrections-F4g0}), a correction to the continuous BF term for the 0-form symmetries can be derived as:
\begin{equation}\label{eq:BF-corrections-F2g2}
\begin{split}
      S_{\text{CS}}^{\text{M}}   \ \supset  &-\frac{1}{6}\int_{\mathcal{M}_{4}\times L_{7}}^{\br{H}}\,(3\cdot 2)\,\left[ (\br{F}_{2}\star \br{\vol}_{2})\star(\br{f}_{3}\star\br{\vol}_{5})\right]
        \\
        & =\,\, \int_{\mathcal{M}_{4}\times L_{7}}^{\br{H}} \,\left[ (\br{\vol}_{2} \star\br{\vol}_{5})\,\times\,(\br{F}_{2} \star \br{f}_{3}) \right]
        \\
         &= (-1)^{\text{dim}(L_{7})}\int_{\mathcal{M}_{4}}^{\br{H}} \,\,\br{F}_{2} \star \br{f}_{3}\,, \qquad \text{with}\ \, \mathscr{F}(\br{f}_{3}) = \frac{1}{2\pi}dg_{2}\,.
\end{split}
\end{equation}
Combining this result with the continuous BF terms, we obtain:
\begin{equation}
    \int_{\mathcal{M}_{4}}\,\frac{F_{2}}{2\pi}\,\wedge\,\left(\frac{h_{2}}{2\pi}+\frac{g_{2}}{2\pi}\right)\,,
\end{equation}
which reflects the structure of the symmetry operator in (\ref{Sym-op-(0)-form-U(1)-(1)}).

Wrapping $G_{4}$-fluxbranes along the free 2-cycles, generate two additional 0-form symmetries. The corresponding topological generators can be expressed as
 \begin{equation}\label{Sym-op-(0)-form-U(1)-(2)}
    \begin{aligned}
        \exp \left\{  i \varphi S^{G_4\text{-flux along }H_{2}} \right\} &= \exp \left\{ i \frac{\varphi}{2\pi} \int_{L_7 \times \Sigma_2} \vol_5^{\neq i} \wedge G_4 \right\} \\
        &= \exp \left\{ i \frac{\varphi}{2\pi} \int_{L_7 \times \Sigma_{2}} \vol_5^{\neq i} \wedge \left[ \sum_{j=\mathrm{f,b}} \widetilde{F}_2^{j} \wedge \vol_2^{j} + \cdots \right] \right\}  \\
        &= \exp \left\{ i \frac{\varphi }{2\pi}\int_{\Sigma_{2}} F_2^{i} \right\} .
\end{aligned}
\end{equation}
These 0-form symmetries are dual to the previously discussed 0-form symmetries, as demonstrated in \cite[(2.34)]{Najjar:2024vmm}. The duality arises from the exchange of M2- and M5-branes, which act as the defects generating these symmetries, under an electromagnetic transformation in M-theory. This transformation also exchanges the $P_{4}$-fluxbrane with the $P_{7}$-fluxbrane accordingly.

\section{The CY4 geometric transition}\label{sec:geometrical-transition}

\subsection{Resolution and deformation phases of \texorpdfstring{$\mc{C}(Q^{\scriptscriptstyle(1,1,1)})$}{C(Q(1,1,1))}}

We discuss the resolution and deformation of the CY4 cone over $Q^{\scriptscriptstyle(1,1,1)}$, which is an isolated singularity $\mc{C}(Q^{\scriptscriptstyle(1,1,1)})$ described with a non-complete-intersection of nine equations in $\mb{C}^8$:
\be
\ba
\label{Q111-eq}
&z_1 z_2-z_3 z_4=0\ ,\ z_5 z_6-z_7 z_8=0\ ,\ z_1 z_7-z_3 z_5=0\cr
&z_4 z_6-z_2 z_8=0\ ,\ z_1 z_4-z_5 z_8=0\ ,\ z_1 z_6-z_3 z_8=0\cr
&z_2 z_3-z_6 z_7=0\ ,\ z_2 z_5-z_4 z_7=0\ ,\ z_1 z_2-z_5 z_6=0\,.
\ea
\ee
One can check that the Jacobian matrix has rank 4 at a generic point on $\mc{C}(Q^{\scriptscriptstyle(1,1,1)})$, and has rank 0 at the origin $z_i=0$ $(i=1,\dots,8)$ where $\mc{C}(Q^{\scriptscriptstyle(1,1,1)})$ is singular. As another comment, the ninth equation of (\ref{Q111-eq}) is actually redundant, and can be ignored.

\paragraph{Resolution $\widetilde{\mc{C}(Q^{\scriptscriptstyle(1,1,1)})}$}

We perform the crepant resolution by introducing a set of $\mb{P}^1$ projective coordinates $[\mu_1:\mu_2]$ and $[\lambda_1:\lambda_2]$, and resolve (\ref{Q111-eq}) into
\be
\ba
\label{CQ111-resol}
\bp z_1 & z_3\\z_4 & z_2\\z_5 & z_7\\z_8 & z_6\ep\bp \lambda_1\\ \lambda_2\ep=\bp 0 \\ 0\\ 0\\ 0\ep\cr
\bp z_1 & z_8\\z_5 & z_4\\z_3 & z_6\\z_7 & z_2\ep\bp \mu_1\\ \mu_2\ep=\bp 0\\ 0\\ 0\\ 0\ep\,.
\ea
\ee
One can check that the CY4 is Jacobian matrix has rank 6 at any point, and is now fully smooth, hence (\ref{CQ111-resol}) is a valid crepant resolution of the singularity (\ref{Q111-eq}). 

The exceptional locus is $\mb{P}^1\times\mb{P}^1$, and the projective coordinates of the two $\mb{P}^1$s are $[\mu_1:\mu_2]$ and $[\lambda_1:\lambda_2]$.

\paragraph{Partial Deformation+Resolution (DR) phase $\overline{\mc{C}(Q^{\scriptscriptstyle(1,1,1)})}$ }

In order to get an exceptional locus of $\mathbb{S}^{2}\times \mathbb{S}^{3}$, we can attempt to do the following partial resolution + deformation. We first do a partial resolution: from (\ref{Q111-eq}) we introduce the projective coordinates $[\mu_1:\mu_2]$ and write down the following equations
\be
\bp z_1 & z_8\\z_5 & z_4\\ z_3 & z_6\\z_7 & z_2\ep\bp\mu_1\\ \mu_2\ep=\bp 0\\0\\0\\0\ep\,.
\ee
Then we do a deformation and finally get the following set of equations:
\be
\ba
\label{Q111-partial-2}
        z_{1}z_{2}-z_{3}z_{4}&=\epsilon\neq 0
        \cr
        z_5 z_6-z_7 z_8&=-\epsilon\cr
        z_{1}\mu_1+z_8\mu_2&=0\cr
        z_{5}\mu_1+z_4\mu_2&=0\cr
        z_{3}\mu_1+z_6\mu_2&=0\cr
        z_{7}\mu_1+z_2\mu_2&=0
        \,.
\ea
\ee
 One can check that the Jacobian matrix of the equations (\ref{Q111-partial-2}) has rank 5 at any point, and hence the non-compact CY4 $\overline{\mc{C}(Q^{\scriptscriptstyle(1,1,1)})}$ is now fully smooth. 

The exceptional $\mathbb{S}^{3}$ is defined from the first equation
\be
\label{z1234-ep}
z_1 z_2-z_3 z_4=\epsilon
\ee
as
\be
\bp z_1 & z_4\\z_3 & z_2\ep=\bp a+bi & c+di\\-c+di & a-bi\ep\quad (a,b,c,d\in\mb{R}\ ,\ a^2+b^2+c^2+d^2=\epsilon)\,,
\ee
along with a particular fixed set of $(z_5,z_6,z_7,z_8,\mu_1,\mu_2)$. In fact, after defining the ratio $p=\mu_1/\mu_2$ (when $\mu_1,\mu_2\neq 0$), we can plug 
\be
z_8=-z_1 p\ ,\ z_5=-z_4 p^{-1}\ ,\ z_6=-z_3 p\ ,\ z_7=-z_2 p^{-1}\,,
\ee
into the second equation of (\ref{Q111-partial-2}), which results in the same equation of (\ref{z1234-ep}). Hence there is only a single exceptional $\mathbb{S}^{3}$ on $\overline{\mc{C}(Q^{\scriptscriptstyle(1,1,1)})}$.

As usual, the exceptional $\mb{P}^1\cong \mathbb{S}^{2}$ is parametrized by the projective coordinates $[\mu_1:\mu_2]$. One may notice that the north pole $[\mu_1:\mu_2]=[1,0]$ would correspond to $z_2,z_4,z_6,z_8\rightarrow\infty$ and the south pole $[\mu_1:\mu_2]=[0,1]$ would correspond to $z_1,z_3,z_5,z_7\rightarrow\infty$. Nonetheless these points on the exceptional $\mb{P}^1$ do not correspond to infinite distance points in respect to the origin, after giving a properly defined metric.

As can be seen from the equations, after one fixes a point $(z_1,z_2,z_3,z_4)$ on the $\mathbb{S}^{3}$ in (\ref{Q111-partial-2}), the point $[\mu_1:\mu_2]$ on the $\mb{P}^1$ can be freely chosen. Hence the exceptional $\mathbb{S}^{3}$ and $\mathbb{S}^{2}$ defined in this way do not mix with each other, and the total exceptional locus is indeed $\mathbb{S}^{3}\times \mathbb{S}^{2}$.

\subsection{On a CY4 metric on the DR-phase}

\paragraph{General discussion: Ricci flat K\"ahler metrics.}

A Calabi-Yau manifold admits a K\"ahler metric that is Ricci-flat. It is known that a K\"ahler metric can be expressed in terms of a K\"ahler potential $\mathcal{F}$ as (e.g., \cite{Candelas:1989js,Hou:1999qc,Nakahara:2003nw})
\begin{equation}
    g_{\mu\bar{\nu}}\,=\, \partial_{\mu}\partial_{\bar{\nu}}\,\mathcal{F}\,,
\end{equation}
with $\mathcal{F}$ is a real-valued function on the manifold.

The isometries of the underlying space constrain the functional form of $\mathcal{F}$. In the cases where the  K\"ahler potential is invariant under the action of some Lie group, $\mathcal{F}$ depends solely on the radial coordinate squared, i.e., 
\begin{equation}\label{eq:F=F(r2)}
    \mathcal{F}\,=\,\mathcal{F}(r^{2})\,.
\end{equation}
For this class of potentials, the K\"ahler metric takes the explicit form:
\begin{equation}\label{eq:kahler-metric-(2)}
   g_{\mu\bar{\nu}}\,=\, (\partial_{\mu}\partial_{\bar{\nu}}r^{2})\,\mathcal{F}' \,+ \,(\partial_{\mu}r^{2})(\partial_{\bar{\nu}}r^{2})\,\mathcal{F}''\,, 
\end{equation}
where the prime means derivative with respect to $r^{2}$. 

The singular cone $\mathcal{C}(Q^{\scriptscriptstyle(1,1,1)})$ possesses an $(SU(2))^3$ isometry group, sufficiently large to enforce the above behaviour on $\mathcal{F}$. The symmetry group extends to the DR-phase through its $\mathbb{S}^{3} \times \mathbb{S}^{2}$ zero-section geometry (or a non-trivial subgroup), and maintains the functional constraint on $\mathcal{F}$.

For a K\"ahler metric $g_{\mu\bar{\nu}}$, the Ricci curvature takes the form (e.g., \cite{Candelas:1989js,Hou:1999qc,Nakahara:2003nw})
\begin{equation}
    R_{\mu\bar{\nu}}\,=\,\partial_{\mu}\partial_{\bar{\nu}}\,\ln(\sqrt{g})\,,
\end{equation}
with $g^{1/2}=\det[g_{\mu\bar{\nu}}]$. A Ricci-flat K\"ahler metric, i.e., a CY metric, therefore requires:
\begin{equation}
   \det[g_{\mu\bar{\nu}}] \, =\, \mathrm{constant}\,. 
\end{equation}

Under the isometry constraint in (\ref{eq:F=F(r2)}), the Ricci-flatness condition reduces to a non-linear ordinary differential equation for a function $f$ defined by
\begin{equation}
    f\,:=\, r^{2}\,\mathcal{F}'\,.
\end{equation}
The existence of a well-behave solution $f$  implies the existence of the CY metric \cite{Candelas:1989js}.

\paragraph{On the structure of the CY4 metric on $\mathcal{C}(Q^{\scriptscriptstyle(1,1,1)})$.}

We can view the CY4 cone $\mathcal{C}(Q^{\scriptscriptstyle(1,1,1)})$ as two interlaced copies of the CY3 conifold $\mathcal{C}(T^{\scriptscriptstyle(1,1)})$ as described in appendix \ref{app:interlacing-geometry}. Each conifold is described by one of the following matrices
\begin{equation}\label{eq:W1-W2}
 W_{(1)}\,=\,   \begin{pmatrix}
        z_{1} & z_{3}
        \\
        z_{4} & z_{2}
\end{pmatrix}
\,,\qquad 
W_{(2)}\,=\,   \begin{pmatrix}
        z_{5} & z_{7}
        \\
        z_{8} & z_{6}
\end{pmatrix}\, ,
\end{equation}
with $z_{1},\cdots, z_{8}\in\C^{8}$. The two matrices satisfy
\begin{equation}\label{eq:det-W1-W2=zero}
    \det(W_{(i)})\,=\,0\,,\quad\mathrm{for} \ \  i=1,2\,.
\end{equation}
which are equivelent to the first two conditions in (\ref{Q111-eq}), togather with the rest of the conditions defined the CY4 cone. 

To deal with the CY4 metric, we define a $4\times4$ matrix:
\begin{equation}
    W \,=\,
    \begin{pmatrix}
        W_{(1)} & 0 
        \\
        0 & W_{(2)}
    \end{pmatrix}\,,
\end{equation}
built from (\ref{eq:W1-W2}). The matrix $W$ enable us to deal with the CY4 cone $\mathcal{C}(Q^{\scriptscriptstyle(1,1,1)})$ space in a similar way to the CY3 conifold of \cite{Candelas:1989js}. We observe that the parameter $r^{2}$, the square of the radial direction for the cone, is obtained as
\begin{equation}
    \tr(W^{\dagger}W)\,=\,r^{2}\,=\, \sum_{I=1}^{8}\,|z_{I}|^{2}\,.
\end{equation}
From (\ref{eq:kahler-metric-(2)}), we learn that the general form of the CY4 metric is given as
\begin{equation}\label{eq:CY4-metric-general}
\begin{split}
    ds^{2}\,&=\,    \mathcal{F}'\,\left(\tr(dW^{\dagger}_{(1)}dW_{(1)}) + \tr(dW^{\dagger}_{(2)}dW_{(2)})\right) 
    \\
    &+ \mathcal{F}''\,\left|\tr(W^{\dagger}_{(1)}dW_{(1)}) + \tr(W^{\dagger}_{(2)}dW_{(2)})\right|^{2} \,.
\end{split}
\end{equation}

With the parametrization given in (\ref{eq:W1-W2}) it is hard to deal with the above metric. To over come this issue, we parametrize the $W_{(i)}$ matrices as   
\begin{equation}\label{eq:W1-W2-a-b}
    W_{(1)}\,=\, 
    \begin{pmatrix}
        b_{1}
        \\
        b_{2}
    \end{pmatrix}
    \begin{pmatrix}
        a_{1} & a_{2}
    \end{pmatrix}
    \,,\qquad 
       W_{(2)}\,=\, 
    \begin{pmatrix}
        b_{3}
        \\
        b_{4}
    \end{pmatrix}
    \begin{pmatrix}
        a_{1} & a_{2}
    \end{pmatrix}\,.
\end{equation}
The relation between the $\{z_{I}\}$ coordinates and the new coordinates $\{a_{1},a_{2},b_{1},b_{2},b_{3},b_{4}\}$ is given by equating the two representations. Furthermore, the conditions in (\ref{Q111-eq}) are now given as
\begin{equation}\label{eq:new-conditions-a-b}
    \begin{split}
        &a_{1}b_{1}a_{2}b_{2}-a_{2}b_{1}a_{1}b_{2}=0\,,\, a_{1}b_{3}a_{2}b_{4}-a_{2}b_{3}a_{1}b_{4}=0\,,\,a_{1}b_{1}a_{2}b_{3}-a_{2}b_{1}a_{1}b_{3}=0\,,
        \\
        &a_{1}b_{2}a_{2}b_{4}-a_{2}b_{2}a_{1}b_{4}=0\,,\,a_{1}b_{1}a_{1}b_{2}-a_{1}b_{3}a_{1}b_{4}=0\,,\,a_{1}b_{1}a_{2}b_{4}-a_{2}b_{1}a_{1}b_{4}=0\,,
        \\
        &a_{2}b_{2}a_{2}b_{1}-a_{2}b_{4}a_{2}b_{3}=0\,,\,a_{2}b_{2}a_{1}b_{3}-a_{1}b_{2}a_{2}b_{3}=0\,,\,a_{1}b_{1}a_{2}b_{2}-a_{1}b_{3}a_{2}b_{4}=0\,.
    \end{split}
\end{equation}
From the above we note the following non-trivial condition on $b_{j}$ (for $j=1,2,3,4$)
\begin{equation}\label{new-condition-on-b1234}
    b_{1}b_{2}-b_{3}b_{4}\,=\,0\,.
\end{equation}

Moreover, we note that in the new parametrization, the radial-square $r^{2}$ is given as
\begin{equation}
    \tr(W^{\dagger}W)\,=\,r^{2}\,=\, (|a_{1}|^{2}+|a_{2}|^{2})\,(|b_{1}|^{2}+|b_{2}|^{2}+|b_{3}|^{2}+|b_{4}|^{2})\,.
\end{equation}

While the derivation of explicit metrics for both the singular and resolved cones would provide valuable geometric insight, we defer this non-trivial analytical task to future investigation.

\paragraph{The (numerical) CY4 metric on the DR-phase.}

The decomposition of the CY4 metric into two interlaced CY3 conifolds (\ref{eq:W1-W2-a-b}) allows us to treat each conifold independently. Specifically, we can consider deforming $W_{(1)}$:
\begin{equation}
     W_{(1)}^{\mathrm{D}}\,=\, 
    \begin{pmatrix}
        b_{1}
        \\
        b_{2}
    \end{pmatrix}
    \begin{pmatrix}
        a_{1} & a_{2}
    \end{pmatrix}\,.
\end{equation}
The conditions in (\ref{eq:new-conditions-a-b}) can be used to set $b_{4}=0$. Furthermore, we consider resolving $W_{(2)}$, i.e., 
\begin{equation}
   W_{(2)}^{\mathrm{R}}\,=\,
    \begin{pmatrix}
        -a_{1}\lambda\,\, & \,\,a_{1}
        \\
        -a_{2}\lambda\,\, & \,\,a_{2}
    \end{pmatrix}\,.
\end{equation}
Here, the parameter $\lambda = \lambda_{1}/\lambda_{2}$ with $(\lambda_{1},\lambda_{2})$ being the coordinate on $\C\P^{1}$ of the resolution. 

In the DR-phase, the $r^{2}$ parameter becomes 
\begin{equation}\label{eq:r2-for-DR-phase}
    r^{2}\,=\, (|a_{1}|^{2}+|a_{2}|^{2})\,(|b_{1}|^{2}+|b_{2}|^{2})\,+ \,(|a_{1}|^{2}+|a_{2}|^{2})(1+|\lambda|^{2})\,. 
\end{equation}

The general CY4 metric structure for the DR-phase (following (\ref{eq:CY4-metric-general})) takes the form
\begin{equation}\label{eq:DR-metric-general}
\begin{split}
    ds^{2}_{\scriptscriptstyle\mathrm{DR}}\,&=\,    \mathcal{F}'\,\left(\tr(d(W^{\mathrm{D}}_{(1)})^{\dagger}dW_{(1)}^{\mathrm{D}}) + \tr(d(W^{\mathrm{R}}_{(2)})^{\dagger}dW_{(2)}^{\mathrm{R}})\right) 
    \\
    &+ \mathcal{F}''\,\left|\tr((W^{\mathrm{D}}_{(1)})^{\dagger}dW_{(1)}^{\mathrm{D}}) + \tr((W^{\mathrm{R}}_{(2)})^{\dagger}dW_{(2)}^{\mathrm{R}})\right|^{2} 
    \\
    \,&+\, 4a^{2}\frac{|d\lambda|^{2}}{\Lambda^{2}}\,.
\end{split}
\end{equation}
The last term is Fubini–Study metric for the $\C\P^{1}\cong \mathbb{S}^{2}$ which results from resolving $W_{(2)}$, see \cite[\S4]{Candelas:1989js}. The parameter $\Lambda$ is defined as
\begin{equation}
    \Lambda\,=\, 1\,+\, |\lambda|^{2}\,,
\end{equation}
which can be represented as $\Lambda \sim r$.

Using (\ref{new-condition-on-b1234}), we may take $(a_{1},a_{2},b_{1})$ as coordinates to describe the conifold given by $W_{(1)}$. The determinant for the DR-metric can be written as
\begin{equation}
\begin{split}
   \Lambda^{2}\, \det(g)\,= \,&\,4a^{2}r^{3}(\mathcal{F}')^{3}\,+\, r^{4}\Lambda^{2}(\mathcal{F}')^{4}\,+\,4a^{2}r^{4} (\mathcal{F}')^{2}\,\mathcal{F}'' (|a_{1}|^{2}+|a_{2}|^{2}+|b_{1}|^{2}) 
    \\
    & + r^{5}\Lambda^{2}(|a_{1}|^{2}+|a_{2}|^{2}+|b_{1}|^{2}+|\lambda|^{2})(\mathcal{F}')^{3}\mathcal{F}''\,.
\end{split}
\end{equation}
Let us define two functions $G$ and $H$ as
\begin{equation}
    G(r;\epsilon)\,=\,|a_{1}|^{2}+|a_{2}|^{2}+|b_{1}|^{2}\,,\quad\mathrm{and}\quad  H(r;\epsilon)\,=\,|a_{1}|^{2}+|a_{2}|^{2}+|b_{1}|^{2}+|\lambda|^{2}\,,
\end{equation}
where they only depend on the radial $r$ variable which we determine shortly. Using the redefinition
\begin{equation}
f\,=\,r^{2}\,\mathcal{F}'\,,
\end{equation}
we can rewrite the metric determinant as:
\begin{equation}
\begin{split}
  r^{3}\Lambda^{2}\,  \det(g)\,= \,&\,  f^{3}\,\left(\, 4a^{2}(1-\frac{G}{r})\,+\,\frac{\Lambda^{2}f}{r^{2}}\,(r\, -\, H) \,    \right)
  \\
  \,&\,+\, f^{2}f'\,\left(\, 4a^{2}\,rG\,+\, \Lambda^{2} Hf    \,\right)\,.
\end{split}
\end{equation}
The deformation can be introduced by shifting $r^{2}$ in (\ref{eq:r2-for-DR-phase}) to $(r^{2}-\epsilon^{2})$. Roughly, the functions $G$ and $H$ can be written as
\begin{equation}
    G\,\simeq\,r\,,\quad\mathrm{and}\,\quad\, H\,\simeq\,r-\epsilon\,,
\end{equation}
such that $r^{n}H\simeq r^{n+1}-\epsilon^{n+1}$.

The above $\det(g)$ equation then can be expressed as
\begin{equation}\label{eq:diff-eq-1}
    r^{5}\,\det(g)\,=\, \frac{\epsilon^{3}f^{4}}{r^{2}}\, + \,f^{2}f'\,\left(4a^{2}r^{2} \,+\, (r^{3}-\epsilon^{3})\,f \right)\,.
\end{equation}

\begin{figure}[H]
\centering
\includegraphics[height=5cm]{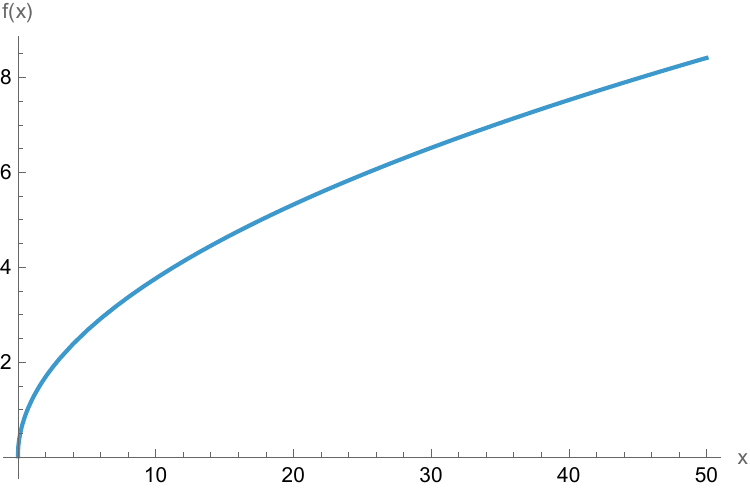}
\caption{The numerical solution of $f(x)=f(r^{2})$ to the differential equation (\ref{eq:diff-eq-1}), in the singular case of $a=\epsilon=0$. We set the constant$=1$, and initial value $f(0.01)=0.01$.}
\label{fig:1-sing}
\end{figure}

\begin{figure}[H]
\centering
\includegraphics[height=5cm]{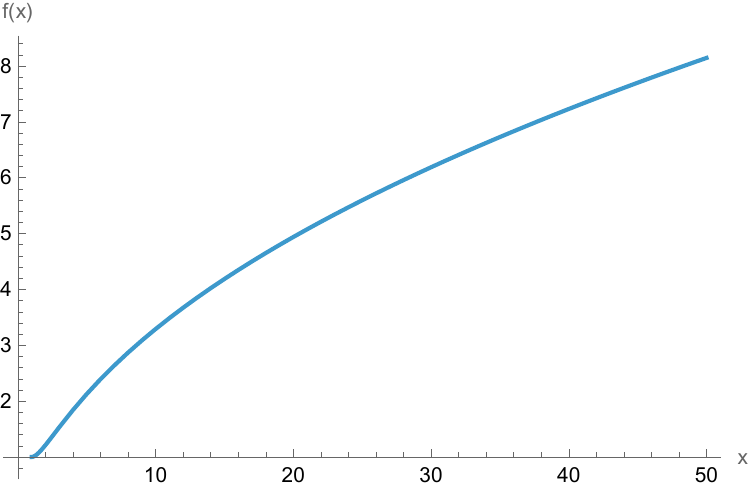}
\caption{The numerical solution of $f(x)=f(r^{2})$ to the differential equation (\ref{eq:diff-eq-1}), in the smooth case of $a=\epsilon=1$. We set the constant$=1$, and initial value $f(1.01)=1$.}
\label{fig:1-smooth}
\end{figure}

We present numerical solution to the above differential equation in figures \ref{fig:1-sing} and \ref{fig:1-smooth}, where we have performed a change of variables from $r$ to $x$, defined by $x=r^{2}$.

First, in the singular case of $a=\epsilon=0$, we solve the differential equation by setting the constant$=1$, and initial value $f(0.01)=0.01$, with the solutions in figure \ref{fig:1-sing}. For the singular case, one can simply solve the differential equation and find that $f\propto r$ as $r\rightarrow\infty$. A redefinition of the radial coordinate can bring the metric into the canonical form of a cone metric.

For the smooth case of $a=\epsilon=1$, we solve the differential equation by setting the constant$=1$, and initial value $f(1.01)=1$, see the solutions in figures \ref{fig:1-smooth}. One can check that $f\propto r$ as $r\rightarrow\infty$, such that the smoothing do not affect the asymptotic behaviour of the metric.

Now we examine the behaviour of the metric near $r\to\epsilon$. Let us write a solution for $W^{\mathrm{D}}_{(1)}$ and $W_{(2)}^{\mathrm{R}}$ as 
\begin{equation}
\begin{split}
        &W_{(1)}\,\simeq\, \epsilon\,L_{(1)}\,\sigma_{3}\,R^{\dagger}\, + \sqrt{r^{2}-\epsilon^{2}}\,L_{(1)}\,Z_{0}^{(1)}\,R^{\dagger}\,,
        \\
        &W_{(2)}\,\simeq\, \sqrt{r^{2}-\epsilon^{2}}\,L_{(2)}Z_{0}^{(2)}\,R^{\dagger}\,,
\end{split}
\end{equation}
where $L_{(1)}$, $L_{(2)}$, and $R$ are elements of the isometry group $(SU(2))^{3}$ of the metric, as exhibited in (\ref{eq:isometry-Q111-ZN-from-metric}). These expressions are constructed to satisfy the constraints (\ref{eq:det-W1-W2=zero}) and (\ref{eq:r2-for-DR-phase}) upon shifting $r^{2}\to r^{2}-\epsilon^{2}$. In the limit $r\to\epsilon$, only the first term of the first equation survives.

Following \cite{Candelas:1989js}, we define a $T$ matrix as
\begin{equation}
    T\,=\,L_{(1)}\,\sigma_{3}\,R^{\dagger}\,\sigma_{3}\,\in\, SU(2)\,.
\end{equation}
Then from the general form of the metric (\ref{eq:DR-metric-general}), in the limit $r\to\epsilon$, we get 
\begin{equation}
    \restr{ds^{2}_{\scriptscriptstyle\mathrm{DR}}}{r\to\epsilon}\,\sim\,  f(\epsilon) \,\tr(dT^{\dagger}dT)\,+\, 4a^{2}\frac{|d\lambda|^{2}}{\Lambda^{2}}\,.
\end{equation}
From the numerical solution, we learn that $f(\epsilon)\sim\epsilon$. The first term above defines a 3-sphere $\mathbb{S}^{3}$, along with the second term, one finds the $\mathbb{S}^{3}\times\mathbb{S}^{2}$ zero-section for the DR-phase.

At this stage, we conclude our analysis of the CY4 metric for the DR-phase. Further details—including other phases, flop transitions, and properties of the proposed metric—will be discussed in a future work.

\subsection{Resolution and deformation phases of \texorpdfstring{$\mc{C}(Q^{\scriptscriptstyle(1,1,1)}/\mb{Z}_N)$}{C(Q(1,1,1)/ZN)}}

In this section we consider the quotient of (\ref{Q111-eq}) by the equivalence relation
\be
\label{ZN-quotient}
(z_1,z_2,z_3,z_4,z_5,z_6,z_7,z_8)\sim (\lambda z_1,\lambda^{-1} z_2,\lambda z_3,\lambda^{-1}z_4,\lambda^{-1}z_5,\lambda z_6,\lambda^{-1}z_7,\lambda z_8)\,,
\ee
where $\lambda=e^{2\pi i/N}$. Now we similarly do a partial deformation + resolution, leading to a smooth geometry $\overline{\mc{C}(Q^{\scriptscriptstyle(1,1,1)}/\mb{Z}_N)}$ with the same set of equations
\begin{equation}
    \begin{split}
    \label{Q111ZN-deformed}
        z_{1}z_{2}-z_{3}z_{4}&=\epsilon\neq 0
        \cr
        z_5 z_6-z_7 z_8&=-\epsilon\cr
        z_{1}\mu_1+z_8\mu_2&=0\cr
        z_{5}\mu_1+z_4\mu_2&=0\cr
        z_{3}\mu_1+z_6\mu_2&=0\cr
        z_{7}\mu_1+z_2\mu_2&=0
        \,.
    \end{split}
\end{equation}
Note that all of the above equations are invariant under the $\mb{Z}_N$ action (\ref{ZN-quotient}). The exceptional $\mb{P}^1$ has projective coordinates $[\mu_1:\mu_2]$.

Similar to before, from the first equation in (\ref{Q111ZN-deformed}) , we can identify a compact 3-cycle with the parametrization
\be
\label{3-cycle-ZN}
z_1=a+bi\ ,\ z_2=a-bi\ ,\ z_3=c+di\ ,\ z_4=-c+di\ (a,b,c,d\in\mb{R}\ ,\ a^2+b^2+c^2+d^2=\epsilon)\,.
\ee
Due to the $\mb{Z}_N$ quotient action (\ref{ZN-quotient}), the $\mathbb{S}^{3}$ defined in (\ref{3-cycle-ZN}) should also be quotiented by $\mb{Z}_N$, which results in a compact 3-cycle with topology $\mathbb{S}^{3}/\mb{Z}_N$.

Similar to the unquotiented case, the $\mathbb{S}^{3}/\mb{Z}_N$ is independent of the $\mb{P}^1$ coordinate, hence the exceptional locus is a direct product space $\mathbb{S}^{3}/\mb{Z}_N\times \mathbb{S}^{2}$.

\subsection{Physics of the new DR-phase}

Here, we examine the physics of M-theory on the new desingularization (DR-phase) $\overline{\mc{C}(Q^{(1,1,1)})}$ and $\overline{\mc{C}(Q^{(1,1,1)}/\mb{Z}_N)}$. Note that $\overline{\mc{C}(Q^{(1,1,1)})}$ corresponds to the case of $N=1$, and there is no compact torsional cycles in $\overline{\mc{C}(Q^{(1,1,1)})}$.

\paragraph{Expansion of the M-theory 3-form.}

In a given geometric compactification background in M-theory, and in the absence of any superpotentials, one may obtain massless states through the expansion of the M-theory $C_{3}$-form field in bases of $L^{2}$-normalizable harmonic forms of the compactification background. Thus, the number of massless degrees of freedom is determined by the dimension of the space of $L^{2}$-normalizable harmonic forms, denoted by $\mathcal{H}^{p}_{L^{2}}(M)$, up to $p=3$.

In our case, the geometric space $M$ is the cone over $Q^{\scriptscriptstyle(1,1,1)}$, up to the $\Z_{N}$ quotient defined in \ref{quotientCQZNonzi}, with an asymptotically conical (AC) metric, that behaves as
\begin{equation}
    g(X) \ \rightarrow\ dr^{2} + r^{2}\  h(Q^{\scriptscriptstyle(1,1,1)})\,,\qquad \text{as}\ r\rightarrow\infty.
\end{equation}
The asymptotic behaviour of the DR-phase matches that of the singular metric as can be seen from figures \ref{fig:1-sing}-\ref{fig:1-smooth}. This justifies the application of the following theorem.

Following \cite[Theorem 1A]{Hausel:2002xg}, the space $\mathcal{H}^{p}_{L^{2}}(M)$ of $L^{2}$-normalizable cohomology is given as
\begin{equation}
    \mathcal{H}^{p}_{L^{2}}(M) \, \cong \,  \left\{
   \begin{aligned}
      &H^{p}(M,\partial M), & p < m/2,
     \\
     & \text{Im}(H^{p}(M,\partial M)\ \rightarrow\ H^{p}(M)), & p = m/2,
     \\
     & H^{p}(M), & p > m/2.
   \end{aligned}
   \right.
\end{equation}
Here, $H^{p}(M)$ is the de Rham cohomology, and $H^{p}(M,\partial M)$ is the relative cohomology with respect to the boundary, i.e. the link space $L_{7}$, of $M$. The above theorem teaches us that the space $\mathcal{H}^{p}_{L^{2}}(M)$ is topological. The discussion in \cite{e0a630bb-2fb9-3d1a-a046-a4c4cc00dcd5,ASNSP_1985_4_12_3_409_0} is also relevant to the above theorem.

We note that the theorem does not require the cycles introduced during the resolution or deformation of the geometry to be supersymmetric. Therefore, we assume the validity of the theorem in the context of the new DR-phase, despite the fact that the 5-cycles introduced in the DR-phase are not supersymmetric.

In the following, we compute the space $\mathcal{H}^{p}_{L^{2}}$ up to $p=3$ though direct computation of the relative cohomology. We have the long exact sequence of relative cohomology ($M=\overline{\mc{C}(Q^{\scriptscriptstyle(1,1,1)})}$)
\begin{equation}
\begin{split}
        &0\to H^{0}(M,\partial M)\to H^{0}(M)\to H^{0}(\partial M) \to H^{1}(M,\partial M) \to H^{1}(M)\to H^{1}(\partial M)
\\
    &\to H^{2}(M,\partial M)\to H^{2}(M)\to H^{2}(\partial M) \to H^{3}(M,\partial M)\to H^{3}(M) \to H^{3}(\partial M) \cdots
\end{split}
\end{equation}
where 
\begin{equation}
    H^{2}(M)=\Z\,,\quad H^{2}(\partial M) = \Z \oplus \Z\,,\quad H^{3}(M)=\Z\,,\quad H^{3}(\partial M)=0
\end{equation}
and we can cut off the long exact sequence into the exact sequence
\be
0 \to H^{2}(M,\partial M)\overset{f}{\to} H^{2}(M)\overset{g}{\to} H^{2}(\partial M) \overset{h}{\to} H^{3}(M,\partial M)\overset{i}{\to} H^{3}(M) \to 0\,.
\ee
Note that after we dualize $H^2(M)\cong H_3(M)$, $H^2(\ptl M)\cong H_5(\ptl M)$, the map $g$ is equivalent to the inclusion map of $\mathbb{S}^{3}\hookrightarrow \mathbb{S}^{3}\times \mathbb{S}^{2}$, where $\mathbb{S}^{3}$ is the generator of $H_3(M)=\mb{Z}$ and $\mathbb{S}^{3}\times \mathbb{S}^{2}$ is the generator of $H_5(\ptl M)=\mb{Z}$. Hence we have $\ker{g}=0$ thus $\im{f}=0$. Because $\ker{f}=0$ as well, we can conclude that $H^2(M,\ptl M)=0$.

Now because $\im(g)=\mb{Z}=\ker{h}$, we can choose to write the $g$ map as $g:a\rightarrow (a,0)$, and the $h$ map can be written as $h:(a,b)\rightarrow (0,b)$. Finally, since $\im(i)=H^3(M)=\mb{Z}$, which is orthogonal to $\im({h})=\ker{i}=\mb{Z}$, we conclude that $H^{3}(M,\partial M)=\mb{Z}^2$.

In the end, we obtained the required relative cohomology groups
\be
H^2(M,\ptl M)=0\ ,\ H^3(M,\ptl M)=\mb{Z}^2\,.
\ee

Based on the above calculations, the M-theory $C_{3}$-form field can be expanded along the basis of $\mathcal{H}^{3}_{L^{2}}(X)$. Explicitly, we write,
\begin{equation}\label{eq:C3-expansion-DR-phase}
    C_{3} \ =\ \sum_{i=1}^{2} \,\left(\phi_{i}\,\lambda_{\scriptscriptstyle(2,1)}^{i} + \phi^{\ast}_{i} \,\lambda_{\scriptscriptstyle(1,2)}^{i}\,\right)\,.
\end{equation}
Here, $\lambda_{\scriptscriptstyle (2,1)}^{i}$ denotes the basis of $\mathcal{H}^{3}_{L^{2}}(X)$, and $\phi_{i}$ are spacetime scalar fields corresponding to the zeroth component of 3d $\N=2$ chiral multiplets. Since $C_{3}$ is a real gauge field, its expansion necessarily includes both the chiral ($\Phi$) and anti-chiral ($\Phi^{\dagger}$) components, as shown. The superpartners of these scalar fields are expected to arise from the dimensional reduction of the M-theory gravitino.

In the absence of a superpotential, the effective theory in the DR-phase consists of two free hypermultiplets $\Phi^{i}$ along with their conjugates. Here, a hypermultiplet is defined as a combination of chiral and anti-chiral multiplets. However, this description encounters a significant limitation due to the presence of a superpotential. Specifically, the superpotential naturally induces a mass term for the chiral supermultiplets $\Phi^{i}$ (and their conjugates). Generally, there is no obstruction to introducing such a mass term for these chiral multiplets. As will be shown, this superpotential can emerge from Euclidean M2-branes wrapping free 3-cycles. As a result, in the deep IR, the kinetic terms of the chiral multiplets vanish, leading to a gapped theory.

\paragraph{Massive states from wrapped M-branes.}

The zero section of the new DR-phase is not supersymmetric, as discussed toward the end of Appendix \ref{app:gauged-8d-sugra} following \cite{Gomis:2001vk}. Consequently, states arising from M-branes wrapping the homological (free) cycles of the zero section are generally not expected to be BPS states. The existence of massive non-BPS states is not an obstruction to consider the deep IR limit, where we integrate out all massive modes, and we conjecture a topological description to emerge.

In the new branch, effective heavy states emerge from M-branes wrapping homological cycles of the zero section $(\mathbb{S}^{3}\times \mathbb{S}^{2})/\Z_{N}\cong \mathbb{S}^{3}/\Z_{N}\times \mathbb{S}^{2}$ and extending in the direction of the normal bundle, which is locally $\R^{3}$, as well as in the spacetime $\R^{1,2}$. The homology groups of the zero section are given by:
\begin{equation}\label{homologyT11ZN}
    H_{\bullet} (\mathbb{S}^{3}/\Z_{N}\times \mathbb{S}^{2};\Z) =( \Z ,   \Z_{N} ,  \Z , \Z\oplus\Z_N ,  0 , \Z ).
\end{equation}
Below, we enumerate the various M-brane wrapping configurations, their physical interpretations, and relevant comments.

First, we begin with the torsional cycles:
\begin{itemize}
\item M2-brane wrapping Tor$H_{1}=\Z_{N}$: These M2-branes extend along two directions of $\R^{1,2}$, leading to $\Z_{N}$ electric (confining) strings. In three dimensions, such states can also be interpreted as domain walls (DW). These configurations correspond to topological defects represented as
\begin{equation}\label{eq:ZN-string}
 U(\Sigma_{2};p) \ =  \ \exp( i p \int_{\Sigma_{2}} a_{2} )\,,\qquad p\in\Z\,.
\end{equation}
Here, $\Sigma_{2}$ is a 2D surface in spacetime, and $a_{2}$ is a 2-form gauge field that couples to these strings, taking values in $H^{2}(\R^{1,2};\Z_{N})$. The defect $U(\Sigma_{2};p)$ is topological, since $a_{2}$ is flat. The charge $p$ is defined modulo $N$.

These $\mathbb{Z}_{N}$ strings can be interpreted as confining strings, which are stable due to their charges being classified by $\text{Tor}H_{1} = \mathbb{Z}_{N}$, or equivalently, by the first homotopy group. The $\mathbb{Z}_{N}$ symmetry corresponds to the centre of the $SU(N)$ gauge group in the deconfinement phase, leading to an unbroken $\mathbb{Z}_{N}$ 1-form symmetry, as expected.  

\item M5-brane wrapping $\text{Tor}H_{1}=\Z_{N}$ and extending along the $\R^{3}$ normal direction: From the 3d perspective, this leads to magnetic $\Z_{N}$ strings. The corresponding topological defect is given by,
\begin{equation}\label{eq:ZN-strings-(2)}
    \widetilde{U}(\widetilde{\Sigma}_{2};m)\,=\,\exp(im\int_{\widetilde{\Sigma}_{2}} \,  c_{2})
\end{equation}
Similar to the previous operator, $\widetilde{\Sigma}_{2}$ is a 2D surface in spacetime, and $b_{2}$ is a 2-form gauge field that couples to these strings, taking values in $H^{2}(\R^{1,2},\Z_{N})$. The defect $\widetilde{U}(\widetilde{\Sigma}_{2};p)$ is topological, as $b_{2}$ is flat. The charge $m$ is defined modulo $N$.   

\item EM2-brane wrapping Tor$H_{3}$: In the 3d effective theory, these states are understood as $\Z_{N}$ instantons. Since these instantons are constructed using M2-branes, we refer to them as $\Z_{N}$ electric instantons. The 3d effective topological operator is written as:
\begin{equation}\label{eq:ZN-instanton}
    V(\wp,q) \ = \ \exp(i q \, \varphi(\wp))\,,\qquad q\in \Z\,.
\end{equation}
Here, $\wp$ is a point in $\R^{1,2}$, $\varphi$ is the gauge field coupling to the instanton, taking values in $H^{0}(\R^{1,2};\Z_{N})$. The charge $q$ is defined modulo $N$.

\item EM5-brane wrapping Tor$H_{3}=\Z_{N}$ and filling the normal directions $\R^{3}$: These M5-branes give rise to $\Z_{N}$ magnetic instantons. The corresponding 3d topological operator is given by
\begin{equation}\label{eq:ZN-instantons-(2)}
    \widetilde{V}(\wp ; n)\,=\,\exp(im\chi(\wp))
\end{equation}
Here, $\chi$ is the gauge field coupling to the magnetic instanton, taking values in $H^{0}(\R^{1,2};\Z_{N})$. The charge $n$ is defined modulo $N$.
\end{itemize}

Next, we consider M-brane wrapping free cycles:
\begin{itemize}
\item M2-brane wrapping the free $H_{2}=\Z$: This configuration corresponds to a massive, possibly BPS, particle. Such states can always be integrated out to obtain an effective description.

\item Euclidean M2-brane (EM2) wrapping the free $H_{3} = \Z$: From the perspective of the lower-dimensional theory, this configuration induces a potential in the effective theory. The general structure of the superpotential is given by \cite{Ganor:1996pe,Witten:1996bn, Harvey:1999as, Braun:2018fdp}:
\begin{equation}
W_{\scriptscriptstyle\text{EM2}}\,\sim\, f(\cdots)\, \exp{2\pi i \int_{\mathbb{S}^{3}} \left( C_{3}\, +i \vol_{3} \right) } + \ \text{Non-BPS contributions.}
\end{equation}
Here, $\vol_{3}$ is a differential form dual to the free 2-cycle of $(\mathbb{S}^{3}\times \mathbb{S}^{2})/\Z_{N}$, such that its integral over the 3-sphere defining its volume.

Plugging the $C_{3}$-field expansion (\ref{eq:C3-expansion-DR-phase}) into the above superpotential generates a mass term for the chiral supermultiplets $\Phi^{i}$. Consequently, in the deep IR, the kinetic terms of these chiral multiplets are lifted. By integrating out these modes, one arrives at a gapped 3d theory.

The precise form of the function $f(\cdots)$ is not well understood. It may receive quantum corrections, or non-BPS corrections. The non-BPS contributions could be absorbed into $f(\cdots)$. Moreover, $f(\cdots)$ could depend on other parameters, or even vanish at certain points in parameter space. Furthermore, the presence of such a potential may break supersymmetry. However, even if supersymmetry is (completely or partially) broken, as we will argue below, the DR-phase results in a gapped theory described by a topological action. Consequently, the role of the superpotential in breaking supersymmetry becomes irrelevant in this phase.

\item M5-brane wrapping the free $H_{5}= \Z$, i.e. the entire $(\mathbb{S}^{3}\times \mathbb{S}^{2})/\Z_{N}$ space: In general, M2-branes and M5-branes can only consistently intersect along one spatial direction \cite{deRoo:1997gq}. To avoid inconsistent overlaps or intersections with the aforementioned M2-brane configurations, this case is excluded from the current analysis.
\end{itemize}

Other states arising from M5-branes wrapping homological cycles of the zero section are more relevant in AdS/CFT constructions. While they play no role in the current work, we list them for completeness:
\begin{itemize}
\item M5-brane wrapping the free $H_{3}=\Z$: Extending along the $\R^{1,2}$ spacetime.
\item M5-brane wrapping Tor$H_{3}=\Z_{N}$: Extending along the $\R^{1,2}$ spacetime.
\end{itemize}

\paragraph{The deep IR limit.} 

As discussed in \cite{Najjar:2023hee}, for 3d $\N=2$ theories, there exists a deep IR limit associated with any given geometrical branch. In this regime, all massive states are integrated out through a Wilsonian approach. Geometrically, this corresponds to taking the large volume limit of all homological cycles. 

For free cycles, this limit is straightforward; however, the situation is fundamentally different for torsional cycles, as these lack a well-defined notion of volume. Consequently, defects generated by M-branes wrapping torsional cycles are not integrated out using the standard Wilsonian procedure. We adopt the perspective that these defects persist in the deep IR limit and are described by a TQFT. 

Although the effect of the large volume limit on the 3d potential is not fully understood--and it most likely lift supersymmetry, as previously discussed--the fact that the deep IR limit is captured by a TQFT indicates that such a lifting does not significantly impact the effective theory in this setting.

\paragraph{Topological $\Z_{N}$ gauge theory.}

Let us determine the $\Z_{N}$ topological theories that correspond to the defects in (\ref{eq:ZN-string}), (\ref{eq:ZN-strings-(2)}), (\ref{eq:ZN-instanton}) and (\ref{eq:ZN-instantons-(2)}).

The analysis begins with the compactification of M-theory on the DR-phase of the CY4 geometry, whose topology is $\R^{3}\times (\mathbb{S}^{3}\times \mathbb{S}^{2})/\Z_{N}$. This compactification proceeds in two distinct steps:
\begin{itemize}
\item Compactification on $(\mathbb{S}^{3}\times \mathbb{S}^{2})/\Z_{N}$:

First, we compactify along the compact 5-cycle $(\mathbb{S}^{3}\times \mathbb{S}^{2})/\Z_{N}$. The resulting theory is defined, at least locally, on a 6-dimensional spacetime $\R^{1,2}\times \R^{3}$, where $\R^{3}$ represents the normal bundle of $(\mathbb{S}^{3}\times \mathbb{S}^{2})/\Z_{N}$.
    
At this stage, we leverage the general framework outlined in \cite{Camara:2011jg,Berasaluce-Gonzalez:2012abm} to identify the possible $\Z_{N}$ gauge theories arising from this compactification.
    
\item Compactification on $\R^{3}$:

Next, we compactify along the normal bundle $\R^{3}$, reducing the system to an effective 3-dimensional field theory on $\R^{1,2}$.
\end{itemize}

In M-theory compactifications on compact manifolds $X_D$, the presence of torsional cycles leads to discrete gauge theories, as discussed in \cite{Camara:2011jg,Berasaluce-Gonzalez:2012abm}. The general approach outlined in these references introduces a pair of forms $(\omega_{p}^{\alpha
},\Theta_{p+1}^{\alpha})$ for each torsional cycle $\Z_{n}^{\alpha}$ in $\text{Tor}H_{p}(X_{\scriptscriptstyle D})$, satisfying the following relation: 
\begin{equation}\label{eq:d-omega-n-Theta}
    d\omega_{p}^{\alpha}\, =\, n\, \Theta_{p+1}^{\alpha}\,.
\end{equation}
The non-trivial linking between torsional cycles in $\text{Tor}H_{p}(X_{\scriptscriptstyle D})$ and their Pontryagin duals in $\text{Tor}H_{D-p-1}(X_{\scriptscriptstyle D})$ is encoded in the integral:
\begin{equation}\label{eq:integral-omega-Theta}
    \int_{X_{\scriptscriptstyle D}}\omega_{p}^{\alpha}\wedge \Theta^{\beta}_{D-p}\,=\,\delta^{\alpha\,\beta}\,.
\end{equation}

We apply this framework to the first step of our compactification, namely, compactifying M-theory on $(\mathbb{S}^{3}\times \mathbb{S}^{2})/\Z_{N}$, to determine the resulting $\Z_{N}$ gauge theories on  $\R^{1,2}\times \R^{3}$. Notably, $(\mathbb{S}^{3}\times \mathbb{S}^{2})/\Z_{N}$ admits a $\Z_{N}$ torsional 1-cycle and its Pontryagin dual 3-cycle, allowing us to introduce the following relations:   
\begin{equation}
    d\omega_{1} \,=\, N\, \Theta_{2}\,,\qquad \qquad d\omega_{3}\, =\, N\, \Theta_{4}
\end{equation}

To arrive at the discrete gauge theories kinetic terms, we use the kinetic term of the $C_{3}$-form field in M-theory \cite[App.C]{Najjar:2024vmm}:
\begin{equation}\label{eq:M-theory-G4-G7}
    S_{\scriptscriptstyle\text{kin}}^{\scriptscriptstyle\text{M}} = \frac{1}{2\pi}\,\int G_{4}\wedge G_{7}\,.
\end{equation}
First, we expand the M-theory 3-form field $C_{3}$ in terms of the pairs $(\omega_{p}^{\alpha},\Theta_{p+1}^{\alpha})$. For the case at hand, the expansion is given by: 
\begin{equation}
        C_{3} = \varphi\,\, \omega_{3} + a_{2}\wedge \omega_{1} + a_{1}\wedge \Theta_{2}
\end{equation}
Using the defining property in (\ref{eq:d-omega-n-Theta}), the 4-form field strength $G_{4}$ becomes:
\begin{equation}\label{eq:G4-flux-expansion}
        G_{4}  = dC_{3} = d\varphi \wedge \omega_{3} + \left(da_{1} +  N a_{2}\right) \wedge \Theta_{2} + \cdots \,.
\end{equation}
Next, we expand the 7-form flux $G_{7}$ in terms of the pairs $(\omega_{p}^{\alpha},\Theta_{p+1}^{\alpha})$ as,
\begin{equation}\label{eq:G7-flux-expansion}
    G_{7} = F_{6}\wedge \omega_{1} + F_{5}\wedge \Theta_{2} + F_{4}\wedge \omega_{3} + F_{3}\wedge\Theta_{4}
\end{equation}
Here, $F_{i}$ may be interpreted as field strengths. The exact form of the relevant field strengths will be given shortly. 

By inserting (\ref{eq:G4-flux-expansion}) and (\ref{eq:G7-flux-expansion}) into (\ref{eq:M-theory-G4-G7}), and applying (\ref{eq:integral-omega-Theta}), we arrive at the following 6d action:
\begin{equation}
    S_{\scriptscriptstyle\text{kin}}^{\scriptscriptstyle\text{6d}} = \frac{1}{2\pi}\, \int_{\R^{1,2}\times\R^{3}}\, \left[ \,\left(da_{1}+Na_{2}\right)\wedge F_{4} + N\varphi\wedge F_{6}\,\right]\,.
\end{equation}
The 3d effective action is obtained by further reduce along the $\R^{3}$ direction as,
\begin{equation}\label{eq:3d-kinetic-term}
    S_{\scriptscriptstyle\text{kin}}^{\scriptscriptstyle\text{3d}} =\frac{1}{2\pi}\,\int_{\R^{1,2}}\, \left[ \,\left(da_{1}+Na_{2}\right)\wedge F_{1} + N\varphi\wedge F_{3}\,\right]\,.
\end{equation}
The interpretation of $F_{1}$ and $F_{3}$ are given as
\begin{equation}
    F_{1} = \ast_{3}\, \left(da_{1} + Na_{2}\right) \,,\qquad  F_{3} = dc_{2}\,.
\end{equation}
As a result of this identification, the second term of (\ref{eq:3d-kinetic-term}) given a topological BF term for a $\Z_{N}$ gauge theory. The first term leads to a kinetic term for another $\Z_{N}$ gauge theory, given as
\begin{equation}
 \left(da_{1} + Na_{2}\right)^{2}
\end{equation}
To arrive at a topological BF description, we dualize the above action in a similar way to that given in \cite[Sec.2.2]{Banks:2010zn}. First, we scale the fields $a_{1}$ and $a_{2}$ as
\begin{equation}
    a_{1}\,\to\, t\,a_{1}\,,\qquad a_{2}\,\to\,t\,a_{2}\,
\end{equation}
to arrive at
\begin{equation}
    t^{2}\,\left(da_{1} + Na_{2}\right)^{2}\,. 
\end{equation}
Next, we dualize the field $a_{1}$ as
\begin{equation}
    da_{1}\, =\, \ast_{3} \, d\chi\,.
\end{equation}
In the dual frame, the above action then takes the form
\begin{equation}
    \frac{1}{t^{2}}\,\,d\chi\,\wedge\, \ast_{3}d\chi \,+\, \frac{N}{2\pi}\,\chi\wedge da_{2}\,.
\end{equation}
By taking the limit $t\to\infty$, we recover another $\Z_{N}$ topological BF term. All in all, the topological 3d theory is given by
\begin{equation}\label{eq:3d-TQFT-dynamical-from-DR-phase}
  S_{\scriptscriptstyle\text{Top}}^{\scriptscriptstyle\text{3d}} \,=\,   \frac{N}{2\pi}\,\chi\,\wedge\,da_{2}\,+\,\frac{N}{2\pi}\,\varphi\,\wedge\,dc_{2}\,.
\end{equation}

In general, the expectation value of the defects operators in (\ref{eq:ZN-string}), (\ref{eq:ZN-strings-(2)}), (\ref{eq:ZN-instanton}) and (\ref{eq:ZN-instantons-(2)}) can be non-trivial only under specific conditions:
\begin{itemize}
    \item The linking number between the spacetime point, where the instanton is located, and the surface, where the strings are located, is non-trivial.
    \item The torsional cycles wrapped by the M-branes must have non-trivial linking numbers, as is the case for Tor$H_{1}$ and Tor$H_{3}$.
    \item There exist a corresponding BF-term that gives non-trivial expectation values \cite{Birmingham:1991ty}.
\end{itemize}

As a result, the expectation value of the operators in (\ref{eq:ZN-string}) and (\ref{eq:ZN-instanton}), is determined using the first BF term above as:
\begin{equation}
\begin{split}
        \expval{U(\Sigma_{2},q)\,V(\wp,p)}  &=  \exp{2\pi i\  p q\  \text{Link}(\Sigma_{2},\wp)\ \text{Link}(\alpha_{(1)},\beta_{(3)})}\, \expval{V(\wp,p)\,U(\Sigma_{2},q)}
        \\
        &=\exp{\frac{2\pi i }{N}\  p q\  \text{Link}(\Sigma_{2},\wp) }\, \expval{V(\wp,p)\,U(\Sigma_{2},q)}\,.
\end{split}
\end{equation}
Here, $\text{Link}(\Sigma_{2},\wp)$ is the linking number between the spacetime point $\wp$ and the surface $\Sigma_{2}$. $\alpha_{(1)} \in  \text{Tor}H_{1}$ and $\beta_{(3)} \in \text{Tor}H_{3}$, with $\text{Link}(\alpha_{(1)},\beta_{(3)})$ representing the linking number between the torsional cycles.

Similarly, the expectation value of the operators in 
(\ref{eq:ZN-strings-(2)}) and (\ref{eq:ZN-instantons-(2)}) is given by the second BF-term as,
\begin{equation}
\begin{split}
        \expval{\widetilde{U}(\widetilde{\Sigma}_{2},m)\,\widetilde{V}(\wp,n)}  &=  \exp{2\pi i\  m n\  \text{Link}(\widetilde{\Sigma_{2}},\wp)\ \text{Link}(\alpha_{(1)},\beta_{(3)})}\, \expval{\widetilde{V}(\wp,n)\,\widetilde{U}(\widetilde{\Sigma}_{2},m)}
        \\
        &=\exp{\frac{2\pi i }{N}\  m n\  \text{Link}(\widetilde{\Sigma}_{2},\wp) }\, \expval{\widetilde{V}(\wp,n)\,\widetilde{U}(\widetilde{\Sigma}_{2},m)}\,.
\end{split}
\end{equation}

\paragraph{A concluding remark.}

It is important to emphasize that the BF theory in (\ref{eq:3d-TQFT-dynamical-from-DR-phase}) does not provide a complete topological description of the deep IR limit. A key reason for this is the 't Hooft UV-IR anomaly matching condition \cite{tHooft:1979rat}. If a theory exhibits a non-trivial 't Hooft anomaly—typically computed in the UV—its IR behaviour cannot be entirely trivial, as there must exist degrees of freedom that match the UV anomaly. Consequently, the full IR TQFT must account for the presence of the electric 1-form anomaly.

As demonstrated in the previous section, in the bulk theory, the 1-form anomaly couples to a scalar field. A similar behaviour is expected in the 3d IR theory, ensuring that the 1-form anomaly is matched. Specifically, the IR TQFT should include a term of the form:
\begin{equation}
    \frac{N^{2}}{4\pi}\,\zeta\, \left(c_{1} \wedge b_{2}\right)\,,
\end{equation}
where, $\zeta$ is a spacetime background 0-form, $c_{1}$ is a 1-form gauge field, and $b_{2}$ is a 2-form gauge field of the electric $\Z_{N}^{\scriptscriptstyle[1]}$ 1-form symmetry.

The structure of the anomaly matching term can be understood through the dimensional reduction of the TQFT description of the confining phase of 4d $\N=1$ SYM theory with an $SU(N)$ gauge group, as presented in \cite[(7.7)]{Gaiotto:2014kfa}:
\begin{equation}\label{eq:4d-TQFT}
    S_{\scriptscriptstyle\text{TQFT}}^{4d}\,=\,\frac{iN}{2\pi}\,\int_{\R^{3,1}}\,\phi\,\left(da_{3}+\frac{N}{4\pi}b_{2}\wedge b_{2}\right)\,.
\end{equation}
The reduction is performed along a compact $S^{1}$ direction. The reduction of the second term reveals the presence of a 't Hooft anomaly associated with the electric $\Z_{N}^{\scriptscriptstyle[1,\,e]}$ 1-form symmetry. To perform the dimensional reduction explicitly, we expand the fields $a_{3},b_{2}$, and $\phi$ in the cohomological basis $\{1\},\{\vol_{1}\}$ of $S^{1}$ as: 
\begin{equation}
    \begin{split}
        \phi\, &= \phi \wedge 1,
        \\
        a_{3} &= a_{3} \wedge 1 + \widetilde{a}_{2}\wedge \vol_{1},
        \\
        b_{2} &= b_{2}\wedge 1 + c_{1} \wedge \vol_{1}.
    \end{split}
\end{equation}
Substituting these expansions into (\ref{eq:4d-TQFT}) and normalizing the integral $\int_{S^{1}}\vol_{1}=1$, we obtain the following 3d TQFT action:
\begin{equation}\label{eq:3d-TQFT-from-4d-TQFT}
    \frac{iN}{2\pi}\,\int_{\R^{2,1}} \phi\,\left(d\,\widetilde{a}_{2}  + \frac{N}{2\pi}  c_{1} \wedge b_{2} \right) \,.
\end{equation}
A similar construction, albeit in a different context, is discussed in \cite[App.C]{Closset:2024sle}.

The general expectation is that one could, in principle, derive the above 't Hooft anomaly matching term—or even the full TQFT description—from the bulk SymTFT theory. However, a precise derivation of this term, along with the complete IR TQFT description, lies beyond the scope of this work.

\section{Conclusion and outlook}

In this paper, we have investigated the occurrence of a new geometric transition in a non-compact Calabi-Yau 4-fold, specifically the cone over the 7d Sasaki-Einstein manifold $Q^{\scriptscriptstyle(1,1,1)}$, denoted as $\CC(Q^{\scriptscriptstyle(1,1,1)})$. One end of the geometric transition corresponds to the usual Crepant resolution, which yields a smooth space with the topology $\R^{4}\times \mathbb{S}^{2}\times \mathbb{S}^{2}$. The other end of the transition corresponds to a phase where partial resolution and partial deformation are applied, referred to as the DR-phase, resulting in a smooth geometry with the topology $\R^{3}\times \mathbb{S}^{3}\times \mathbb{S}^{2}$.

We have applied a specific $\Z_{N}$ quotient that acts on the $U(1)$ bundle of the $Q^{\scriptscriptstyle(1,1,1)}$. In the resolved case, the $\Z_{N}$ action introduces a codimension-4 singularity. In the context of M-theory geometric engineering, this singularity gives rise to a 3d $\N=2$ $SU(N)$ gauge theory. On the other end of the geometric transition, however, the $\Z_{N}$  quotient acts freely on the Hopf fibers of $\mathbb{S}^{3}$. The absence of a singularity in this phase implies the absence of the 3d $SU(N)$ gauge theory. Furthermore, we have demonstrated that the deep IR limit of the DR-phase can be described by a gapped TQFT. Although the full IR TQFT description remains beyond the scope of this work, the DR-phase provides a clear notion of confinement for 3d $\N=2$ $SU(N)$ gauge theories. The above discussion is summarized in the figure below. 

Furthermore, we have investigated the 4d SymTFT bulk theory corresponding to the link space $Q^{\scriptscriptstyle(1,1,1)}/\Z_{N}$, analysing the possible 't Hooft anomalies, BF terms, and the associated defects and symmetry topological operators. 

\begin{figure}[H]
\centering
\begin{tikzpicture}[node distance=1cm, auto]
    \node (M1) {M-theory on $\mathbb{R}^{3}\times \frac{\mathbb{S}^{3}}{\mathbb{Z}_{N}}\times \mathbb{S}^{2}$};
    \node (IIA1) [below=of M1] {A gapped TQFT};
    \node (M2) [right=5cm of M1] {M-theory on $\frac{\R^{4}}{\mathbb{Z}_{N}} \times \mathbb{S}^{2}\times \mathbb{S}^{2}$};
    \node (IIA2) [below=of M2] {3d $\N=2$ $SU(N)$ SYM theory};

    \draw[<->] (M1) -- node {Geometric Transition} (M2);

    \draw[<->] (M1) -- (IIA1);
    \draw[<->] (M2) -- (IIA2);
\end{tikzpicture}
\label{Fig:conclusion-geometric-transition}
\end{figure}

Our analysis opens up several promising directions for future research:
\begin{itemize}
    \item The explicit non-compact CY4 metric for the DR-phase remains unknown. Determining this metric is an intriguing problem for both physicists and mathematicians. Techniques from gauged 8d supergravity may prove useful in this endeavour. 
    \item Exploring other possible geometric transitions for CY4 cones over different 7D Sasaki-Einstein manifolds, particularly investigating potential DR-phases in these contexts. 
    \item A natural generalization involves the inclusion of O-planes to construct $SO/Sp$ 3d $\N=2$ field theories. This would entail studying the geometry of the link space in such cases, the possible $p$-form symmetries, and the role played by the DR-phase. 
    \item How to relate the confinement/deconfinement phase transition triggered by the geometric transition with the usual notion of phase transition, whose order parameters are temperature or chemical potential?
\end{itemize}

\acknowledgments
We would like to thank Hao N. Zhang for the early stage collaboration. The author MN thanks Osama Khlaif, Leonardo Santilli, and Yi Zhang for helpful discussions. The work is supported by National Natural Science Foundation of China under Grant No. 12175004, No. 12422503 and by Young Elite Scientists Sponsorship Program by CAST (2023QNRC001, 2024QNRC001).

\appendix 
\section{Additional arguments for the CY4 geometric transition}\label{app:arguments-for-the-DR-phase}

\subsection{The gauged 8d supergravity}\label{app:gauged-8d-sugra}

\paragraph{Setting the stage.} 

The phenomenon of geometric transitions in M-theory can be addressed within the framework of 8-dimensional gauged supergravity. This framework has been explored in various contexts; see, for instance, \cite{Edelstein:2001pu,Acharya:2004qe,Acharya:2020vmg}.

Following the seminal work by A. Salam and E. Sezgin \cite{Salam:1984ft}, 8-dimensional gauged supergravity can be derived by compactifying M-theory on an $SU(2)\cong \mathbb{S}^{3}$ manifold. Alternatively, the gauged supergravity can be constructed by compactifying M-theory on $SO(3)\cong SU(2)/\Z_{2}$, as demonstrated in \cite{LassoAndino:2016lwl}. This latter case proves particularly useful for our purposes.

In addition, we consider 8d supergravity solutions with zero cosmological constant, as these settings are relevant for geometric engineering.  

In certain compactification scenarios, such as those discussed in \cite{Edelstein:2001pu,Acharya:2004qe,Acharya:2020vmg}, the 8-dimensional theory can be interpreted as the worldvolume theory on a Type IIA D6-brane, or a stack of $N$ D6-branes. These branes may wrap some compact $(7-d)$-dimensional cycles, while the transverse space includes a radial direction $\R_{+}$. The spacetime topology is schematically given by:
\begin{equation} 
\R^{1,d-1}\ \times (7-d)\text{-cycle}\ \times \R_{+}.
\end{equation}
The corresponding wrapped metric solution takes the form:
\begin{equation}
    ds^{2}_{8d} = e^{2f(r)}\,dx^{2}_{1,d-1} + e^{2h(r)}\,d\Omega_{7-d}^{2} + dr^{2},
\end{equation}
with $f(r)$ and $h(r)$ are functions of the radial coordinate $r\in\R_{+}$.

In general, the $(7-d)$-cycles persist in both asymptotic limits of $\R_{+}$, i.e., as $r\to 0$ and $r\to \infty$ \cite{Edelstein:2001pu}. The resulting d-dimensional effective theory typically reduces to a pure $\mathfrak{su}(N)$ gauge theory, arising from the stack of the $N$ D6-branes.

In the following, we revisit the famous phenomenon of the CY3 conifold transition \cite{Candelas:1989js} within the framework of 8d gauged supergravity. Understanding the conifold transition in this context provides insights for generalizing such transitions to CY4. 

\paragraph{The CY3 conifold transition.}

Consider M-theory on a particular $\Z_{N}$ quotient of the resolved conifold, referred to as (ladder) hyperconifold in \cite{Davies:2011is,Davies:2013pna, Acharya:2020vmg, Najjar:2022eci}, which is given as
\begin{equation}
    \CC(T^{\scriptscriptstyle(1,1)}/\Z_{N}) \ \cong  \ \frac{\mathcal{O}(-1)\,\oplus\,\mathcal{O}(-1)}{\Z_{N}} \,  \hookrightarrow \,  \C\mathbb{P}^{1}\,.
\end{equation}
In M-theory geometric engineering, the effective 5d theory is given by (see, e.g., \cite{Najjar:2022eci,Acharya:2021jsp}) 
\begin{equation}
    \mathcal{T}^{\text{M}}_{\scriptscriptstyle\text{5d}}(\CC(T^{\scriptscriptstyle(1,1)}/\Z_{N})) \ \, \simeq \ \,  \text{5d $\N=1$ pure $\mathfrak{su}(N)$ SYM}.
\end{equation}

The corresponding gauged 8d supergravity can be understood through $N$ D6-branes wrapping the base $\C\P^{1}$, with a transverse $\R_{+}$ direction as described above. In the Type IIA uplift of this configuration, the $\C\P^{1}$ is viewed as the zero section of the non-compact $T^{\ast}\mathbb{S}^{2}$, a Calabi-Yau 2-fold \cite[Sec.4]{Edelstein:2001pu}. The link of the $T^{\ast}\mathbb{S}^{2}$ space is well-known to be $\mathbb{S}^{3}/\Z_{2}$ \cite{EGUCHI197982,Acharya:2004qe}. This is precisely the compact space used in M-theory to construct the $SO(3)$-gauged 8d supergravity \cite{LassoAndino:2016lwl}.

Concerning supersymmetry, it can be explicitly verified that both the M-theory and Type IIA descriptions preserve the same amount of supersymmetry \cite{Taylor:1999ii,Edelstein:2001pu}.

On the other side of the conifold transition, the geometry is described by M-theory on the $\Z_{N}$ quotient of the deformed conifold, $T^{\ast}(\mathbb{S}^{3}/\Z_{N})$ \cite{Acharya:2020vmg, Acharya:2024bnt}. In the corresponding dual gauged 8d supergravity configuration, there are $N$ units of $F_{2}$ RR-flux threading in the compact $\C\P^{1}$ \cite{Edelstein:2001pu}. In the Type IIA picture, the two-sphere $\C\P^{1}$ is again interpreted as the zero section of $T^{\ast}\mathbb{S}^{2}$ \cite{Vafa:2000wi}. The uplift from Type IIA to M-theory involves incorporating the $N$-units of two-form flux as the Hopf fiber $U(1)/\Z_{N}$ of $\mathbb{S}^{3}/\Z_{N}$ \cite{Curio:2001dz,Minasian:2001sq}. This Hopf fiber is identified with the M-theory circle. In this limit, the geometric setup $U(1)/\Z_{N}\hookrightarrow \C\P^{1}$ is interpreted in M-theory as the zero section of $T^{\ast}(\mathbb{S}^{3}/\Z_{N})$, unifying the perspectives from both sides of the conifold transition.

The discussion above reveals that the Calabi-Yau 3-fold (CY3) conifold transition, when interpreted within the framework of gauged 8d supergravity, can be reformulated in terms of a brane/flux transition. Specifically, this transition replaces $N$ D6-branes with $N$ quanta of 2-form RR flux, or vice versa. This equivalence arises because D6-branes couple magnetically to the RR 2-form flux.

\begin{figure}[H]
\centering
\begin{tikzpicture}[node distance=2cm, auto]

    \node (M1) {M-theory on $T^{\ast}(\mathbb{S}^{3}/\Z_{N})$};
    \node (IIA1) [below=of M1] {$N$ $F_{2}$ through $\mathbb{S}^{2}$ of $T^{\ast}\mathbb{S}^{2}$\,\,};
    \node (M2) [right=5cm of M1] {M theory on $\frac{\R^{4}}{\Z_{N}} \times \mathbb{S}^{2}$};
    \node (IIA2) [below=of M2] {\,\,$N$ D6 brane wrapping $\mathbb{S}^{2}$ of $T^{\ast}\mathbb{S}^{2}$};

    \draw[<->] (M1) -- node {Geometric Transition} (M2);
    \draw[<->] (IIA1) -- node {Brane/Flux Transition} (IIA2);

    \draw[<->] (M1) -- (IIA1);
    \draw[<->] (M2) -- (IIA2);

\end{tikzpicture}
\caption{The known CY3 geometric transition. Both sides of the M-theory geometry have the link space $L_{5} = T^{\scriptscriptstyle(1,1)}/\Z_{N}$.}
\label{Fig:conifold-transition}
\end{figure}

\paragraph{Generalization.}

This insight allows for a natural generalization to scenarios involving geometric transitions in M-theory between two spaces, $\widetilde{\mathbb{X}}$ and $\widehat{\mathbb{X}}$. Within the gauged 8d supergravity framework, the geometric transition can be understood through the following argument:
\begin{enumerate}[label=\Roman*.]
    \item From M-theory to gauged 8d supergravity: 

    Assume that M-theory compactified on $\widetilde{\mathbb{X}}$ has a corresponding gauged 8d supergravity description. In this description, $N$ D6-branes wrap the zero section, denoted by $\widetilde{\mathbb{B}}$, of the geometry $ \widetilde{\mathbb{X}}$.
  
    \item Brane/flux transition:

    Replace the $N$ D6-branes wrapping $\widetilde{\mathbb{B}}$ with $N$ quanta of $F_{2}$ RR flux passing through a $\C\P^{1}\subseteq \widetilde{\mathbb{B}}$. This transition reformulates the geometric deformation of the compactification in terms of flux data in the gauged supergravity.

    \item Uplift to M-theory:

    In the language of M-theory, the $N$ quanta of $F_{2}$ flux on $\widetilde{\mathbb{B}}$ can be reinterpreted geometrically as a $U(1)/\Z_{N}$ bundle over $\widetilde{\mathbb{B}}$. This is expressed as:
    \begin{equation}
       \frac{U(1)}{\Z_{N}}\ \hookrightarrow \ \widehat{\mathbb{B}}\ \to \ \widetilde{\mathbb{B}}\,.
    \end{equation}
    Under the uplift to M-theory, $\widehat{\mathbb{B}}$ becomes the zero section of the geometry $\widehat{\mathbb{X}}$.
\end{enumerate}

Several well-known examples align with this prescription:
\begin{itemize}
    \item CY3 Conifold Transition \cite{Candelas:1989js}: For completeness, we present the transition in the following equation: 
    \begin{equation}\label{eq:coifold-geometric-transition}
    \text{M-theory on} \ \mathbb{R}^3 \times \mathbb{S}^{3} \xleftrightarrow{\text{Geometric Transition}} \text{M-theory on}\ \mathbb{R}^4 \times \mathbb{S}^{2}\,.
\end{equation}
    \item G2-Flop Transition: As studied in \cite{Edelstein:2001pu,Atiyah:2001qf,Acharya:2004qe}, this involves a transition in the context of G2-holonomy spaces. For instance, for the spin bundle over 3-sphere, the G2-flop is given as
    \begin{equation}\label{eq:G2-geometric-transition}
    \text{M-theory on} \ \mathbb{R}^4 \times \mathbb{S}^{3} \xleftrightarrow{\text{\,G2-Flop\,}} \text{M-theory on}\ \mathbb{S}^{3}\times \mathbb{R}^4 \,.
\end{equation}
    \item $Spin(7)$-Transition: Another example, described in \cite{Gukov:2002zg,Acharya:2004qe}, provides a parallel transition framework in $Spin(7)$-holonomy spaces. The geometric transition in this case is summarized by
\begin{equation}\label{eq:Spin7-geometric-transition}
    \text{M-theory on} \ \mathbb{R}^3 \times \mathbb{S}^5 \xleftrightarrow{\text{Geometric Transition}} \text{M-theory on}\ \mathbb{R}^4 \times \mathbb{CP}^2\,.
\end{equation}
\end{itemize}

Building on this framework, we will now apply the above argument to explore the geometric transition in the context of CY4.   

\paragraph{The geometric transition for the cone $\mathcal{C}(Q^{\scriptscriptstyle(1,1,1)}/\Z_{N})$.} 

As we have argued in section \ref{sec:physics-Q111-ZN}, putting M-theory on $\widetilde{\mathcal{C}(Q^{\scriptscriptstyle(1,1,1)}/\Z_{N})}$ is interpreted as
\begin{equation}
    \mathcal{T}^{\text{M}}_{\scriptscriptstyle\text{3d}}(\widetilde{\CC(Q^{\scriptscriptstyle(1,1,1)}/\Z_{N})}) \ \, \simeq \ \,  \text{3d $\N=2$ pure $\mathfrak{su}(N)$ SYM}\,,
\end{equation}
along with massive order parameters.

The gauged 8d supergravity description for this setup arises from $N$ D6-branes wrapping the 4-cycle $\C\P^{1}\times \C\P^{1}$, which is transverse to $\R_{+}$. The 4-cycle $\C\P^{1}\times \C\P^{1}$ serves as the zero section of the resolved cone over $Q^{\scriptscriptstyle(1,1,1)}/\Z_{N}$.

Parallel to the CY3 conifold transition discussed earlier. The type IIA limit, the compact 4-cycle $\C\P^{1}\times \C\P^{1}$ is interpreted as the zero section of a CY3 space. Since we are working with $SO(3)$-gauged 8d supergravity (instead of $SU(2)$-gauging), the link of the CY3 geometry must involve $\mathbb{S}^{3}/\Z_{2}$ as a component. With the zero section identified as the trivial bundle $\C\P^{1}\times \C\P^{1}$, the Calabi-Yau 3-fold is concluded to be the canonical bundle over $\C\P^{1}\times \C\P^{1}$ \cite{PandoZayas:2001iw,Hosomichi:2005ja}. This geometry can be understood as a $\Z_{2}$ quotient of the CY3 conifold \cite{Davies:2011is,Davies:2013pna,Hosomichi:2005ja,Najjar:2022eci}. 

Using the brane/flux transition mechanism described earlier, we can alternatively describe this setup in terms of $N$ quanta of 2-form flux threading the 4-cycle $\C\P^{1}\times \C\P^{1}$. This flux configuration is geometrically interpreted as a $U(1)$-bundle over $\C\P^{1}\times \C\P^{1}$ \cite{Curio:2001dz}. Extending this to a $(U(1)/\Z_{N})$-bundle, the M-theory circle is identified with the $U(1)$-fiber, and the corresponding pure M-theory geometric description becomes a non-compact CY4 space. The zero section of this CY4 is given by $(\mathbb{S}^{3}\times\mathbb{S}^{2})/\Z_{N}$, interpreted as the $U(1)/\Z_{N}$ bundle over $\C\P^{1}\times \C\P^{1}$.

Thus, the topology of the new CY4 geometry can be expressed as:
\begin{equation}
    \R^{3}\times \mathbb{S}^{2}\times (\mathbb{S}^{3}/\Z_{N})
\end{equation}

The above discussion is summarized in Figure \ref{Fig:CY4-geometric-transition}.

\begin{figure}[ht]
\centering
\begin{tikzpicture}[node distance=2cm, auto]

    \node (M1) {M-theory on $\mathbb{R}^{3}\times \frac{\mathbb{S}^{3}\times\mathbb{S}^{2}}{\mathbb{Z}_{N}}$};
    \node (IIA1) [below=of M1] {$N$ $F_{2}$ through $\mathbb{S}^{2}\times \mathbb{S}^{2}$ of CY3\,\,};
    \node (M2) [right=5cm of M1] {M-theory on $\frac{\R^{4}}{\mathbb{Z}_{N}} \times \mathbb{S}^{2}\times \mathbb{S}^{2}$};
    \node (IIA2) [below=of M2] {\,\,$N$ D6-brane wrapping $\mathbb{S}^{2}\times \mathbb{S}^{2}$ of CY3};

    \draw[<->] (M1) -- node {Geometric Transition} (M2);
    \draw[<->] (IIA1) -- node {Brane/Flux Transition} (IIA2);

    \draw[<->] (M1) -- (IIA1);
    \draw[<->] (M2) -- (IIA2);

\end{tikzpicture}
\caption{The proposed CY4 geometric transition. Both sides of the M-theory geometry have the link space $L_{7} = Q^{\scriptscriptstyle(1,1,1)}/\mathbb{Z}_{N}$.}
\label{Fig:CY4-geometric-transition}
\end{figure}

\paragraph{Cycles and irreducible holonomy.}

For non-compact conical spaces with special holonomy, such as $SU(n)$, G2, and $Spin(7)$, a crucial question arises: whether the cycles at the zero section, after resolving or deforming the geometry, are supersymmetric. That is, whether these cycles preserve some amount of supersymmetry. These supersymmetric cycles, and their properties, are summarized in Table \ref{Tab:cycles-holonomy}; for further discussion, see \cite{Gomis:2001vk}. The significance of such cycles lies in the fact that when D-branes or M-branes are wrapped on them, they give rise to BPS states.

Interestingly, generic manifolds with $Spin(7)$ holonomy lack calibrated 5-cycles, as noted in Table \ref{Tab:cycles-holonomy}. Despite this, the geometric transition described by (\ref{eq:Spin7-geometric-transition}) remains valid. This is because 3d $\N=1$ theories inherently do not support BPS particle states \cite{Ferrara:1997tx,Gukov:2002zg}.

For the proposed CY4 geometric transition, a 5-cycle emerges, as illustrated in Figure \ref{Fig:CY4-geometric-transition}. While generic conical spaces with $SU(4)$ holonomy (i.e., CY4) do not admit supersymmetric 5-cycles, states arising from branes wrapping this 5-cycle, or its components, are not expected to be BPS. However, this does not necessarily preclude the occurrence of the geometric transition.

Nevertheless, since our focus lies in the deep IR regime, where all heavy states are integrated out, the effective description is governed by a topological theory. In this context, whether this phase preserves the four supercharges or not is irrelevant for the deep IR topological description.

\begin{table}[H]\label{cyclesandholonomy}
\centering
\begin{tabular}{|c|c|c|c|c|c|}
\hline
\( p+1 \) & SU(2) & SU(3) & G2 & SU(4) & Spin(7) \\ \hline
2 & divisor/SLag & holomorphic & - & holomorphic & - \\ \hline
3 & - & SLag & associative & - & - \\ \hline
4 & X & divisor & coassociative & Cayley & Cayley \\ \hline
5 &  & - & - & - & - \\ \hline
6 &  & X & - & divisor & - \\ \hline
7 &  &  & X & - & - \\ \hline
8 &  &  &  & X & X \\ \hline
\end{tabular}
\caption{Supersymmetric cycles in irreducible holonomy manifolds. The table is borrowed from \cite{Gomis:2001vk}.}
\label{Tab:cycles-holonomy}
\end{table}

\subsection{Isometry and isotropy of \texorpdfstring{$\CC(Q^{\scriptscriptstyle(1,1,1)})$}{C(Q(1,1,1))}}

As mentioned in section \ref{sec:CY4-Q111-quotient-physics}, the Sasaki-Einstein 7-manifold $Q^{\scriptscriptstyle(1,1,1)}$, (\ref{defQ111space}), is defined by the quotient space
\begin{equation}
    Q^{\scriptscriptstyle(1,1,1)} \ \cong \ \frac{SU(2)\times SU(2) \times SU(2)}{U(1)\times U(1)}.
\end{equation}
The cone over $Q^{\scriptscriptstyle(1,1,1)}$ is an example of asymptotically conical spaces (AC). Asymptotically, on such spaces the metric takes the form
\begin{equation}
    ds^{2}_{d}(\CC(L_{d-1})) = dr^{2} + r^{2} ds^{2}(L_{d-1}),
\end{equation}
with $ds^{2}(L_{d-1})$ being the metric on cone's link.

The $Q^{\scriptscriptstyle(1,1,1)}$ manifold can be characterized as a cohomogeneity-one manifold, meaning it possesses a high degree of symmetry. Specifically, a Riemannian manifold $(M,g)$ is known as a cohomogeneity-one metric if the isometric group $G$ is a Lie group and the manifold's principle bundle is given by $G/K$ of co-dimension one.

Since the link is a cohomogeneity-one space, then the AC space can be defined through the group system \cite{Acharya:2004qe},
\begin{equation}\label{groupdataKHG}
    K \subset H \subset G.
\end{equation}
Here, $G$ is the isometry group of the AC space, $H$ and $K$ are the isotropy/stabilizer groups. In particular, the link $L_{d-1}$ is the homogeneous quotient space
\begin{equation}
    L_{d-1} = \frac{G}{K}.
\end{equation}
Whereas a non-trivial zero section, $B$, is defined as
\begin{equation}
    B = \frac{G}{H}.
\end{equation}

The topology of the normal bundle over the zero section $B$ can be determined by considering the quotient $H/K$. In general, this quotient gives a $k$-sphere $S^{k}$, which can be understood as the sphere at infinity of the Euclidean space $\R^{k+1}$. Thus, the topology of the resolved or deformed cone over a cohomogeneity-one link space is given by $\R^{k+1}\times B$. 

In the case of resolved $\CC(Q^{\scriptscriptstyle(1,1,1)})$, the isotropy group of it's zero section, $\C\P^{1}\times \C\P^{1}$, is given as $U(1)\times U(1)$. To describe the resolved cone, we need to enlarge this to $SU(2)\times U(1)\times U(1)$, determining the isotropy $H$. Hence, the group system that describes the resolved cone shall be given as
\begin{equation}
    U(1)\times U(1) \subset SU(2)\times U(1)\times U(1) \subset  SU(2) \times SU(2) \times SU(2).
\end{equation}
For consistency, the zero section is given by
\begin{equation}
    \begin{split}
        \frac{SU(2) \times SU(2) \times SU(2)}{SU(2)\times U(1)\times U(1)} &=\, \frac{SU(2)\times SU(2)}{U(1)\times U(1)} 
        \\
        &=\, \C\P^{1}\times \C\P^{1}
        \\
        &\simeq\, \mathbb{S}^{2}\times \mathbb{S}^{2}.
    \end{split}
\end{equation}
It is noteworthy that there exist three distinct possibilities for such a space, depending on the embedding of $SU(2)\subset H$ in $G$. This is reflected in the three possible flops that can be performed on the toric diagram in Figure \ref{Fig:toricQ111}.   

According to the general discussion presented above, the topology of this phase is given as 
\begin{equation}
    \R^{4}\times \mathbb{S}^{2} \times \mathbb{S}^{2}.
\end{equation}

The zero section of the proposed deformed + resolved (DR) phase is given as $\mathbb{S}^{3}\times\mathbb{S}^{2}$. Its isotropy group is given as $SU(2)\times U(1)$. This isotropy group can be identified with $H$ without further enlargement, as it satisfies the condition in (\ref{groupdataKHG}). Therefore, the group data for the new DR-phase is 
\begin{equation}
    U(1)\times U(1) \subset SU(2)\times U(1) \subset  SU(2) \times SU(2) \times SU(2).
\end{equation}
To verify, we have the following relation for the zero section,
\begin{equation}
    \begin{split}
        \frac{SU(2) \times SU(2) \times SU(2)}{SU(2)\times U(1)} &= \, \frac{SU(2)\times SU(2)}{U(1)} 
        \\
        &= \, U(1)\hookrightarrow \C\P^{1}\times \C\P^{1}\,.
    \end{split}
\end{equation}
This defines a 5-manifold $\mathbb{S}^{2}\times\mathbb{S}^{3}$.

The topology of the deformed and resolved (DR) phase of the cone is given by
\begin{equation}
    \R^{3}\times \mathbb{S}^{3}\times \mathbb{S}^{2}\,.
\end{equation}

Note that, the embedding of $SU(2)\subset H$ into $G$ can be done in three different ways. This suggests that there exist three different DR-phases. 

\paragraph{Including the $\Z_{N}$ quotient.}

Here, we consider the description of the DR-phase for the quotient presented in (\ref{quotientCQZNonzi}). To do that, we pick up the discussion from (\ref{eq:ZN-action-Bi}) onward. Recall that the $\Z_{N}$ action on the symmetry group $SU(2)_{\scriptscriptstyle A}\times SU(2)_{\scriptscriptstyle B}\times SU(2)_{\scriptscriptstyle C}$ is defined as:
\begin{equation}
    \begin{split}
        (A_{1}, A_{2}) \sim (A_{1}, A_{2})\,,   \qquad  (B_{1}, B_{2}) \sim (\lambda^{-1} B_{1}, \lambda B_{2})\,,  \qquad (C_{1}, C_{2}) \sim (C_{1},C_{2})\,.
    \end{split}
\end{equation}
To describe the link space $Q^{\scriptscriptstyle(1,1,1)}$, the doublets must satisfy the constraints \cite{Oh:1998qi,Fabbri:1999hw,Herzog:2000rz}:
\begin{equation}
|A_{1}|^{2} + |A_{2}|^{2}\,=\, 1\,,\qquad |B_{1}|^{2} + |B_{2}|^{2}\,=\, 1\,,\qquad |C_{1}|^{2} + |C_{2}|^{2}\,=\, 1\,,
\end{equation}
reflecting the property $SU(2) \cong \mathbb{S}^{3}$.

Recall that the isometry group of the local metric (\ref{eq:isometry-Q111-ZN-from-metric}) is 
\begin{equation}\label{isometry3}
    SU(2)_{\scriptscriptstyle A}\times SU(2)_{\scriptscriptstyle B}\times SU(2)_{\scriptscriptstyle C}
\end{equation}
We now apply the arguments from this subsection to the quotient space, describing the resolved phase and the DR-phase:
\begin{itemize}
    \item The resolved geometry is given by: 
\begin{equation}\label{S2S2free}
    \frac{SU(2)_{\scriptscriptstyle A}\times SU(2)_{\scriptscriptstyle B} \times SU(2)_{\scriptscriptstyle C}}{SU(2)_{\scriptscriptstyle B}\times U(1)_{\scriptscriptstyle A}\times U(1)_{\scriptscriptstyle C}} \ \cong \ \mathbb{S}^{2}\times \mathbb{S}^{2}
\end{equation}
Here, the two 2-spheres are smooth. To determine the topology, consider
\begin{equation}
    \begin{split}
        \frac{SU(2)_{\scriptscriptstyle B}\times U(1)_{\scriptscriptstyle A}\times U(1)_{\scriptscriptstyle C}}{U(1)_{\scriptscriptstyle A}\times U(1)_{\scriptscriptstyle C}} \ &\cong \  SU(2)_{\scriptscriptstyle B} \ 
        \\
        &\cong  \mathbb{S}^{3}/\Z_{N}\,.
    \end{split}
\end{equation}
By viewing this $\mathbb{S}^{3}/\Z_{N}$ as the sphere at infinity of the normal space $\R^{4}/\Z_{N}$, we find:
\begin{equation}
    (\R^{4}/\Z_{N}) \times \mathbb{S}^{2}\times \mathbb{S}^{2}.
\end{equation}
This embedding of the system $K \subset H \subset G$ is consistent with the geometry described in (\ref{GeoofConeoverQZN}) and (\ref{realcoordZNresolved}).

\item The new DR-phase is described as:
\begin{equation}\label{eq:DR-phase-isometry-approach}
    \begin{split}
        \frac{SU(2)_{\scriptscriptstyle A}\times SU(2)_{\scriptscriptstyle C} \times SU(2)_{\scriptscriptstyle B}}{SU(2)_{\scriptscriptstyle A}\times U(1)_{\scriptscriptstyle C}} \ & \cong  \ (\mathbb{S}^{3}\times \mathbb{S}^{2})/\Z_{N}
    \end{split}
\end{equation}
To determine the topology of the normal direction, we compute
\begin{equation}
        \frac{SU(2)_{\scriptscriptstyle A}\times U(1)_{\scriptscriptstyle C}}{U(1)_{\scriptscriptstyle A}\times U(1)_{\scriptscriptstyle C}} \ \cong \  \mathbb{S}^{2}\,.
\end{equation}
Thus, the topology of the normal direction of $L(N,1)\times \mathbb{S}^{2}$ is $\R^{3}$, leading to the topology of the total CY4 space as: 
\begin{equation}
    \R^{3}\,\times\, L(N,1)\,\times\, \mathbb{S}^{2}\,.
\end{equation}
\end{itemize}

Other choices of the embedding $K\subset H\subset G$ may lead to distinct branches or phases. For instance, modifying the embedding in (\ref{S2S2free}) can introduce singularities along the $\mathbb{S}^{2} \times \mathbb{S}^{2}$ base. However, such cases fall beyond the scope of this paper.

Before concluding this section, we observe that the space $T^{\scriptscriptstyle(1,1)}$ can be described as a hypersurface in  $Q^{\scriptscriptstyle(1,1,1)}$. For instance, consider the matrix
\begin{equation}\label{eq:W-T11-hypersurface}
    W =
    \begin{pmatrix}
        B_{1}
        \\
        B_{2} 
    \end{pmatrix}
    \begin{pmatrix}
        C_{1}  \ C_{2}
    \end{pmatrix}\,, \qquad \text{such that}\ \,   \det(W)=0\,. 
\end{equation} 
For this choice, following the analysis in \cite{Evslin:2007ux}, the $\Z_{N}$ quotient acts on $\mathbb{S}^{3}\times\mathbb{S}^{2}$ as:
\begin{equation}\label{ZNonT11}
    L(N,1)\times \mathbb{S}^{2}\ \cong \ (\mathbb{S}^{3}/\Z_{N}) \times \mathbb{S}^{2}\,,
\end{equation}
with $L(N,1)$ denotes the lens space.
Notably, the  $C_{i}$ doublet in the construction can be replaced with the  $A_{i}$ doublet, yielding an equivalent result.

We observe that the $\mathbb{S}^{3}\times\mathbb{S}^{2}$ hypersurface can be identified with the one described in (\ref{eq:DR-phase-isometry-approach}), as both are characterized by the doublets $B_{i}$ and $C_{i}$.

Aside note, an alternative description of (\ref{eq:W-T11-hypersurface}) can be expressed in terms of complex coordinates in $\C^{4}$,
\begin{equation}
    \begin{split}
        &\widetilde{z}_{1} = B_{1}C_{1}, \qquad \widetilde{z}_{2} = B_{2}C_{2},
        \\
        &\widetilde{z}_{3} = B_{1}C_{2}, \qquad \widetilde{z}_{4} = B_{2}C_{1}\,,
    \end{split}
\end{equation}
satisfying the conifold equation:
\begin{equation}
    \widetilde{z}_{1}\widetilde{z}_{2}-\widetilde{z}_{3}\widetilde{z}_{4} = 0\,.
\end{equation}
This quotient on the conifold coincides with the one considered in \cite{Davies:2011is, Davies:2013pna}, with its physical interpretation explored in works such as \cite{Closset:2018bjz, Acharya:2020vmg, Najjar:2022eci}. Consequently, the base of the conifold remains given by (\ref{ZNonT11}).

\section{The \texorpdfstring{$\CC(Q^{\scriptscriptstyle(1,1,1)})$}{C(Q(1,1,1))} as an interlaced geometry}\label{app:interlacing-geometry}

The toric diagram of the cone $\mathcal{C}(Q^{\scriptscriptstyle(1,1,1)})$ is shown in Figure \ref{Fig:toricQ111}. From a toric perspective, its interlacing structure is evident, as outlined in \cite[App.B]{Najjar:2023hee} and references therein. Specifically, the diagram in Figure \ref{Fig:toricQ111} can be constructed from two distinct $\mathcal{C}(T^{\scriptscriptstyle(1,1)})$ cones embedded in orthogonal sublattices $\Z^{2}\subset \Z^{3}$. There are three possible configurations; for example,
\begin{equation}\label{interlacingCQ111}
    \text{Toric}\  \mathcal{C}(Q^{\scriptscriptstyle(1,1,1)})|_{\Z^{3}} \,\cong\,  \text{Toric}\  \mathcal{C}(T^{\scriptscriptstyle(1,1)})|_{\Z^{2}}\, \cap\, \text{Toric} \  \mathcal{C}(T^{\scriptscriptstyle(1,1)})|_{\widetilde{\Z}^{2}}.
\end{equation}
With $\Z^{2}\subset \Z^{3}$ and $\widetilde{\Z}^{2}\subset \Z^{3}$ are orthogonal 2d sub-lattices.

\paragraph{The $\mathcal{C}(Q^{\scriptscriptstyle(1,1,1)})$ geometry.}

The resolved conifold $ \widetilde{\mathcal{C}(T^{\scriptscriptstyle(1,1)})}$ is given by the total space of the line bundle \cite{Candelas:1989js},
\begin{equation}
    \widetilde{\mathcal{C}(T^{\scriptscriptstyle(1,1)})} \,\cong\, \O(-1)\oplus\O(-1)\hookrightarrow \C\P^{1}.
\end{equation}
Extending this to ${\CC(Q^{\scriptscriptstyle(1,1,1)})}$, the resolved cone $\widetilde{\CC(Q^{\scriptscriptstyle(1,1,1)})}$ can be obtained by combining two resolved CY3 conifolds. Distinguishing the two CY3 conifolds with labels $a$ and $b$, we have:  
\begin{equation}\label{OOfiberoverPP}
  \begin{split}
        \widetilde{\CC(Q^{\scriptscriptstyle(1,1,1)})}\, \cong \, &(\O(-1)_{a}\oplus \O(-1)_{1} \hookrightarrow  \C\P^{1}_{a}  )\  \otimes \ ( \O(-1)_{b}\oplus \O(-1)_{b} \hookrightarrow \C\P^{1}_{b} ) 
         \\
         \qquad \qquad &=  \O_{\C\P^{1}_{a}\times\C\P^{1}_{b}}(-1,-1) \ \oplus\  \O_{\C\P^{1}_{a}\times\C\P^{1}_{b}}(-1,-1) \hookrightarrow \C\P^{1}_{a}\times\C\P^{1}_{b}.
  \end{split}
\end{equation}
The symbol ``$\otimes$'' denotes the product of line bundles:
\begin{equation}\label{defOtimes}
    \O(-1)_{a}\otimes \O(-1)_{b} \,:=\, \O(-1,-1)_{a,b}.
\end{equation}
The zero section of this resolved geometry corresponds to the 4-cycle $\C\P^{1}\times \C\P^{1}$.

In real coordinates, the topology of $\widetilde{\CC(Q^{\scriptscriptstyle(1,1,1)})}$ is:
\begin{equation}
    \R^{4} \times \mathbb{S}^{2}\times \mathbb{S}^{2}.
\end{equation}

The zero section is the trivial bundle $\C\P^{1}_{a}\times\C\P^{1}_{b}$, classified by $\pi_{2}(SO(3))\simeq \Z_{2}$. Therefore, $\widetilde{\CC(Q^{\scriptscriptstyle(1,1,1)})}$ corresponds to the identity element of $\pi_{2}(SO(3))$, and its geometry can be schematically written as: 
\begin{equation}
     \left(\widetilde{\CC(T^{\scriptscriptstyle(1,1)})} \otimes (\O(-1)\oplus\O(-1))\right)\ \hookrightarrow \ \widetilde{\CC(Q^{\scriptscriptstyle(1,1,1)})} \ \to \  \C\P^{1}.
\end{equation}

Note that this geometry is different from that described in \cite{Intriligator:2012ue}, where they have taken the bundle of $\CC(T^{\scriptscriptstyle(1,1)})\hookrightarrow \Sigma_{(2)}^{g}$, with $\Sigma_{(2)}^{g}$ is a Riemannian surface with genus $g$ such that $g\geq 1$. Their four cycle is then given as $\C\P^{1}\hookrightarrow\Sigma_{(2)}^{g}$.

\paragraph{The $\mathcal{C}(Q^{\scriptscriptstyle(1,1,1)}/\Z_{N})$ geometry.}

The interlacing structure can be extended to the quotient geometry $\CC(Q^{\scriptscriptstyle(1,1,1)}/\Z_{N})$ given in (\ref{quotientCQZNonzi}).  Specifically, the cone $\CC(Q^{\scriptscriptstyle(1,1,1)}/\Z_{N})$  can be represented as an interlacing of two ladder hyperconifolds $\CC(T^{\scriptscriptstyle(1,1)}/\Z_{N})$. The ladder hyperconifolds are discussed in \cite{Davies:2011is,Davies:2013pna, Acharya:2020vmg,Najjar:2022eci}. Parallel to (\ref{interlacingCQ111}), we write
\begin{equation}\label{interlacingCQ111ZN}
    \CC(Q^{\scriptscriptstyle(1,1,1)}/\Z_{N})\  \,\cong\,   \ \text{Toric}\  (\mathcal{C}(T^{\scriptscriptstyle(1,1)}/\Z_{N}))|_{\Z^{2}}\, \cap\, \text{Toric} \  (\mathcal{C}(T^{\scriptscriptstyle(1,1)}/\Z_{N}))|_{\widetilde{\Z}^{2}}.
\end{equation}
Here, $\Z^{2}\subset \Z^{3}$ and $\widetilde{\Z}^{2}\subset \Z^{3}$ are orthogonal 2d sub-lattices. A ladder hyperconifold is a toric CY3 given as a particular $\Z_{N}$ quotient of the conifold. The non-compact toric divisors of the (ladder) hyperconifold are located at
\begin{equation}
    v_{1} = (0,0,1), \quad v_{2}=(1,0,1), \quad v_{3} = (1,N,1), \quad v_{4} = (2,N,1).
\end{equation}

Geometrically, the $\Z_{N}$ ladder hyperconifold can be written as \cite{Davies:2011is,Davies:2013pna,Najjar:2022eci}
\begin{equation}
    \frac{\O(-1) \ \oplus\  \O(-1)}{\Z_{N}} \hookrightarrow \ \CC(T^{1,1})/\Z_{N} \ \to \ \C\P^{1}.
\end{equation}

Following the discussion around (\ref{OOfiberoverPP}) and the interlacing picture above, the fiber bundle description of the quotient geometry $\CC(Q^{\scriptscriptstyle(1,1,1)})/\Z_{N}$ is then given as
\begin{equation}\label{GeoofConeoverQZN}
 \begin{split}
      \CC(Q^{\scriptscriptstyle(1,1,1)}/\Z_{N})\, &\cong\    \ \left(\frac{\O(-1) \ \oplus\  \O(-1)}{\Z_{N}} \hookrightarrow \C\P^{1}\right)\ \otimes \left(\frac{\O(-1) \ \oplus\  \O(-1)}{\Z_{N}}  \hookrightarrow \C\P^{1}\right)
      \\
      &\simeq \  \left[\frac{\O(-1) \ \oplus\  \O(-1)}{\Z_{N}}\right] \ \otimes \left[\frac{\O(-1) \ \oplus\  \O(-1)}{\Z_{N}}\right]\, \hookrightarrow\, \C\P^{1}\times \C\P^{1}
      \\
      &\simeq \   \frac{\O(-1,-1) \ \oplus\  \O(-1,-1)}{\Z_{N}} \,\,\hookrightarrow \,\,\C\P^{1}\times\C\P^{1}.
 \end{split}
\end{equation}
Here, the definition of ``$\otimes$'' is given as in (\ref{defOtimes}).

In real coordinates, the singular geometry is described by 
\begin{equation}\label{realcoordZNresolved}
    (\R^{4}/\Z_{N}) \ \hookrightarrow \mathbb{S}^{2}\times \mathbb{S}^{2}.
\end{equation}
Here, $\mathbb{R}^4 / \mathbb{Z}_N$ fibers over both copies of the 2-sphere, reflecting the line bundle $\frac{\mathcal{O}(-1,-1) \oplus \mathcal{O}(-1,-1)}{\mathbb{Z}_N}$.


\bibliographystyle{JHEP}
\bibliography{F-ref}

\providecommand{\href}[2]{#2}\begingroup\raggedright\begin{thebibliography}{10}

\bibitem{Acharya:2000gb}
B.~S. Acharya, \emph{{On Realizing N=1 superYang-Mills in M theory}},  \href{http://arxiv.org/abs/hep-th/0011089}{{\tt hep-th/0011089}}.

\bibitem{Atiyah:2001qf}
M.~Atiyah and E.~Witten, \emph{{M theory dynamics on a manifold of G(2) holonomy}}, \href{http://dx.doi.org/10.4310/ATMP.2002.v6.n1.a1}{\emph{Adv. Theor. Math. Phys.} {\bf 6} (2003) 1--106}, [\href{http://arxiv.org/abs/hep-th/0107177}{{\tt hep-th/0107177}}].

\bibitem{Acharya:2001hq}
B.~S. Acharya, \emph{{Confining strings from G(2) holonomy space-times}},  \href{http://arxiv.org/abs/hep-th/0101206}{{\tt hep-th/0101206}}.

\bibitem{Acharya:2004qe}
B.~S. Acharya and S.~Gukov, \emph{{M theory and singularities of exceptional holonomy manifolds}}, \href{http://dx.doi.org/10.1016/j.physrep.2003.10.017}{\emph{Phys. Rept.} {\bf 392} (2004) 121--189}, [\href{http://arxiv.org/abs/hep-th/0409191}{{\tt hep-th/0409191}}].

\bibitem{Acharya:2024bnt}
B.~S. Acharya, \emph{{Confinement in Five Dimensions}},  \href{http://arxiv.org/abs/2407.03171}{{\tt 2407.03171}}.

\bibitem{Oh:2001bf}
K.-h. Oh and R.~Tatar, \emph{{Duality and confinement in N=1 supersymmetric theories from geometric transitions}}, \href{http://dx.doi.org/10.4310/ATMP.2002.v6.n1.a3}{\emph{Adv. Theor. Math. Phys.} {\bf 6} (2003) 141--196}, [\href{http://arxiv.org/abs/hep-th/0112040}{{\tt hep-th/0112040}}].

\bibitem{Dasgupta:2001fg}
K.~Dasgupta, K.~Oh and R.~Tatar, \emph{{Open / closed string dualities and Seiberg duality from geometric transitions in M theory}}, \href{http://dx.doi.org/10.1088/1126-6708/2002/08/026}{\emph{JHEP} {\bf 08} (2002) 026}, [\href{http://arxiv.org/abs/hep-th/0106040}{{\tt hep-th/0106040}}].

\bibitem{Diaconescu:1998ua}
D.-E. Diaconescu and S.~Gukov, \emph{{Three-dimensional N=2 gauge theories and degenerations of Calabi-Yau four folds}}, \href{http://dx.doi.org/10.1016/S0550-3213(98)00597-5}{\emph{Nucl. Phys. B} {\bf 535} (1998) 171--196}, [\href{http://arxiv.org/abs/hep-th/9804059}{{\tt hep-th/9804059}}].

\bibitem{Gukov:1999ya}
S.~Gukov, C.~Vafa and E.~Witten, \emph{{CFT's from Calabi-Yau four folds}}, \href{http://dx.doi.org/10.1016/S0550-3213(00)00373-4}{\emph{Nucl. Phys. B} {\bf 584} (2000) 69--108}, [\href{http://arxiv.org/abs/hep-th/9906070}{{\tt hep-th/9906070}}].

\bibitem{Intriligator:2012ue}
K.~Intriligator, H.~Jockers, P.~Mayr, D.~R. Morrison and M.~R. Plesser, \emph{{Conifold Transitions in M-theory on Calabi-Yau Fourfolds with Background Fluxes}}, \href{http://dx.doi.org/10.4310/ATMP.2013.v17.n3.a2}{\emph{Adv. Theor. Math. Phys.} {\bf 17} (2013) 601--699}, [\href{http://arxiv.org/abs/1203.6662}{{\tt 1203.6662}}].

\bibitem{Najjar:2023hee}
M.~Najjar, J.~Tian and Y.-N. Wang, \emph{{3d $ \mathcal{N} $ = 2 theories from M-theory on CY4 and IIB brane box}}, \href{http://dx.doi.org/10.1007/JHEP05(2024)038}{\emph{JHEP} {\bf 05} (2024) 038}, [\href{http://arxiv.org/abs/2312.17082}{{\tt 2312.17082}}].

\bibitem{Sangiovanni:2024nfz}
A.~Sangiovanni and R.~Valandro, \emph{{M-theory geometric engineering for rank-0 3d $\mathcal{N}=2$ theories}},  \href{http://arxiv.org/abs/2410.13943}{{\tt 2410.13943}}.

\bibitem{DAuria:1983sda}
R.~D'Auria, P.~Fre and P.~van Nieuwenhuizen, \emph{{$N=2$ Matter Coupled Supergravity From Compactification on a Coset $G/H$ Possessing an Additional Killing Vector}}, \href{http://dx.doi.org/10.1016/0370-2693(84)92018-5}{\emph{Phys. Lett. B} {\bf 136} (1984) 347--353}.

\bibitem{Oh:1998qi}
K.~Oh and R.~Tatar, \emph{{Three-dimensional SCFT from M2-branes at conifold singularities}}, \href{http://dx.doi.org/10.1088/1126-6708/1999/02/025}{\emph{JHEP} {\bf 02} (1999) 025}, [\href{http://arxiv.org/abs/hep-th/9810244}{{\tt hep-th/9810244}}].

\bibitem{caibar1999minimal}
M.~Caibar, \emph{Minimal models of canonical singularities and their cohomology}.
\newblock PhD thesis, Ph. D. thesis, University of Warwick, 1999.

\bibitem{Berasaluce-Gonzalez:2012abm}
M.~Berasaluce-Gonzalez, P.~G. Camara, F.~Marchesano, D.~Regalado and A.~M. Uranga, \emph{{Non-Abelian discrete gauge symmetries in 4d string models}}, \href{http://dx.doi.org/10.1007/JHEP09(2012)059}{\emph{JHEP} {\bf 09} (2012) 059}, [\href{http://arxiv.org/abs/1206.2383}{{\tt 1206.2383}}].

\bibitem{vanBeest:2022fss}
M.~van Beest, D.~S.~W. Gould, S.~Schafer-Nameki and Y.-N. Wang, \emph{{Symmetry TFTs for 3d QFTs from M-theory}}, \href{http://dx.doi.org/10.1007/JHEP02(2023)226}{\emph{JHEP} {\bf 02} (2023) 226}, [\href{http://arxiv.org/abs/2210.03703}{{\tt 2210.03703}}].

\bibitem{Garcia-Valdecasas:2023mis}
E.~Garc\'\i{}a-Valdecasas, \emph{{Non-invertible symmetries in supergravity}}, \href{http://dx.doi.org/10.1007/JHEP04(2023)102}{\emph{JHEP} {\bf 04} (2023) 102}, [\href{http://arxiv.org/abs/2301.00777}{{\tt 2301.00777}}].

\bibitem{Najjar:2024vmm}
M.~Najjar, L.~Santilli and Y.-N. Wang, \emph{{(-1)-form symmetries from M-theory and SymTFTs}},  \href{http://arxiv.org/abs/2411.19683}{{\tt 2411.19683}}.

\bibitem{Nilsson:1984bj}
B.~E.~W. Nilsson and C.~N. Pope, \emph{{Hopf Fibration of Eleven-dimensional Supergravity}}, \href{http://dx.doi.org/10.1088/0264-9381/1/5/005}{\emph{Class. Quant. Grav.} {\bf 1} (1984) 499}.

\bibitem{Sorokin:1984ca}
D.~P. Sorokin, V.~I. Tkach and D.~V. Volkov, \emph{{KALUZA-KLEIN THEORIES AND SPONTANEOUS COMPACTIFICATION MECHANISMS OF EXTRA SPACE DIMENSIONS}},  in \emph{{3rd Seminar on Quantum Gravity}}, 1984.

\bibitem{Sorokin:1985ap}
D.~P. Sorokin, V.~I. Tkach and D.~V. Volkov, \emph{{ON THE RELATIONSHIP BETWEEN COMPACTIFIED VACUA OF D = 11 AND D = 10 SUPERGRAVITIES}}, \href{http://dx.doi.org/10.1016/0370-2693(85)90766-X}{\emph{Phys. Lett. B} {\bf 161} (1985) 301--306}.

\bibitem{DUFF19861}
M.~Duff, B.~Nilsson and C.~Pope, \emph{Kaluza-klein supergravity}, \href{http://dx.doi.org/https://doi.org/10.1016/0370-1573(86)90163-8}{\emph{Physics Reports} {\bf 130} (1986) 1--142}.

\bibitem{Gauntlett:2004hh}
J.~P. Gauntlett, D.~Martelli, J.~F. Sparks and D.~Waldram, \emph{{A New infinite class of Sasaki-Einstein manifolds}}, \href{http://dx.doi.org/10.4310/ATMP.2004.v8.n6.a3}{\emph{Adv. Theor. Math. Phys.} {\bf 8} (2004) 987--1000}, [\href{http://arxiv.org/abs/hep-th/0403038}{{\tt hep-th/0403038}}].

\bibitem{Franco:2009sp}
S.~Franco, I.~R. Klebanov and D.~Rodriguez-Gomez, \emph{{M2-branes on Orbifolds of the Cone over $Q^{1,1,1}$}}, \href{http://dx.doi.org/10.1088/1126-6708/2009/08/033}{\emph{JHEP} {\bf 08} (2009) 033}, [\href{http://arxiv.org/abs/0903.3231}{{\tt 0903.3231}}].

\bibitem{FRIEDRICH1997259}
T.~Friedrich, I.~Kath, A.~Moroianu and U.~Semmelmann, \emph{On nearly parallel g2-structures}, \href{http://dx.doi.org/https://doi.org/10.1016/S0393-0440(97)80004-6}{\emph{Journal of Geometry and Physics} {\bf 23} (1997) 259--286}.

\bibitem{Acharya:1998db}
B.~S. Acharya, J.~M. Figueroa-O'Farrill, C.~M. Hull and B.~J. Spence, \emph{{Branes at conical singularities and holography}}, \href{http://dx.doi.org/10.4310/ATMP.1998.v2.n6.a2}{\emph{Adv. Theor. Math. Phys.} {\bf 2} (1999) 1249--1286}, [\href{http://arxiv.org/abs/hep-th/9808014}{{\tt hep-th/9808014}}].

\bibitem{Fabbri:1999hw}
D.~Fabbri, P.~Fre', L.~Gualtieri, C.~Reina, A.~Tomasiello, A.~Zaffaroni et~al., \emph{{3-D superconformal theories from Sasakian seven manifolds: New nontrivial evidences for AdS(4) / CFT(3)}}, \href{http://dx.doi.org/10.1016/S0550-3213(00)00098-5}{\emph{Nucl. Phys. B} {\bf 577} (2000) 547--608}, [\href{http://arxiv.org/abs/hep-th/9907219}{{\tt hep-th/9907219}}].

\bibitem{EGUCHI197982}
T.~Eguchi and A.~J. Hanson, \emph{Self-dual solutions to euclidean gravity}, \href{http://dx.doi.org/https://doi.org/10.1016/0003-4916(79)90282-3}{\emph{Annals of Physics} {\bf 120} (1979) 82--106}.

\bibitem{Eguchi:1980jx}
T.~Eguchi, P.~B. Gilkey and A.~J. Hanson, \emph{{Gravitation, Gauge Theories and Differential Geometry}}, \href{http://dx.doi.org/10.1016/0370-1573(80)90130-1}{\emph{Phys. Rept.} {\bf 66} (1980) 213}.

\bibitem{DallAgata:1999ivu}
G.~Dall'Agata, \emph{{N=2 conformal field theories from M2-branes at conifold singularities}}, \href{http://dx.doi.org/10.1016/S0370-2693(99)00642-5}{\emph{Phys. Lett. B} {\bf 460} (1999) 79--86}, [\href{http://arxiv.org/abs/hep-th/9904198}{{\tt hep-th/9904198}}].

\bibitem{Herzog:2000rz}
C.~P. Herzog and I.~R. Klebanov, \emph{{Gravity duals of fractional branes in various dimensions}}, \href{http://dx.doi.org/10.1103/PhysRevD.63.126005}{\emph{Phys. Rev. D} {\bf 63} (2001) 126005}, [\href{http://arxiv.org/abs/hep-th/0101020}{{\tt hep-th/0101020}}].

\bibitem{Candelas:1989js}
P.~Candelas and X.~C. de~la Ossa, \emph{{Comments on Conifolds}}, \href{http://dx.doi.org/10.1016/0550-3213(90)90577-Z}{\emph{Nucl. Phys. B} {\bf 342} (1990) 246--268}.

\bibitem{Closset:2009sv}
C.~Closset, \emph{{Toric geometry and local Calabi-Yau varieties: An Introduction to toric geometry (for physicists)}},  \href{http://arxiv.org/abs/0901.3695}{{\tt 0901.3695}}.

\bibitem{Davies:2013pna}
R.~Davies, \emph{{Classification and Properties of Hyperconifold Singularities and Transitions}},  \href{http://arxiv.org/abs/1309.6778}{{\tt 1309.6778}}.

\bibitem{cox2011toric}
D.~Cox, J.~Little and H.~Schenck, \emph{Toric Varieties}.
\newblock Graduate studies in mathematics. American Mathematical Society, 2011.

\bibitem{Martelli:2008rt}
D.~Martelli and J.~Sparks, \emph{{Notes on toric Sasaki-Einstein seven-manifolds and AdS(4) / CFT(3)}}, \href{http://dx.doi.org/10.1088/1126-6708/2008/11/016}{\emph{JHEP} {\bf 11} (2008) 016}, [\href{http://arxiv.org/abs/0808.0904}{{\tt 0808.0904}}].

\bibitem{Leung:1997tw}
N.~C. Leung and C.~Vafa, \emph{{Branes and toric geometry}}, \href{http://dx.doi.org/10.4310/ATMP.1998.v2.n1.a4}{\emph{Adv. Theor. Math. Phys.} {\bf 2} (1998) 91--118}, [\href{http://arxiv.org/abs/hep-th/9711013}{{\tt hep-th/9711013}}].

\bibitem{Acharya:2020vmg}
B.~S. Acharya, L.~Foscolo, M.~Najjar and E.~E. Svanes, \emph{{New G$_{2}$-conifolds in M-theory and their field theory interpretation}}, \href{http://dx.doi.org/10.1007/JHEP05(2021)250}{\emph{JHEP} {\bf 05} (2021) 250}, [\href{http://arxiv.org/abs/2011.06998}{{\tt 2011.06998}}].

\bibitem{Apruzzi:2021nmk}
F.~Apruzzi, F.~Bonetti, I.~n. Garc\'\i{}a~Etxebarria, S.~S. Hosseini and S.~Schafer-Nameki, \emph{{Symmetry TFTs from String Theory}}, \href{http://dx.doi.org/10.1007/s00220-023-04737-2}{\emph{Commun. Math. Phys.} {\bf 402} (2023) 895--949}, [\href{http://arxiv.org/abs/2112.02092}{{\tt 2112.02092}}].

\bibitem{Cordova:2019uob}
C.~C\'ordova, D.~S. Freed, H.~T. Lam and N.~Seiberg, \emph{{Anomalies in the Space of Coupling Constants and Their Dynamical Applications II}}, \href{http://dx.doi.org/10.21468/SciPostPhys.8.1.002}{\emph{SciPost Phys.} {\bf 8} (2020) 002}, [\href{http://arxiv.org/abs/1905.13361}{{\tt 1905.13361}}].

\bibitem{Gaiotto:2014kfa}
D.~Gaiotto, A.~Kapustin, N.~Seiberg and B.~Willett, \emph{{Generalized Global Symmetries}}, \href{http://dx.doi.org/10.1007/JHEP02(2015)172}{\emph{JHEP} {\bf 02} (2015) 172}, [\href{http://arxiv.org/abs/1412.5148}{{\tt 1412.5148}}].

\bibitem{Kapustin:2014gua}
A.~Kapustin and N.~Seiberg, \emph{{Coupling a QFT to a TQFT and Duality}}, \href{http://dx.doi.org/10.1007/JHEP04(2014)001}{\emph{JHEP} {\bf 04} (2014) 001}, [\href{http://arxiv.org/abs/1401.0740}{{\tt 1401.0740}}].

\bibitem{Heckman:2022muc}
J.~J. Heckman, M.~H\"ubner, E.~Torres and H.~Y. Zhang, \emph{{The Branes Behind Generalized Symmetry Operators}}, \href{http://dx.doi.org/10.1002/prop.202200180}{\emph{Fortsch. Phys.} {\bf 71} (2023) 2200180}, [\href{http://arxiv.org/abs/2209.03343}{{\tt 2209.03343}}].

\bibitem{Cvetic:2023plv}
M.~Cveti\v{c}, J.~J. Heckman, M.~H\"ubner and E.~Torres, \emph{{Fluxbranes, generalized symmetries, and Verlinde\textquoteright{}s metastable monopole}}, \href{http://dx.doi.org/10.1103/PhysRevD.109.046007}{\emph{Phys. Rev. D} {\bf 109} (2024) 046007}, [\href{http://arxiv.org/abs/2305.09665}{{\tt 2305.09665}}].

\bibitem{DelZotto:2015isa}
M.~Del~Zotto, J.~J. Heckman, D.~S. Park and T.~Rudelius, \emph{{On the Defect Group of a 6D SCFT}}, \href{http://dx.doi.org/10.1007/s11005-016-0839-5}{\emph{Lett. Math. Phys.} {\bf 106} (2016) 765--786}, [\href{http://arxiv.org/abs/1503.04806}{{\tt 1503.04806}}].

\bibitem{Albertini:2020mdx}
F.~Albertini, M.~Del~Zotto, I.~n. Garc\'\i{}a~Etxebarria and S.~S. Hosseini, \emph{{Higher Form Symmetries and M-theory}}, \href{http://dx.doi.org/10.1007/JHEP12(2020)203}{\emph{JHEP} {\bf 12} (2020) 203}, [\href{http://arxiv.org/abs/2005.12831}{{\tt 2005.12831}}].

\bibitem{Page:1983mke}
D.~N. Page, \emph{{Classical Stability of Round and Squashed Seven Spheres in Eleven-dimensional Supergravity}}, \href{http://dx.doi.org/10.1103/PhysRevD.28.2976}{\emph{Phys. Rev. D} {\bf 28} (1983) 2976}.

\bibitem{Bandos:1997ui}
I.~A. Bandos, K.~Lechner, A.~Nurmagambetov, P.~Pasti, D.~P. Sorokin and M.~Tonin, \emph{{Covariant action for the superfive-brane of M theory}}, \href{http://dx.doi.org/10.1103/PhysRevLett.78.4332}{\emph{Phys. Rev. Lett.} {\bf 78} (1997) 4332--4334}, [\href{http://arxiv.org/abs/hep-th/9701149}{{\tt hep-th/9701149}}].

\bibitem{Intriligator:2000eq}
K.~A. Intriligator, \emph{{Anomaly matching and a Hopf-Wess-Zumino term in 6d, N=(2,0) field theories}}, \href{http://dx.doi.org/10.1016/S0550-3213(00)00148-6}{\emph{Nucl. Phys. B} {\bf 581} (2000) 257--273}, [\href{http://arxiv.org/abs/hep-th/0001205}{{\tt hep-th/0001205}}].

\bibitem{Aharony:2013hda}
O.~Aharony, N.~Seiberg and Y.~Tachikawa, \emph{{Reading between the lines of four-dimensional gauge theories}}, \href{http://dx.doi.org/10.1007/JHEP08(2013)115}{\emph{JHEP} {\bf 08} (2013) 115}, [\href{http://arxiv.org/abs/1305.0318}{{\tt 1305.0318}}].

\bibitem{Pantev:2005rh}
T.~Pantev and E.~Sharpe, \emph{{Notes on gauging noneffective group actions}},  \href{http://arxiv.org/abs/hep-th/0502027}{{\tt hep-th/0502027}}.

\bibitem{Pantev:2005wj}
T.~Pantev and E.~Sharpe, \emph{{String compactifications on Calabi-Yau stacks}}, \href{http://dx.doi.org/10.1016/j.nuclphysb.2005.10.035}{\emph{Nucl. Phys. B} {\bf 733} (2006) 233--296}, [\href{http://arxiv.org/abs/hep-th/0502044}{{\tt hep-th/0502044}}].

\bibitem{Pantev:2005zs}
T.~Pantev and E.~Sharpe, \emph{{GLSM's for Gerbes (and other toric stacks)}}, \href{http://dx.doi.org/10.4310/ATMP.2006.v10.n1.a4}{\emph{Adv. Theor. Math. Phys.} {\bf 10} (2006) 77--121}, [\href{http://arxiv.org/abs/hep-th/0502053}{{\tt hep-th/0502053}}].

\bibitem{Hellerman:2006zs}
S.~Hellerman, A.~Henriques, T.~Pantev, E.~Sharpe and M.~Ando, \emph{{Cluster decomposition, T-duality, and gerby CFT's}}, \href{http://dx.doi.org/10.4310/ATMP.2007.v11.n5.a2}{\emph{Adv. Theor. Math. Phys.} {\bf 11} (2007) 751--818}, [\href{http://arxiv.org/abs/hep-th/0606034}{{\tt hep-th/0606034}}].

\bibitem{Sharpe:2022ene}
E.~Sharpe, \emph{{An introduction to decomposition}},  \href{http://arxiv.org/abs/2204.09117}{{\tt 2204.09117}}.

\bibitem{Seiberg:2010qd}
N.~Seiberg, \emph{{Modifying the Sum Over Topological Sectors and Constraints on Supergravity}}, \href{http://dx.doi.org/10.1007/JHEP07(2010)070}{\emph{JHEP} {\bf 07} (2010) 070}, [\href{http://arxiv.org/abs/1005.0002}{{\tt 1005.0002}}].

\bibitem{Tachikawa:2013hya}
Y.~Tachikawa, \emph{{On the 6d origin of discrete additional data of 4d gauge theories}}, \href{http://dx.doi.org/10.1007/JHEP05(2014)020}{\emph{JHEP} {\bf 05} (2014) 020}, [\href{http://arxiv.org/abs/1309.0697}{{\tt 1309.0697}}].

\bibitem{Sharpe:2014tca}
E.~Sharpe, \emph{{Decomposition in diverse dimensions}}, \href{http://dx.doi.org/10.1103/PhysRevD.90.025030}{\emph{Phys. Rev. D} {\bf 90} (2014) 025030}, [\href{http://arxiv.org/abs/1404.3986}{{\tt 1404.3986}}].

\bibitem{Tanizaki:2019rbk}
Y.~Tanizaki and M.~\"Unsal, \emph{{Modified instanton sum in QCD and higher-groups}}, \href{http://dx.doi.org/10.1007/JHEP03(2020)123}{\emph{JHEP} {\bf 03} (2020) 123}, [\href{http://arxiv.org/abs/1912.01033}{{\tt 1912.01033}}].

\bibitem{Najjar:2025htp}
M.~Najjar, \emph{{Modified instanton sum and 4-group structure in 4d $\mathcal{N}=1$$SU(M)$ SYM from holography}},  \href{http://arxiv.org/abs/2503.17108}{{\tt 2503.17108}}.

\bibitem{Hou:1999qc}
B.-Y. Hou and B.-Y. Hou, \emph{{Differential geometry for physicists}}, vol.~6.
\newblock 1999.

\bibitem{Nakahara:2003nw}
M.~Nakahara, \emph{{Geometry, topology and physics}}.
\newblock 2003.

\bibitem{Hausel:2002xg}
T.~Hausel, E.~Hunsicker and R.~Mazzeo, \emph{{\hyperlink{https://arxiv.org/pdf/math/0207169.pdf}{Hodge cohomology of gravitational instantons}}},  \href{http://arxiv.org/abs/math/0207169}{{\tt math/0207169}}.

\bibitem{e0a630bb-2fb9-3d1a-a046-a4c4cc00dcd5}
J.~Cheeger, \emph{On the spectral geometry of spaces with cone-like singularities}, {\emph{Proceedings of the National Academy of Sciences of the United States of America} {\bf 76} (1979) 2103--2106}.

\bibitem{ASNSP_1985_4_12_3_409_0}
R.~B. Lockhart and R.~C. Mc~Owen, \emph{Elliptic differential operators on noncompact manifolds}, {\emph{Annali della Scuola Normale Superiore di Pisa - Classe di Scienze} {\bf Ser. 4, 12} (1985) 409--447}.

\bibitem{Gomis:2001vk}
J.~Gomis, \emph{{D-branes, holonomy and M theory}}, \href{http://dx.doi.org/10.1016/S0550-3213(01)00247-4}{\emph{Nucl. Phys. B} {\bf 606} (2001) 3--17}, [\href{http://arxiv.org/abs/hep-th/0103115}{{\tt hep-th/0103115}}].

\bibitem{Ganor:1996pe}
O.~J. Ganor, \emph{{A Note on zeros of superpotentials in F theory}}, \href{http://dx.doi.org/10.1016/S0550-3213(97)00311-8}{\emph{Nucl. Phys. B} {\bf 499} (1997) 55--66}, [\href{http://arxiv.org/abs/hep-th/9612077}{{\tt hep-th/9612077}}].

\bibitem{Witten:1996bn}
E.~Witten, \emph{{Nonperturbative superpotentials in string theory}}, \href{http://dx.doi.org/10.1016/0550-3213(96)00283-0}{\emph{Nucl. Phys. B} {\bf 474} (1996) 343--360}, [\href{http://arxiv.org/abs/hep-th/9604030}{{\tt hep-th/9604030}}].

\bibitem{Harvey:1999as}
J.~A. Harvey and G.~W. Moore, \emph{{Superpotentials and membrane instantons}},  \href{http://arxiv.org/abs/hep-th/9907026}{{\tt hep-th/9907026}}.

\bibitem{Braun:2018fdp}
A.~P. Braun, M.~Del~Zotto, J.~Halverson, M.~Larfors, D.~R. Morrison and S.~Sch\"afer-Nameki, \emph{{Infinitely many M2-instanton corrections to M-theory on G$_{2}$-manifolds}}, \href{http://dx.doi.org/10.1007/JHEP09(2018)077}{\emph{JHEP} {\bf 09} (2018) 077}, [\href{http://arxiv.org/abs/1803.02343}{{\tt 1803.02343}}].

\bibitem{deRoo:1997gq}
M.~de~Roo, \emph{{Intersecting branes and supersymmetry}}, \href{http://dx.doi.org/10.1007/BFb0105225}{\emph{Lect. Notes Phys.} {\bf 509} (1998) 18}, [\href{http://arxiv.org/abs/hep-th/9703124}{{\tt hep-th/9703124}}].

\bibitem{Camara:2011jg}
P.~G. Camara, L.~E. Ibanez and F.~Marchesano, \emph{{RR photons}}, \href{http://dx.doi.org/10.1007/JHEP09(2011)110}{\emph{JHEP} {\bf 09} (2011) 110}, [\href{http://arxiv.org/abs/1106.0060}{{\tt 1106.0060}}].

\bibitem{Banks:2010zn}
T.~Banks and N.~Seiberg, \emph{{Symmetries and Strings in Field Theory and Gravity}}, \href{http://dx.doi.org/10.1103/PhysRevD.83.084019}{\emph{Phys. Rev. D} {\bf 83} (2011) 084019}, [\href{http://arxiv.org/abs/1011.5120}{{\tt 1011.5120}}].

\bibitem{Birmingham:1991ty}
D.~Birmingham, M.~Blau, M.~Rakowski and G.~Thompson, \emph{{Topological field theory}}, \href{http://dx.doi.org/10.1016/0370-1573(91)90117-5}{\emph{Phys. Rept.} {\bf 209} (1991) 129--340}.

\bibitem{tHooft:1979rat}
G.~'t~Hooft, \emph{{Naturalness, chiral symmetry, and spontaneous chiral symmetry breaking}}, \href{http://dx.doi.org/10.1007/978-1-4684-7571-5_9}{\emph{NATO Sci. Ser. B} {\bf 59} (1980) 135--157}.

\bibitem{Closset:2024sle}
C.~Closset, E.~Furrer and O.~Khlaif, \emph{{One-form symmetries and the 3d $\mathcal{N}=2$ $A$-model: Topologically twisted indices and CS theories}},  \href{http://arxiv.org/abs/2405.18141}{{\tt 2405.18141}}.

\bibitem{Edelstein:2001pu}
J.~D. Edelstein and C.~Nunez, \emph{{D6-branes and M theory geometrical transitions from gauged supergravity}}, \href{http://dx.doi.org/10.1088/1126-6708/2001/04/028}{\emph{JHEP} {\bf 04} (2001) 028}, [\href{http://arxiv.org/abs/hep-th/0103167}{{\tt hep-th/0103167}}].

\bibitem{Salam:1984ft}
A.~Salam and E.~Sezgin, \emph{{d=8 supergravity}}, \href{http://dx.doi.org/10.1016/0550-3213(85)90613-3}{\emph{Nucl. Phys. B} {\bf 258} (1985) 284--304}.

\bibitem{LassoAndino:2016lwl}
O.~Lasso~Andino and T.~Ort\'\i{}n, \emph{{On gauged maximal $d = 8$ supergravities}}, \href{http://dx.doi.org/10.1088/1361-6382/aaafa9}{\emph{Class. Quant. Grav.} {\bf 35} (2018) 075011}, [\href{http://arxiv.org/abs/1605.09629}{{\tt 1605.09629}}].

\bibitem{Davies:2011is}
R.~Davies, \emph{{Hyperconifold Transitions, Mirror Symmetry, and String Theory}}, \href{http://dx.doi.org/10.1016/j.nuclphysb.2011.04.010}{\emph{Nucl. Phys. B} {\bf 850} (2011) 214--231}, [\href{http://arxiv.org/abs/1102.1428}{{\tt 1102.1428}}].

\bibitem{Najjar:2022eci}
M.~A.~M. Najjar, \emph{{Field Theory Dynamics from M-theory on Special Holonomy Manifolds}}.
\newblock PhD thesis, King's Coll. London, 2022.

\bibitem{Acharya:2021jsp}
B.~Acharya, N.~Lambert, M.~Najjar, E.~E. Svanes and J.~Tian, \emph{{Gauging discrete symmetries of T$_{N}$-theories in five dimensions}}, \href{http://dx.doi.org/10.1007/JHEP04(2022)114}{\emph{JHEP} {\bf 04} (2022) 114}, [\href{http://arxiv.org/abs/2110.14441}{{\tt 2110.14441}}].

\bibitem{Taylor:1999ii}
T.~R. Taylor and C.~Vafa, \emph{{R R flux on Calabi-Yau and partial supersymmetry breaking}}, \href{http://dx.doi.org/10.1016/S0370-2693(00)00005-8}{\emph{Phys. Lett. B} {\bf 474} (2000) 130--137}, [\href{http://arxiv.org/abs/hep-th/9912152}{{\tt hep-th/9912152}}].

\bibitem{Vafa:2000wi}
C.~Vafa, \emph{{Superstrings and topological strings at large N}}, \href{http://dx.doi.org/10.1063/1.1376161}{\emph{J. Math. Phys.} {\bf 42} (2001) 2798--2817}, [\href{http://arxiv.org/abs/hep-th/0008142}{{\tt hep-th/0008142}}].

\bibitem{Curio:2001dz}
G.~Curio, B.~Kors and D.~Lust, \emph{{Fluxes and branes in type II vacua and M theory geometry with G(2) and spin(7) holonomy}}, \href{http://dx.doi.org/10.1016/S0550-3213(02)00404-2}{\emph{Nucl. Phys. B} {\bf 636} (2002) 197--224}, [\href{http://arxiv.org/abs/hep-th/0111165}{{\tt hep-th/0111165}}].

\bibitem{Minasian:2001sq}
R.~Minasian and D.~Tsimpis, \emph{{Hopf reductions, fluxes and branes}}, \href{http://dx.doi.org/10.1016/S0550-3213(01)00387-X}{\emph{Nucl. Phys. B} {\bf 613} (2001) 127--146}, [\href{http://arxiv.org/abs/hep-th/0106266}{{\tt hep-th/0106266}}].

\bibitem{Gukov:2002zg}
S.~Gukov, J.~Sparks and D.~Tong, \emph{{Conifold transitions and five-brane condensation in M theory on spin(7) manifolds}}, \href{http://dx.doi.org/10.1088/0264-9381/20/4/306}{\emph{Class. Quant. Grav.} {\bf 20} (2003) 665--706}, [\href{http://arxiv.org/abs/hep-th/0207244}{{\tt hep-th/0207244}}].

\bibitem{PandoZayas:2001iw}
L.~A. Pando~Zayas and A.~A. Tseytlin, \emph{{3-branes on spaces with R x S**2 x S**3 topology}}, \href{http://dx.doi.org/10.1103/PhysRevD.63.086006}{\emph{Phys. Rev. D} {\bf 63} (2001) 086006}, [\href{http://arxiv.org/abs/hep-th/0101043}{{\tt hep-th/0101043}}].

\bibitem{Hosomichi:2005ja}
K.~Hosomichi and D.~C. Page, \emph{{G(2) holonomy, mirror symmetry and phases of N=1 SYM}}, \href{http://dx.doi.org/10.1088/1126-6708/2005/05/041}{\emph{JHEP} {\bf 05} (2005) 041}, [\href{http://arxiv.org/abs/hep-th/0501195}{{\tt hep-th/0501195}}].

\bibitem{Ferrara:1997tx}
S.~Ferrara and M.~Porrati, \emph{{Central extensions of supersymmetry in four-dimensions and three-dimensions}}, \href{http://dx.doi.org/10.1016/S0370-2693(97)01586-4}{\emph{Phys. Lett. B} {\bf 423} (1998) 255--260}, [\href{http://arxiv.org/abs/hep-th/9711116}{{\tt hep-th/9711116}}].

\bibitem{Evslin:2007ux}
J.~Evslin and S.~Kuperstein, \emph{{Trivializing and Orbifolding the Conifold's Base}}, \href{http://dx.doi.org/10.1088/1126-6708/2007/04/001}{\emph{JHEP} {\bf 04} (2007) 001}, [\href{http://arxiv.org/abs/hep-th/0702041}{{\tt hep-th/0702041}}].

\bibitem{Closset:2018bjz}
C.~Closset, M.~Del~Zotto and V.~Saxena, \emph{{Five-dimensional SCFTs and gauge theory phases: an M-theory/type IIA perspective}}, \href{http://dx.doi.org/10.21468/SciPostPhys.6.5.052}{\emph{SciPost Phys.} {\bf 6} (2019) 052}, [\href{http://arxiv.org/abs/1812.10451}{{\tt 1812.10451}}].

\end{thebibliography}\endgroup

\end{document}